\documentclass[12pt]{article} 
\usepackage{epsfig}
\usepackage{a4}
\usepackage{latexsym}
\usepackage{cite}
\usepackage{bm}

\usepackage{pslatex}
\usepackage[latin1]{inputenc}
\usepackage[T1]{fontenc}

\usepackage{color,colordvi}

\def\muRs{\mu_{\mbox{\scriptsize \sc r}}^{\:\!2}}
\def\muFs{\mu_{\mbox{\scriptsize \sc f}}^{\:\!2}}

\def\arp#1{a_{\rm s}^{\,#1}}
\def\bb#1{\beta_{#1}^{}}
\def\gam#1{\gamma_{\:\!#1}^{}}

\def\tI#1{\widetilde{I}_{\:\!#1}(z)}

\def\colour4colour#1{\Blue{#1}}
\def\colour2colour#1{\Red{#1}}

\def\x1{1\!-\!x}

\renewcommand{\theequation}{\thesection.\arabic{equation}}

\newcommand{\hspn}{{\hspace{-4mm}}}
\newcommand{\hspp}{{\hspace{4mm}}}
\newcommand{\beq}{\begin{equation}}
\newcommand{\eeq}{\end{equation}}
\newcommand{\bea}{\begin{eqnarray}}
\newcommand{\eea}{\end{eqnarray}}
\newcommand{\nn}{\nonumber}
\newcommand{\nin}{\noindent}
\newcommand{\MSb}{$\overline{\mbox{MS}}$}
\newcommand{\Nb}{\overline{N}}
\newcommand{\as}{\alpha_{\rm s}}
\newcommand{\ass}{\alpha_{\rm s}^{\:\!2}}
\newcommand{\asth}{\alpha_{\rm s}^{\:\!3}}
\newcommand{\asfo}{\alpha_{\rm s}^{\:\!4}}
\newcommand{\asn}{\alpha_{\rm s}^{\,n}}
\newcommand{\ar}{a_{\rm s}}
\newcommand{\ars}{a_{\rm s}^{\,2}}

\newcommand{\arn}{a_{\rm s}^{\,n}}
\newcommand{\ra}{\rightarrow}
\newcommand{\ep}{\epsilon}
\newcommand{\eps}{\epsilon^{\:\!2}}

\newcommand{\xx}{1\!-\!x}

\def\binom#1#2{{#1 \choose #2}}

\def\frct#1#2{\mbox{\small{$\displaystyle\frac{#1}{#2}$}}}
\def\frkt#1#2{\mbox{\large{$\frac{#1}{#2}\:\!$}}}

\def\ncx#1{{n^{\: #1}_c}}
\def\cax#1{{C^{\: #1}_A}}
\def\cfx#1{{C^{\:#1}_F}}
\def\nfx#1{{n^{\:#1}_{\! f}}}

\def\NMx#1{{N^{\:\!-#1}}}
\def\NPx#1{{N^{\:\!#1}}}

\begin{document}

\setlength{\parskip}{0.3cm}
\setlength{\baselineskip}{0.53cm}

\def\Qs{Q^{\:\! 2}}
\def\qs{q^{\:\! 2}}

\def\mus{\mu^{\:\! 2}}
\def\nc{{n_c}}
\def\ncs{{n_c^{\: 2}}}
\def\nct{{n_c^{\: 3}}}
\def\ncp#1{{n_c^{\: #1}}}
\def\ca{{C_A}}
\def\cas{{C^{\: 2}_A}}
\def\cath{{C^{\: 3}_A}}
\def\cafo{{C^{\: 4}_A}}
\def\cf{{C_F}}
\def\cfs{{C^{\: 2}_F}}
\def\cft{{C^{\: 3}_F}}
\def\cff{{C^{\: 4}_F}}
\def\cfi{{C^{\: 5}_F}}
\def\nf{{n^{}_{\! f}}}
\def\nfs{{n^{\,2}_{\! f}}}
\def\nft{{n^{\,3}_{\! f}}}
\def\nff{{n^{\,4}_{\! f}}}
\def\bo{\beta_0}
\def\bos{\beta^{\: 2}_0}
\def\bot{\beta^{\: 3}_0}

\def\Lx#1{\ln^{\,#1\!} x}

\def\lix{\ln\frct{1}{x}}
\def\lixp#1{\ln^{\,#1\!} \frct{1}{x}}
\def\z#1{{\zeta_{\,#1}^{}}}
\def\zts{{\zeta_{\,3}^{\,2}}}

\def\arp#1{a_{\rm s}^{\,#1}}

\def\fhat#1{\widehat{F}_{#1}}

\def\cat#1{{\mathcal C}_{#1}}
\def\dat#1{{\mathcal D}_{#1}}
\def\floor#1{\lfloor#1\rfloor}

\def\lx{\ln{x}}
\def\pqq{p_{qq}}
\def\ooomx{(1-x)^{-1}}
\def\H(#1){{\rm{H}}_{#1}^{}}
\def\Hh(#1,#2){{\rm{H}}_{#1,#2}^{}}
\def\Hhh(#1,#2,#3){{\rm{H}}_{#1,#2,#3}^{}}
\def\Hhhh(#1,#2,#3,#4){{\rm{H}}_{#1,#2,#3,#4}^{}}
\def\Hhhhh(#1,#2,#3,#4,#5){{\rm{H}}_{#1,#2,#3,#4,#5}^{}}
\def\xSpaceOuterLB{\nn\\&&\vphantom{\Bigg\{}{}}
\def\xSpaceInnerLB{\nn\\[-0.5mm]&&\vphantom{\Bigg\{}\hphantom{+ \lx^n \Bigg[}{}}

%
\begin{titlepage}
%
%
\nin
LTH 1289    \hfill February 2022\\
DESY 21-225 
\vspace{2.0cm}
\begin{center}
\Large
{\bf \boldmath Resummation of small-$\bm x$ double logarithms in QCD: 
 \\[1.5mm]
 inclusive deep-inelastic scattering}\\
\vspace{1.5cm}
\large
J. Davies$^{\, a}$, C.-H. Kom, S. Moch$^{\, b}$ and A. Vogt$^{\, c}$\\
\vspace{1.0cm}
\normalsize
{\it $^a$Department of Physics and Astronomy, University of Sussex \\
\vspace{0.1cm}
Brighton BN1 9HQ, United Kingdom}\\
\vspace{0.6cm}
{\it $^b$II.~Institute for Theoretical Physics, Hamburg University\\
\vspace{0.1cm}
Luruper Chaussee 149, D-22761 Hamburg, Germany}\\
\vspace{0.6cm}
{\it $^c$Department of Mathematical Sciences, University of Liverpool \\
\vspace{0.1cm}
Liverpool L69 3BX, United Kingdom}
\\[2.0cm]
\large
{\bf Abstract}
\vspace{-0.2cm}
\end{center}
%

\nin
We present a comprehensive study of high-energy double logarithms in inclusive 
DIS. They appear parametrically as $\alpha_s^{\,n}\, \ln^{\,2n-k} x$ at the 
$n$-th order in perturbation theory in the splitting functions for the parton 
evolution and the coefficient functions for the hard scattering process, and 
represent the leading corrections at small $x$ in the flavour non-singlet case.
We perform their resummation, in terms of modified Bessel functions, to all 
orders in full QCD up to NNLL accuracy, and partly to N$^3$LL and beyond in the 
large-$n_c$ limit, and provide fixed-order expansions up to five loops. In~the 
flavour-singlet sector, where these double logarithms are sub-dominant at small
$x$ compared to single-logarithmic $\alpha_s^{\,n}\, x^{\,-1}\,\ln^{\,n-k} x$ 
BFKL contributions, we construct fixed-order expansions up to five loops at 
NNLL accuracy in full QCD. The results elucidate the analytic small-$x$ 
structure underlying inclusive DIS results in fixed-order perturbation theory 
and provide important information for present and future numerical and analytic 
calculations of these quantities.

\vspace{7mm}
\end{titlepage}
 
%
\section{Introduction}
\label{sec:intro}
%

Inclusive deep-inelastic lepton-hadron scattering (DIS) is an experimental and 
theoretical reference process for Quantum Chromodynamics (QCD), the theory of 
the strong interaction.
Important information on the parton (quark and gluon) distribution functions
(PDFs) of the proton, in particular, is provided by the dependence of the 
corresponding cross sections or the structure functions $F_a(x,\Qs)$ on the 
Bjorken variable $x$ and on the scale $\Qs = -\:\!\qs$ set by the momentum~$q$ 
of the exchanged (gauge) boson. 
Moreover the scaling violations, i.e., the $\Qs$-dependence of the structure 
functions $F_2$ and $F_3$, facilitate high-precision determinations of the 
strong coupling constant~$\as\:\!$. 
 
Due to their relation to propagator-type Feynman integrals via forward Compton 
amplitudes and the light-cone operator-product expansion (OPE), see, e.g., 
refs.~\cite{3loopN1,3loopN2,Bierenbaum:2009mv,BCKrev15} 
and references therein, structure functions in DIS are particularly well 
suited for analytical high-order computations in massless perturbative QCD. 
Indeed,~the complete third-order contributions to the (initial-state) splitting 
functions governing the evolution of the PDFs were obtained more than fifteen 
years 
ago \cite{MVV3,MVV4} in computations that also provided the 
third-order cross-section projections (\mbox{coefficient} functions) for the 
most important structure functions in spin-averaged DIS \cite{MVV5,MVV6,MVV10}. 
During the past five years those computations have been extended, if only for 
a limited number of Mellin moments, to the fourth order in $\as$ 
\cite{avLL2016,Pij3lowN,4loopN1,4loopN2}; for the lowest moments see also 
refs.~\cite{BKN2nf3,VelizN2,VelizN34,SumRule1,SumRule2}.

The perturbation series for the splitting functions and coefficient functions 
appear to be very well behaved (except for the longitudinal structure 
function $F_L$ \cite{MVV5,avLL2016}) outside the threshold region 
$\xx \ll 1$ and the high-energy limit $x \ra 0$.
With the exception of the diagonal (quark-quark and gluon-gluon) splitting
functions in the standard \MSb\ scheme \cite{PvsCusp,AlBall01,DMS05}, the
splitting and coefficient functions include threshold double
logarithms at all powers of $\,\xx$.
The dominant \mbox{$(\xx)^{-1} \ln^{\,\ell} (\xx)$} contributions to the 
coefficient functions in DIS have been resummed to a high accuracy 
\cite{MVV7,DMV1}, see also ref.~\cite{Rav2006}, in the framework of the 
soft-gluon exponentiation \cite{sgeSt,sgeCT1,sgeMg,sgeCT2,sgeCMNP,sgeCLS}.
The~resummation of the double logarithms has been extended to 
non-negative  powers of $\xx\,$ by analyzing the physical evolution kernels of 
the structure functions \cite{MV3,MV5,SMVV1}, see also ref.~\cite{Grnb2009}, 
and the structure of the `raw' (unfactorized) expressions in dimensional 
regularization \cite{AV2010,ASV,ALPV2}.

The latter approach can be applied also to high-energy double logarithms 
$\sim \asn \: x^{\:\!p} \ln^{\,2n - n_0^{} - k} x\,$ in splitting functions 
and coefficient functions, albeit, at least in its present form, not at all 
\mbox{powers} $p$ of $x$. 
In~particular, the resummation of the dominant $p=-1$ contributions to the 
splitting functions for the final-state parton fragmentation functions and to 
the coefficient functions for semi-inclusive electron-positron annihilation 
(SIA) \cite{Mueller81,Mueller83} have been extended to the $k=2$ 
next-to-next-to-leading logarithmic (N$^2$LL) accuracy in 
refs.~\cite{AV2011,KVY}, see also refs.~\cite{BKKprl,BKKnpb}.
Corresponding $p=0$ results for inclusive DIS were obtained at about the same 
time. While being formally analogous to their SIA counterparts, these results
were not of direct phenomenological relevance, and only one example expression 
was presented at the time \cite{avLL2012}.

Such results become relevant, however, for approximate or exact reconstructions
of higher-order splitting functions and coefficient functions, if they can be 
combined with a sufficient amount of other information, such as a sufficiently 
large number of Mellin moments. 
Due to the development of the {\sc Forcer} program \cite{Forcer} for four-loop 
propagator-type Feynman integrals, this point has now been reached for 
fourth-order corrections; see refs.~\cite{DRUVV,MRUVV,Pij3lowN} for published
results on the splitting functions. 
In fact, in the latter two articles fourth-order predictions of the  
resummations discussed above and of a complementary proposal of 
ref.~\cite{VelizL0} have already been employed and, where feasible, confirmed. 
Hence it is now timely to present, in sufficient detail to assist future 
research, the status of the resummation of small-$x$ double logarithms for 
inclusive DIS.  

The remainder of this article is organized as follows: 
In section 2 we specify our notation and discuss the available formalisms, 
and their limitations, to the resummation of small-$x$ (double) logarithms.
We also briefly indicate how the calculations have been performed.
The results for the splitting function for the evolution of flavour 
differences of sums of quark and antiquark PDFs are presented in section 3.
This is the case for which the two complementary approaches overlap. 
The results include another striking illustration of the phenomenological 
inadequacy of representing the splitting functions, at any relevant $x$, 
solely by a N$^\ell$LL small-$x$ approximation at some fixed $\ell$.

In sections 4 and 5 we present the N$^2$LL predictions for the corresponding 
non-singlet coefficient functions and for the flavour-singlet splitting 
and coefficient functions. 
In view of the findings in section 3, we focus in these sections on 
fourth- and fifth-order predictions and the all-order structure of the 
$x^{\:\!0} \ln^{\:\!k\!} x$ contributions. 
We expect that the former results will become useful in combination with  
large-$x$ information on these functions, while the leading-logarithmic
all-order expressions may provide useful `data' for future research into the 
small-$x$ structure of splitting functions and coefficient functions in DIS. 
We briefly summarize our findings in section~6.
Some additional material that may be useful to future research can be found
in the appendix.

%
\section{Notation, formalism and calculations}
\label{sec:calculation}
%

Disregarding $1/\Qs$ power corrections, the structure functions in DIS can be
generically written as
\beq
\label{FaCP}
  F_a(x,\Qs) \;=\; 
  \left[ \, C_{a,\rm i\:\!}(\ar)\,\otimes\, f_{\rm i}^{}(\Qs) \right]\!(x)
\eeq
in terms of the coefficient functions $\,C_{a,\rm i}(x,\ar)\,$ and the 
corresponding (combinations of) parton distributions $f_{\rm i}^{}(x,\Qs)$. 
Here and below we identify the renormalization and mass-factorization scales
$\muRs$ and $\muFs$ with the physical scale $\Qs$; the dependence on $\muRs$
and $\muFs$ can be readily reconstructed a posteriori, 
see, e.g., sections 2 of refs.~\cite{NV2,NV4}.
$\otimes$ represents the Mellin convolution, given by
\beq
\label{Mtrf}
  [\, a \otimes b \,](x) \;=\; 
  \int_x^1\frct{dz}{z}\: a(z)\, b\left(\frct{x}{z}\right)
\eeq
and its generalization for plus-distributions, which corresponds to a simple 
product in Mellin space.
The scale dependence of the PDFs $f_{\rm i}^{}$ is given by the 
renormalization-group evolution equations
\bea
\label{Evol}
  \frct{d}{d\,\ln \Qs}\:f_{\rm i}^{}(x,\Qs) \;=\; 
  \left[ \, P_{\,\rm ik}^{}(\ar) \,\otimes\, f_{\rm k}^{}(\Qs) \right]\! (x)
\; .
\eea
The coefficient functions $C_a$ in eq.~(\ref{FaCP}) and the splitting functions 
$P_{\,\rm ik}$ in eq.~(\ref{Evol}) can be expanded in powers of the strong 
coupling constant, which we normalize as $\,\ar=\as(\Qs)/(4\pi)$,
\bea
\label{Pexp}
  P\,(x,\ar)\, &\!=\!& \ar\,P^{\,(0)}(x) 
             +\, \arp2\,P^{\,(1)}(x) 
             +\, \arp3\,P^{\,(2)}(x)
           \,+\, \arp4\,P^{\,(3)}(x)
           \;+\; \ldots\;,
\\
\label{Cexp}
  C_a\,(x,\ar) &\!=\!& \quad\!
                       c_a^{\,(0)}(x)
          \;+\; \ar\,  c_a^{\,(1)}(x)
          \;+\; \arp2\,c_a^{\,(2)}(x)
          \;+\; \arp3\,c_a^{\,(3)}(x)
          \;+\; \ldots\;
\eea
with
\beq
\label{c23L0}
  c_{a,\rm i}^{\,(0)}(x) \:=\: \delta_{\:\rm iq} \, \delta(\xx)
  \;\; \mbox{ for }  \;\;  a = 2,\,3
\; , \quad 
  c_{L,\rm i}^{\,(0)}(x) \:=\: 0 
\; ,
\eeq
where $\rm i \,=\, q,\, g$.
Consequently, the terms up to $c^{\:\!(n)}$ and $P^{\:\!(n)}$ form the 
(next-to)$^{\:\!n}$-leading order (N$^{\:\!n}$LO) approximation of perturbative
QCD for $F_2$ and $F_3$, while $c_L^{\:\!(n+1)}$ and $P^{\,(n)}$ are required 
for this accuracy for the longitudinal structure function $F_L\:\!$.

The coefficients in eqs.~(\ref{Pexp}) and (\ref{Cexp}) include high-energy 
double logarithms, with contributions up to $x^{\:\!p} \ln^{\,2n} x\,$ for 
$P^{\,(n)}$ and $c_L^{\,(n+1)}$, and terms up to $x^{\:\!p} \ln^{\,2n+1} x\,$ 
for $c_2^{\,(n+1)\!}$ and $c_3^{\,(n+1)\!}$ at $\,p \geq 0\,$%
\footnote
{These quantities do not include small-$x$ double logarithms at $p = -1$, 
 see refs.~\cite{Jarosz82,CataniFM90,CataniH94} and references therein.}.
Our main approach to the resummation of these logarithms, i.e., to the 
determination of the coefficients of 
$\arn \: x^{\:\!p} \ln^{\,2n - n_0^{} - k} x\,$ contributions to 
eqs.~(\ref{Pexp}) and (\ref{Cexp}), is analogous to that presented in 
ref.~\cite{AV2011} for the case of $p=-1$ in semi-inclusive $e^+ e^-$ 
annihilation.

The primary objects of this resummation are the unfactorized partonic 
structure functions
\beq
\label{FhatFact}
  \widehat{F}(x,\ar,\ep) \;=\; 
  [\, \widetilde{C}(\ar,\ep) \,\otimes\, Z(\ar,\ep) \,](x) \;
\eeq
in dimensional regularization (we use $D = 4-2\:\!\ep$) where, for simplicity, 
the indices labelling different structure functions and parton distributions 
have been suppressed. The functions $\widetilde{C}$ are given by Taylor series 
in $\ep$; the $\ep^{\:\!m}$ terms $c_a^{\,(n,m)}$ including $m$ more powers 
in $\ln x$ than the 4-dimensional ($\ep = 0\:\!$) coefficient 
functions~(\ref{Cexp}). 
The transition functions $Z$ consist only of negative powers of $\ep$ and can
be written in terms of the splitting functions (\ref{Pexp}) and the
expansion coefficients $\beta_n$ of the $D$-dimensional beta function,
\beq
\label{bfctD}
\beta_D(\ar) \;=\; -\,\ep\,\ar     
                 \,-\, \beta_0\,\arp2 \,-\,\beta_1\,\arp3 
                 \,-\, \beta_2\,\arp4 \:-\: \ldots \; ,
\eeq
with \cite{beta0a,beta0b,beta1a,beta1b}
\bea
  \bb0 &\!=\!& \frct{11}{3}\; \ca - \frct{2}{3}\: \nf \; ,
\quad
  \bb1  \;=\; \frct{34}{3}\; \cas - \frct{10}{3}\: \ca \nf - 2\,\cf\nf
\eea
etc where $\,\ca = n_c = 3\,$ and $\,\cf = (n_c^{\,2}-1)/(2n_c) = 4/3\,$ in~QCD.
Here and below, $\nf$ denotes the number of light flavours.

In Mellin $N$-space, the transition functions are related to the splitting
functions by
\beq
\label{PofZ}
  P\;=\; \frct{d\,Z}{d\,\ln\Qs}\:Z^{\,-1}
   \:=\;\beta_{D}(\ar) \:\frct{d\,Z}{d\:\!\ar}\: Z^{\,-1} \; .
\eeq
Using this relation, $Z(N,\ar)$ can be expressed order-by-order in terms of 
the anomalous dimensions
$\gamma_{\,n}(N)\,=\,-P^{\,(n)}(N)$ and~$\beta_{n}$.
The first four orders in $\ar$, allowing for $Z$ and $\gamma_{\,n}$ to be 
matrices, read
\bea
\label{ZofP}
  Z &\!\!=\!\!& 
  1 
  \,+\, \ar \,\frct{1}{\ep}\, \gam0 
  \,+\, \arp2 \Big\{ \frct{1}{2\ep^2}\, (\gam0-\bb0) \gam0 
    + \frct{1}{2\ep}\, \gam1 \Big\} 
\nn\\ && \mbox{\hspn}
  \,+\, \arp3 \Big\{ \frct{1}{6\:\!\ep^3}\, (\gam0-\bb0)(\gam0-2\bb0)\gam0 
    + \frct{1}{6\:\!\ep^2}\Big[(\gam0-2\bb0)\gam1 + 2 (\gam1-\bb1)\gam0\Big]
    +\frct{1}{3\ep}\, \gam2 \Big\}
\nn\\ && \mbox{\hspn}
  \,+\, \arp4 \Big\{ \frct{1}{24\,\ep^4}\,(\gam0-\bb0)(\gam0-2\bb0)(\gam0-3\bb0)
      \gam0 
    + \frct{1}{24\,\ep^3}\, \Big[(\gam0-2\bb0)(\gam0-3\bb0)\gam1
\nn\\ && \mbox{\hspp\hspp}
    + 2 (\gam0-3\bb0)(\gam1-\bb1)\gam0 + 3(\gam1-2\bb1)(\gam0-\bb0)\gam0\Big] 
\nn\\[-0.5mm] && \mbox{\hspp} 
  \,+\, \frct{1}{24\,\ep^2}\, \Big[ 2 (\gamma_0-3\bb0)\gam2 
    + 3 (\gam1-2\bb1)\gam1 + 6 (\gam2-\bb2)\gam0\Big] 
    + \frct{1}{4\ep}\, \gam3\Big\} 
  \:+\: \ldots \:\:.
\eea
The higher-order contributions have been generated using {\sc Form} and 
{\sc TForm}\cite{FORM3,TFORM,FORM4} to a sufficiently high order for the
computations of this paper.
At order $\arn$, the dependence of $Z$ on $\beta_m$ and $\gamma_{\,m}^{}$ can 
be summarized as
\beq
\label{ZofNkLO}
  \ep^{\,-n}:\: \gam0 ,\: \beta_0 \:, \quad
  \ep^{\,-n+1}: \:\ldots\,,\, \gam1 ,\: \beta_1 \:, \quad 
  \ep^{\,-n+2}: \:\ldots\,,\, \gam2 ,\: \beta_2 \:, \quad
  \; \ldots \; , \quad
  \ep^{\,-1}:\: \gamma_{\,n-1} \:.
\eeq
Hence fixed-order knowledge at N$^m$LO (i.e., of the splitting functions to 
$P^{\,(m)}$, beta function to $\beta_m$ and the corresponding coefficient 
functions) fixes the first $m\!+\!1$ coefficients in the $\ep$-expansion of 
$\widehat{F}$ at all orders in~$\ar$. 
Furthermore, the property  $P^{\,(n)}\sim\,\ln^{\:\!2n\!}x \;\Leftrightarrow
\; \gamma_{\,n}^{} \sim N^{\,-2n-1}$ means that $\beta_0$ and 
$\beta_0^2$ enter eq.~(\ref{ZofP}) at the next-to-leading logarithmic (NLL) 
and N$^2$LL level, while $\beta_1$ contributes only from the fourth (N$^3$LL)
logarithms. 
Similarly, $\beta_2$ enters only at N$^5$LL accuracy and beyond.

If the above N$^m$LO knowledge can be extended to all powers in $\ep$ at a
given logarithmic accuracy, e.g., to all coefficients of 
  $\arn\, \ep^{-n+k}\, N^{\,-n-k+\ell}$ 
or 
  $\arn\, \ep^{-n+k}\, \ln^{\,n+k-1-\ell\!}x\,$ 
for $\widehat{F}_2$ in eqs.~(\ref{FhatFact}) and (\ref{ZofP}) for a fixed 
$\ell$, then we arrive at an all-order resummation of these terms, e.g., of 
the N$^\ell$LL $x^{\,0}$~contributions to the splitting functions and 
coefficient functions contributing to $F_2$.

This situation is completely analogous to that of the large-$x$ double
logarithms in refs.~\cite{AV2010,ASV,ALPV2}. In that case, the structure that 
allows the extensions to all $\ep$ has been inferred from the calculations of 
inclusive DIS via suitably projected gauge-boson parton cross sections as 
carried out at two loops in refs.~\cite{ZvN1,ZvN2,ZvN4}.
The same strategy can be applied here.
The maximal ($\:\!2 \ra n+1$ particles) phase space for these processes at 
order $\as^{\,n}$ can be schematically written as
\cite{Matsuura:1987wt,Matsuura:1988sm}
\beq
\label{PSn}
 \left(\, \frct{\x1}{x} \,\right)^{\!-n\,\ep} 
 {\int_0^1} d(3\:\!n\!-\!1 \mbox{ other variables}) \: f(x,\,\ldots) \:\:.
\eeq
If the integrals for the $n$-th order purely real (tree graph) contributions
$\widehat{F}_{a,\rm i}^{\:\!(n)\rm R}$ do not lead to any further factors 
$x^{\,\ep}$, their expansion (for $a \neq L$) around $x=0$ can be written as
\beq
\label{FhatReal}
  \widehat{F}_{a,\rm i}^{\,(n)\rm R} \;=\; x^{\,n\,\ep}\, \sum_{p}\:
  x^{\:\!p} \:{1 \over \ep^{2n-1} } \:
  \Big\{
           R_{a,{\rm i},p}^{\,(n)\rm LL} \:+\:
    \ep\,  R_{a,{\rm i},p}^{\,(n)\rm NLL} \:+\:
    \eps\, R_{a,{\rm i},p}^{\,(n)\rm NNL} \:+\: \ldots
  \Big\} \; .
\eeq
If, furthermore, the mixed real-virtual contributions ($\:\!2 \ra r+1$ particles
with $\,n-r \geq 1$ loops) include no more than $n-r$ additional factors of 
$x^{\,\ep}$ from the loop integrals, then we arrive at 
\beq
\label{FhatAllx}
  \widehat{F}_{a,\rm i}^{\,(n)}(x) {\Big|}_{\mbox{$\:\!x^{\:\!p}$}} \;=\; 
  \frac{1}{\ep^{2n-n_a}}\: 
  \sum_{l=1}^{l_a} \, x^{\,\ep l} \:
  \Big\{
           A_{a,{\rm i},p}^{(n,l)} \:+\: 
    \ep\,  B_{a,{\rm i},p}^{(n,l)} \:+\: 
    \eps\, C_{a,{\rm i},p}^{(n,l)} + \:\ldots 
  \Big\} 
\eeq
with $n_a=1$ and $l_a=n$ for all structure functions considered here, with
the exception of $F_L$ for which $n_a=3$ and $l_a=n-1$. 
By expanding $x^{\,\ep l}$ in powers of $\ep l\ln x$, it can be seen that 
$A$, $B$ and $C$ are the coefficients of the LL, NLL and NNLL 
($\equiv$ N$^2$LL)
contributions, respectively.
For a given value of $p$, the $N$-space counterpart of eq.~(\ref{FhatAllx})
reads 
\beq
\label{FhatAllN}
  \widehat{F}_{a,\rm i}^{\,(n)}(N,p) \;=\;
  \frac{1}{\ep^{2n-n_a}}\:
  \sum_{l=1}^{l_a} \: \frac{1}{N+p+\ep l} \:
  \Big\{
           A_{a,{\rm i},\,p}^{(n,l)} \:+\: 
    \ep\,  B_{a,{\rm i},\,p}^{(n,l)} \:+\: 
    \eps\, C_{a,{\rm i},\,p}^{(n,l)} + \:\ldots 
  \Big\} \;.
\eeq

Since eq.~(\ref{ZofP}) includes only poles up to $\ep^{-n}$ at order $\arn$,
the terms  with $\ep^{\,-2n+1}, \:\dots\,,\: \ep^{\,-n-1}$ have to cancel
(for $a\,\neq\,L$) in the sums (\ref{FhatAllx}) and (\ref{FhatAllN}).
Hence there are $\,n\!-\!1$ `zero' relations between the LL coefficients 
$A^{(n,l)}$, $\,n\!-\!2$ relations between the NLL coefficients $B^{(n,l)}$ 
etc. Moreover, as mentioned above, the N$^m$LO results provide the 
(non-vanishing) coefficients of $\ep^{\,-n}, \:\dots\,,\: \ep^{\,-n+m}$ at all 
orders $n$, and thus $m+1$ additional relations between the coefficients in
eqs.~(\ref{FhatAllx}) and (\ref{FhatAllN}). 
Consequently the highest $m\!+\!1$ double logarithms, i.e. the N$^m$LL 
approximation, can be determined and, except for the N$^m$LL terms at order 
$a_{\rm s}^{\,m+1}$, over-constrained order-by-order from the N$^m$LO results. 
This feature also holds for $a\,=\,L$ but (due to $c_{L,\rm i}^{\,(0)}(x) = 0$) 
with only $\,n\!-\!2$ `zero' relations but also one term fewer in the above 
sums.

Using the known unfactorized N$^2$LO expressions for the structure functions, 
it is now possible to establish, to all orders, for which cases
eqs.~(\ref{FhatAllx}) and (\ref{FhatAllN}) hold.
It~turns out that these equations, and hence the resulting resummation of
small-$x$ double logarithms, are applicable at  
\bea
\label{evenp}
  \mbox{even } p : && \hspn 
  \frkt{1}{x}\, F_{2,L} \; \mbox{ for~ e.m., $\nu+\bar{\nu}\,$ DIS} 
\,,\:\:\:
  F_3 \; \mbox{ for~$\, \nu-\bar{\nu}\,$ DIS} 
\,,\:\:\:
  F_\phi
\; ,
\\
\label{oddp}
  \mbox{odd~ } p : && \hspn
  \frkt{1}{x}\, F_{2,L} \; \mbox{ for~$\,\nu-\bar{\nu}\,$ DIS}
\;, \hspace*{1.1cm}
  F_3 \; \mbox{ for~$\,\nu+\bar{\nu}\,$ DIS}
\,,\:\:\;
  g_1^{} \;\: \mbox{for~e.m.~DIS etc}
\; .
\eea
Here `e.m.' (electromagnetic) denotes photon exchange, and $F_\phi$ is the 
structure function for DIS via the exchange of a scalar that, like the Higgs 
boson in the heavy-top limit, couples directly only to gluons. 
The~third-order coefficient functions for this structure function, which is 
experimentally irrelevant but theoretically useful, have been presented in 
ref.~\cite{SMVV1}; the second-order results have also been obtained in 
ref.~\cite{Daleo:2009yj}. For the coefficient functions for $\nu-\bar{\nu}$
charged-current DIS see refs.~\cite{MochRogal07,RogalMV,DMVV-DIS16}. For
completeness, we have included the most important structure function $g_1^{}$
in spin-dependent DIS, for its coefficient functions and splitting functions
at NNLO see refs.~\cite{ZvNg1,RMVV,MVVpol,Bpol1,Bpol2}.

It is worthwhile to note that all structure functions in eq.~(\ref{evenp}) 
are accessible only at even~$N$ via forward Compton amplitudes or the OPE; 
conversely all structure functions in eq.~(\ref{oddp})  are odd-$N$ based 
(for a detailed discussion see, e.g., ref.~\cite{MochRogal07}) 
-- a fact that can hardly be a mere coincidence. 
Moreover, the differences between the splitting functions and coefficient
functions for the flavour non-singlet $\nu+\bar{\nu}$ and $\nu-\bar{\nu}$
structure functions vanish in the limit of a large number of colours~$n_c$,
see refs.~\cite{MVV3,Broadhurst:2004jx,RogalMV,DRUVV,MRUVV}.

The structural difference between these two cases, and its suppression at
large $n_c$, can already be seen from the $p=0\,$ LL resummation of the 
corresponding even-$N$ and odd-$N$ splitting functions $P_{\rm ns}^{\,+}$ and 
$P_{\rm ns}^{\,-}$ for the flavour differences of quark-antiquark sums and 
differences \cite{Kirschner:1983di,BV1995,BV1996a}:
\bea
\label{PnsPto0LL}
  P_{\rm ns,LL}^{\,+} (N,\ar) &\!=\!& \frct{N}{2}\:
  \Big\{ 1 -
    \Big( 1 - \frct{8\:\! \ar C_F}{N^2}
    \Big)^{\!-1/2\,}
  \Big\}
\;\; ,
\\[1mm]
\label{PnsMto0LL}
  P_{\rm ns,LL}^{\,-} (N, \ar) &\!=\!& \frct{N}{2}\:
  \Big\{ 1 -
    \Big( 1 - \frct{8\:\! \ar C_F}{N^2}
      \Big[ 1 - \frct{8\:\! \ar n_c}{N} \: \frct{d}{dN} \,
        \ln \Big ( e^{\:z^{\,2\!}/4} \, D_{-1/[2n_c^{\,2}]}(z) \Big )
            \Big]
    \Big)^{\!-1/2\,}
  \Big\}
\quad
\eea
where $z = N \,(2\, \ar n_c)^{-1/2}$, and $D_{\rm p}(z)$ denotes a parabolic
cylinder function~\cite{Gradsht}. Note that the expansion of 
eq.~(\ref{PnsMto0LL}) in powers of $\ar$ is an asymptotic series, in contrast 
to eq.~(\ref{PnsPto0LL}). 

In ref.~\cite{VelizL0} a surprisingly simple generalization has been proposed
of the equation, first derived in ref.~\cite{Kirschner:1983di}, that leads to 
eq.~(\ref{PnsPto0LL}). This generalization can be stated as
\beq
\label{Pnsto0SL}
  P_{\rm ns}^{\,+}(N,\ar) \,\left( P_{\rm ns}^{\,+}(N,\ar)
  - N + \beta(\ar) / \ar \right) \;=\;  O(1)
\eeq
up to terms that are large-$n_c$ suppressed and include even-$n$ values 
$\zeta_n^{}$ of Riemann's $\zeta$-function.%
\footnote{$\,$This form of the limitation is a conservative all-order 
  extension of that given in ref.~\cite{VelizL0}. 
  See refs.~\cite{JM-no-pi2,DV-no-pi2,BC-no-pi2} and references therein for 
  another context in which the even-$n$ values $\zeta_n^{}$, i.e., 
  powers of $\pi^2$, play a special role.}
Inserting the Laurent expansion
\beq
\label{PexpN0}
  P_{\rm ns}^{\,(n)+}(N) \:=\;\: 
  \sum_{k\,=\,0}^{k_{{\:\!\rm max}_{}}} \: N^{-2n-1+k} \, p_{n,k}^{\,+}  
\eeq
about $N=0$ into eq.~(\ref{Pnsto0SL}), one can readily solve this relation to
`any' desired order $n$ for the coefficients $p_{n,k}^{\,+}$ with $k \leq 2n-1$
that correspond to powers of $\,\ln x\,$ in the small-$x$ expansion.
Specifically, the coefficients $p_{n,0}^{\,+}$ to $p_{n,2\:\!m+1}^{\,+}$ can be
predicted (with the above restriction) at all $n > m$ from eq.~(\ref{Pnsto0SL})
with $k_{\:\!\rm max} = 2\:\!m+1$ if $P_{\rm ns}^{\,+}(N)$ is completely known 
to N$^{\:\!m}$LO. 
So far these predictions have been verified for the $\nfs$ and $\nft$ terms
and the complete large-$n_c$ limit at N$^3$LO \cite{DRUVV,MRUVV}.

The above two approaches to the resummation of small-$x$ terms overlap for the 
$x^{\:\!0} \ln^{\:\!k\!}x$ double logarithms of $P_{\rm ns}^{\,+}(x,\as)$, but 
are largely complementary otherwise. The predictions of eq.~(\ref{FhatAllN})
cover far more than just $x^{\:\!0}$ part of $P_{\rm ns}^{\,+}(x,\as)$, 
while eq.~(\ref{Pnsto0SL}) is very powerful in this specific case, in 
particular in the large-$n_c$ limit. 

Both approaches require Laurent expansions of the fixed-order input 
quantities, including non-negative powers of $N+p$, as written down at 
$p=0$ for $P_{\rm ns}^{\,(n)+}(N)$ in eq.~(\ref{PexpN0}).
These expansions can be obtained, for example, by expanding the exact 
$x$-space expressions in terms of harmonic polylogarithms (HPLs) \cite{HPLs}, 
using the {\sc Harmpol} package for {\sc Form} \cite{FORM3} together with
\beq
\label{Mlog}
  {\rm M} \left[ \,x^{\:\!p}\: \ln^{\,k\!} \left(\, \frct{1}{x} \right) 
  \right](N) 
  \;\: \equiv \;\:
  \int_0^1 \! dx\: x^{\,N-1+p}\, \ln^{\,k\!} \left(\, \frct{1}{x} \right) 
  \;\: = \;\:
  \frac { k! }{(N+p)^{k+1} } 
\:\;\; .
\eeq
An easy extension to the coefficients of non-negative powers of $N\!+\!p$ is 
to transform the functions to $N$-space harmonic sums \cite{HSums1,HSums2}, 
multiply by a sufficiently large power $s$ of $1/(N\!+\!p)$, transform back 
to $x$-space, proceed as above, and finally multiply by $(N\!+\!p)^{s}$. 
Routines for the Mellin transform of the HPLs and its inverse are also 
provided by the {\sc Harmpol} package.
For the convenience of the reader, the $p=0$ coefficients employed in this
article are collected in appendix A.

The resummation predictions for the splitting and coefficient functions can 
then be computed order by order in $\as$. Using {\sc Form}
and {\sc TForm} \cite{FORM3,TFORM,FORM4}, this has been done up to order
$\as^{\:\!30}$ and $\as^{\:\!60}$, respectively, for the flavour singlet
and non-singlet cases. Using the formal similarity to the SIA cases 
covered in refs.~\cite{AV2011,KVY}, these results can then be employed to 
infer their generating functions via over-constrained systems of linear 
equations, thus arriving at all-order expressions.

%
\setcounter{equation}{0}
\section{Results for the non-singlet splitting functions}
\label{sec:resultPns}
%

Non-singlet quantities are dominated at small $x$ by their $x^{\:\!p}\: 
\ln^{\,k\!}x$ contributions with $p=0$ and $k \geq 0$ which correspond to poles 
at $N=0$ in Mellin space. These terms can be resummed via eq.~(\ref{FhatAllN}) 
for the splitting function $P_{\rm ns}^{\,(n)+}$ which enters the structure 
functions in eq.~(\ref{evenp}). The resummed $N$-space expressions can be
expressed in terms of 
\beq
\label{Sdef}
   S \,=\, (1-4\,\xi)^{1/2}
\:\: \mbox{ with } \quad 
   \xi \;=\; \frct{2\:\!\cf\:\!\ar}{N^2} 
       \;\equiv\; \frct{\cf\:\!\as}{2 \pi\, N^{\:\!2}} 
\;\: .
\eeq
The N$^{\,2}$LL result for $P_{\rm ns}^{\,(n)+}$, already presented in 
ref.~\cite{avLL2012}, can be written as 
\bea
\label{PnsPres}
  P_{\,\rm ns}^{\,+}(N,\ar) \!\!&\!=\!& \!\!\mbox{}
  - \frct{1}{2}\,\* N \*(S-1) \,+\,
  \frct{1}{2} \,\* \ar \* (2\,\*\cf-\bo) \* (S^{\,-1}-1)
\nn\\ & & \!\!\mbox{}
  + \frct{1}{96\,\*\cf}\, \* \ar \* N\*\, \Big\{ \big( 
    [156 - 960\*\,\z2]\,\*\cfs
  - [80 - 1152\*\,\z2]\,\*\ca\*\cf - 360\*\,\z2\,\*\cas 
\quad \nn\\[-1mm] && \mbox{\hspp} \vphantom{\frct{1}{1}}
  - 100\,\*\bo\*\cf + 3\*\bos \big) \*(S-1)
  \,+\, 2\,\* \big( [12 - 576\*\,\z2]\,\*\cfs + [40 + 576\*\,\z2]\,\*\ca\*\cf
\nn\\[-0.5mm] && \mbox{\hspp}
  - 180\*\,\z2\,\*\cas
  + 56\,\*\bo\,\*\cf - 3\*\bos \big) \* (S^{\,-1}-1)
  \,+\, 3\,\*(2\,\*\cf - \bo)^2\*(S^{\,-3}-1)\Big\}
\;, \qquad
\eea
where $\bo$ in eq.~(\ref{bfctD}) has been used instead of $\nf$ for a more 
compact representation.
The two terms in the first line of eq.~(\ref{PnsPres}) provide the LL and 
NLL parts; the former agrees, of course, with the earlier result in
eq.~(\ref{PnsPto0LL}) above. 
The remaining three lines represent the N$^{\,2}$LL contribution.

The expansion of eq.~(\ref{PnsPres}) in powers of $\ar$ yields the N$^3$LO 
and N$^4$LO contributions
\bea
\label{PnsP3nn}
  P_{\,\rm ns}^{\,+(3)}(N) \!&\!=\!&
    80\,\*\,\cff\,\*N^{\,-7} 
  + 80\,\*\cft\* \left( 2\,\*\cf - \bo \right)\* N^{\,-6}
  + 8\,\*\cfs\,\* \Big( [16-200\*\,\z2]\*\cfs 
\nn\\[-0.5mm] && \mbox{\hspn}
    + 10\,\*\cf\*\bo + [20 + 192\*\,\z2]\*\cf\*\ca 
    + 3\*\bos - 60\*\,\z2\,\*\cas \Big)\,\* N^{\,-5}
 \:+\: {\cal O} (N^{\,-4\,})
%
\eea
and
\bea
\label{PnsP4nn}
  P_{\,\rm ns}^{\,+(4)}(N) \!&\!=\!&
    448\,\*\cfi\,\*N^{\,-9} 
  + 560\,\*\cff\* \left( 2\,\*\cf - \bo \right)\* N^{\,-8} 
  + 80\,\*\cft\* \left( \vphantom{\frct{1}{1}}
     \left[ 16 - 148\*\z2 \right]
    \*\cfs\right.
\nn\\[-0.5mm] && \left. \mbox{\hspn} 
    + \frct{8}{3}\,\*\cf\*\bo + \left[ \frct{40}{3} + 144\,\*\z2 \right]
    \*\cf\*\ca + 3\*\bos - 45\,\*\z2\,\*\cas\right)\*N^{\,-7}
 \:+\: {\cal O} (N^{\,-6\,})
\eea
to the moments of eq.~(\ref{Pexp}). 
The $\nfs$ part of eq.~(\ref{PnsP3nn}) and its complete large-$n_c$ limit
have been employed in refs.~\cite{DRUVV,MRUVV}, respectively, as constraints
in the determination of the all-$N$ expressions from a limited number of 
moments and endpoint constraints. 
Conversely, the verification of those all-$N$ expressions 
-- via results at 
 higher $N$ and independent form-factor calculations 
 \cite{cusp4a,cusp4b,cusp4c,cusp4d,cusp4e}
 that include the large-$N$ limit of $P_{\,\rm ns}^{\,+(3)}$, the (light-like) 
 four-loop cusp anomalous dimension \cite{PvsCusp,AlBall01} --
provides a stringent check of eq.~(\ref{PnsP3nn}).

Using the N$^3$LO results \cite{DRUVV,MRUVV} together with the corresponding 
coefficient function for $F_2$ in ref.~\cite{MVV6}, eqs.~(\ref{PnsPres})
-- (\ref{PnsP4nn}) can be extended to N$^3$LL small-$x$ accuracy for the
next-to-leading large-$\nf$ terms and in the large-$n_c$ limit.
The latter results will be presented below, together with the predictions of
eq.~(\ref{Pnsto0SL}) at this and higher orders in the small-$x$ expansion.

The structure of the closed-form $N$-space expression (\ref{PnsPres}) is 
similar to the `non-singlet' $p=-1$ part of the `time-like' splitting function 
$P^{\,T}_{gg}(N)$ for final-state fragmentation functions.  
The crucial difference is the sign of $\xi$ in eq.~(\ref{Sdef}) which leads 
to qualitative differences. 
In $x$-space, the latter splitting function can be expressed \cite{KVY} in 
terms of Bessel functions which exhibit an oscillatory behaviour in the 
small-$x$ limit. 
In fact, the resummation is found to completely remove the huge small-$x$ 
spikes present in the fixed-order results for the time-like splitting functions
\cite{PTnnlo1,PTnnlo2}.

In the present `space-like' (initial-state) case, on the other hand, the 
N$^2$LL $x$-space expression is given by
\bea
\label{pns-x}
  P_{\,\rm ns}^{\,+}(x,\as) &\!=\!&
  2\,\*\ar\,\*\cf\,\* 
  \Big\{ 
    1 \,+\, (2\,\*\cf-\bo) \,\* \ar\,\*\lix 
    \,+\, \frct{1}{2}\,\* (2\,\*\cf-\bo)^2 \*\ars \* \lixp2\, 
  \Big\} 
  \*\: \tI1
\nn \\[0.5mm] & & \mbox{\hspn}
  + 2\,\*\ar\,\*\cf \* \Big\{\, 
      \frct{1}{3} \,\* (11\*\bo + 10\,\*\ca - 6\,\*\cf) 
    - 4\,\*\cf\*\z2\Big\}\*\,\ar \*\: \tI0 
\nn \\[0.5mm] & & \mbox{\hspn}
  + 2\,\*\ar\,\*\cf\* \Big\{\, 
      8\,\*\cfs 
    - 2\,\*\z2 \* (\,15\,\*\cas - 48\,\*\cf\*\ca + 44\,\*\cfs \,)
  \Big\}\,\* \ars\*\lixp2 \;\*  \tI2
\eea
with 
\beq
\label{zdef}
   z \;=\; ( 8\,\cf\,\ar )^{1/2}\, \ln \frct{1}{x}
\eeq
and
\beq
\label{Idef}
   \tI{n} \;\equiv\; \left( \,\frct{2}{z}\, \right)^{\!n} I_n (z)
   \; = \; \sum_{k=0}^\infty\: \frct{1}{k! \, (n+k)!}\, 
   \left( \,\frct{z}{2}\, \right)^{\!2\:\!k}
\eeq
in terms of modified Bessel functions $I_n(z)$, see section 9.6 of 
ref.~\cite{AbrSteg}. 
The first two terms in the curly bracket in the first line are the LL and 
NLL results, respectively; the remaining terms provide the N$^{\,2}$LL 
contribution.  The latter can be written in different ways due to the 
recurrence relation expressing $I_{n+1}$ in terms of $I_n$ and $I_{n-1}$. 
The form chosen above yields the most compact coefficients and is in line with
our basis choice for the higher-accuracy large-$n_c$ expressions below.
 
The functions $I_n(z)$ have the exponential form 
  $(2\:\!\pi\,z)^{-1/2} \,e^{\,z}\, \left(1+{\mathcal O}(z^{\,-1})\right)$
in the large-$z$ limit, and thus dwarf the fixed-order small-$x$ behaviour in a 
very unstable manner: 
the LL result is positive, the NLL contribution is negative (for the physically
relevant case $\bb0 > 2\:\!\cf)$, the N$^2$LL `correction' is positive etc.
It is interesting to note that the coefficients of the (for $x \ra 0$ at 
fixed $\ar$) asymptotically dominant 
   $ \ar(\ar\ln x)^\ell\: \tI1$ 
terms in the first line of eq.~(\ref{pns-x}), which correspond to the
 $S^{-2\ell+1}$ parts of the N$^{\:\!\ell\:\!}$LL contributions in 
eq.~(\ref{PnsPres}), seem to point towards a `second resummation', a feature 
already observed in ref.~\cite{KVY} for the time-like splitting functions. 
We~will return to this issue with more `data' below, see eqs.~(\ref{pnsL-x})
and (\ref{pnsL-xr}).

As discussed in section 2, the contributions up to N$^5$LL 
-- with the exception of large-$n_c$ suppressed terms containing $\z2$ -- 
can be obtained from eq.~(\ref{Pnsto0SL}) \cite{VelizL0} and
$P_{\,\rm ns}^{\,+}$ up to three loops \cite{MVV3}.
The resulting coefficients at N$^3$LO (four loops), defined in 
eq.~(\ref{PexpN0}), are given by
\bea
\label{p33p}
  p_{3,3}^{\,+} &\!=\!&
         \cfx4 \, \*  \left[\, 212 + 640\, \* \z3 \right]
       - \ca\, \* \cfx3 \, \*  \left[\, 20 + 288\, \* \z3 \right]
       - \frct{ 13060 }{ 9 } \, \*  \cax2\, \* \cfx2
       - \frct{ 2662 }{ 27 } \, \*  \cax3\, \* \cf
\nn \\[0.5mm] && \mbox{\hspn}
       + 192\, \* \z2 \, \*  \ncx3\, \* \cf
       + 32 \, \*  \nf\, \* \cfx3
       + \frct{ 4184 }{ 9 } \, \*  \nf\, \* \ca\, \* \cfx2
       + \frct{ 484 }{ 9 } \, \*  \nf\, \* \cax2\, \* \cf
       - 48\, \* \z2 \, \*  \nf\, \* \ncx2\, \* \cf
\nn \\[0.5mm] && \mbox{\hspn}
       - \frct{ 304 }{ 9 } \, \*  \nfx2\, \* \cfx2
       - \frct{ 88 }{ 9 } \, \*  \nfx2\, \* \ca\, \* \cf
       + \frct{ 16 }{ 27 } \, \*  \nfx3\, \* \cf
\; , \\
\label{p34p}
  p_{3,4}^{\,+} &\!=\!& \mbox{}
       - \cfx4 \, \*  \left[\, 224 + 512\, \* \z3 \right]
       + \ca\, \* \cfx3 \, \*  \left[ 196 + \frct{ 832 }{ 3 }\, \* \z3 \right]
       + \cax2\, \* \cfx2 \*  \left[\, \frct{ 229480 }{ 81 } 
         + 64\, \* \z3 \right]
\nn \\[0.5mm] && \mbox{\hspn}
       + \frct{ 50006 }{ 81 } \, \*  \cax3\, \* \cf
       - \ncx3\, \* \cf \*  \left[\, \frct{ 7682 }{ 9 }\, \* \z2 
         - 236\, \* \z4 \right]
       - \nf\, \* \cfx3 \*  \left[\, \frct{ 340 }{ 3 } 
         - \frct{ 512 }{ 3 }\, \* \z3 \right]
\nn \\[0.5mm] && \mbox{\hspn}
       - \nf\, \* \ca\, \* \cfx2 \*  \left[\, \frct{ 65936 }{ 81 }
         + 128\, \* \z3 \right]
       - \frct{ 2780 }{ 9 } \, \*  \nf\, \* \cax2\, \* \cf
       + \frct{ 1552 }{ 9 }\, \* \z2 \, \*  \nf\, \* \ncx2\, \* \cf
\nn \\[0.5mm] && \mbox{\hspn}
       + \frct{ 4288 }{ 81 } \, \*  \nfx2\, \* \cfx2
       + \frct{ 1288 }{ 27 } \, \*  \nfx2\, \* \ca\, \* \cf
       - \frct{ 80 }{ 9 }\, \* \z2 \, \*  \nfx2\, \* \nc\, \* \cf
       - \frct{ 176 }{ 81 } \, \*  \nfx3\, \* \cf
\; , \\[3mm]
\label{p35p}
  p_{3,5}^{\,+} &\!=\!&
         \cfx4 \, \*  \left[\, 130 + 944\, \* \z3 - 1920\, \* \z5 \right]
       + \ca\, \* \cfx3  \*  \left[ \frct{ 2761 }{ 3 } 
         + \frct{ 12448 }{ 3 }\, \* \z3 + 960\, \* \z5 \right]
\nn \\[0.5mm] && \mbox{\hspn}
       - \cax2\, \* \cfx2 \*  \left[\, \frct{ 254225 }{ 81 } 
         + \frct{ 8984 }{ 3 }\, \* \z3 - 240\, \* \z5 \right]
       - \cax3\, \* \cf \*  \left[\, \frct{ 146482 }{ 81 } 
         - 264\, \* \z3 \right]
\nn \\[0.5mm] && \mbox{\hspn}
       + \ncx3\, \* \cf \*  \left[\, \frct{ 12221 }{ 9 }\, \* \z2 
         - \frct{ 1312 }{ 3 }\, \* \z4 - 48\, \* \z2\, \* \z3 \right]
       - \nf\, \* \cfx3 \*  \left[\, \frct{ 500 }{ 3 } 
         + \frct{ 2080 }{ 3 }\, \* \z3 \right]
\nn \\[0.5mm] && \mbox{\hspn}
       + \nf\, \* \ca\, \* \cfx2 \*  \left[\, \frct{ 90538 }{ 81 } 
         + \frct{ 1328 }{ 3 }\, \* \z3 \right]
       + \nf\, \* \cax2\, \* \cf \, \*  \left[\, \frct{ 64481 }{ 81 } 
         + \frct{ 32 }{ 3 }\, \* \z3 \right]
\nn \\[0.5mm] && \mbox{\hspn}
       - \nf\, \* \ncx2\, \* \cf \*  \left[\, \frct{ 4006 }{ 9 }\, \* \z2 
         - \frct{ 328 }{ 3 }\, \* \z4 \right]
       - \frct{ 7736 }{ 81 } \, \*  \nfx2\, \* \cfx2
       - \nfx2\, \* \ca\, \* \cf \*  \left[\, \frct{ 7561 }{ 81 } 
         + \frct{ 32 }{ 3 }\, \* \z3 \right]
\nn \\[0.5mm] && \mbox{\hspn}
       + \frct{ 272 }{ 9 }\, \* \z2\, \* \nfx2\, \* \nc\, \* \cf 
       + \frct{ 64 }{ 27 } \, \*  \nfx3\, \* \cf
\eea
up to large-$n_c$ suppressed contributions with (powers of) $\z2$ which may
include quartic group invariants.
The large-$n_c$ limits of eqs.~(\ref{p33p}) -- (\ref{p35p}) have been employed
(and verified, recall the discussion below eq.~(\ref{PnsP4nn})) together with
that of eq.~(\ref{PnsP3nn}) in ref.~\cite{MRUVV}.
The $\nfs$ terms agree with eq.~(4.14) of ref.~\cite{DRUVV} which, unlike 
eqs.~(\ref{p33p}) -- (\ref{p35p}), includes all $\z2$ contributions. 
The $\nft$ terms in eqs.~(\ref{p33p}) -- (\ref{p35p}), which are part of the 
leading large-$\nf$ limit derived to all orders in refs.~\cite{Lnf1,Lnf2}, 
agree with eq.~(4.17) of ref.~\cite{DRUVV}.
 
The corresponding N$^4$LO (five-loop) coefficients read
\bea
\label{p43p}
  p_{4,3}^{\,+} &\!=\!&
         \cfx5 \, \*  \left[\, 1840 + 4480\, \* \z3 \right]
       + \ca\, \* \cfx4 \*  \left[\, \frct{ 7120 }{ 9 } - 1920\, \* \z3 \right] 
       - \frct{ 112000 }{ 9 } \, \*  \cax2\, \* \cfx3
\nn \\[0.5mm] && \mbox{\hspn}
       - \frct{ 53240 }{ 27 } \, \*  \cax3\, \* \cfx2
       + 960\, \* \z2 \, \*  \ncx4\, \* \cf
       + \frct{ 35360 }{ 9 } \, \*  \nf\, \* \ca\, \* \cfx3
       + \frct{ 9680 }{ 9 } \, \*  \nf\, \* \cax2\, \* \cfx2
\nn \\[0.5mm] && \mbox{\hspn}
       + \frct{ 1280 }{ 9 } \, \*  \nf\, \* \cfx4
       - 240\, \* \z2 \, \*  \nf\, \* \ncx3\, \* \cf
       - \frct{ 2560 }{ 9 } \, \*  \nfx2\, \* \cfx3
       - \frct{ 1760 }{ 9 } \, \*  \nfx2\, \* \ca\, \* \cfx2
       + \frct{ 320 }{ 27 } \, \*  \nfx3\, \* \cfx2
\; , \qquad \\[3mm]
\label{p44p}
  p_{4,4}^{\,+} &\!=\!& \mbox{}
       - \cfx5 \, \*  \left[\, 656 + 512\, \* \z3 \right]
       + \ca\, \* \cfx4 \*  \left[\, \frct{ 9008 }{ 9 } 
         - \frct{ 13376 }{ 3 }\, \* \z3 \right]
       + \frct{ 324896 }{ 27 } \, \*  \cax3\, \* \cfx2
\nn \\[0.5mm] && \mbox{\hspn}
       + \cax2\, \* \cfx3 \*  \left[\, \frct{ 559624 }{ 27 } 
         + 2496\, \* \z3 \right]
       + \frct{ 29282 }{ 81 } \, \*  \cax4\, \* \cf
       - \ncx4\, \* \cf \*  \left[\, \frct{ 43478 }{ 9 }\, \* \z2 
         - 1240\, \* \z4 \right]
\nn \\[0.5mm] && \mbox{\hspn}
       - \nf\, \* \cfx4 \*  \left[\, \frct{ 4376 }{ 9 } 
         - \frct{ 5888 }{ 3 }\, \* \z3 \right]
       - \nf\, \* \ca\, \* \cfx3 \*  \left[\, \frct{ 164816 }{ 27 } 
         + 1152\, \* \z3 \right]
       - \frct{ 54304 }{ 9 } \, \*  \nf\, \* \cax2\, \* \cfx2
\nn \\[0.5mm] && \mbox{\hspn}
       - \frct{ 21296 }{ 81 } \, \*  \nf\, \* \cax3\, \* \cf
       + \frct{ 10544 }{ 9 }\, \* \z2 \, \*  \nf\, \* \ncx3\, \* \cf
       + \frct{ 11776 }{ 27 } \, \*  \nfx2\, \* \cfx3
       + 960 \, \*  \nfx2\, \* \ca\, \* \cfx2
\nn \\[0.5mm] && \mbox{\hspn}
       + \frct{ 1936 }{ 27 } \, \*  \nfx2\, \* \cax2\, \* \cf
       - \frct{ 256 }{ 3 }\, \* \z2 \, \*  \nfx2\, \* \ncx2\, \* \cf
       - \frct{ 1280 }{ 27 } \, \*  \nfx3\, \* \cfx2
       - \frct{ 704 }{ 81 } \, \*  \nfx3\, \* \ca\, \* \cf
       + \frct{ 32 }{ 81 } \, \*  \nfx4\, \* \cf
\; , \\[3mm]
\label{p45p}
  p_{4,5}^{\,+} &\!=\!&
         \cfx5 \, \*  \left[\, 500 + 5280\, \* \z3 - 12800\, \* \z5 \right]
       + \ca\, \* \cfx4 \*  \left[\, \frct{ 85946 }{ 9 } 
         + \frct{ 379168 }{ 9 }\, \* \z3 + 6400\, \* \z5 \right]
\nn \\[0.5mm] && \mbox{\hspn}
       - \cax2\, \* \cfx3 \*  \left[\, \frct{ 204556 }{ 9 } 
         + \frct{ 73984 }{ 3 }\, \* \z3 - 1440\, \* \z5 \right]
       - \cax3\, \* \cfx2 \*  \left[\, \frct{ 2795072 }{ 81 } 
         - 1232\, \* \z3 \right]
\nn \\[0.5mm] && \mbox{\hspn}
       - \frct{ 624118 }{ 243 } \, \*  \cax4\, \* \cf
       + \ncx4\, \* \cf \*  \left[\, \frct{ 321290 }{ 27 }\, \* \z2
         - 256\, \* \z3\, \* \z2 - 3948\, \* \z4 \right]
\nn \\[0.5mm] && \mbox{\hspn}
       - \nf\, \* \cfx4 \, \*  \left[ \frct{ 16676 }{ 9 } 
         + \frct{ 56128 }{ 9 }\, \* \z3 \right]
       + \nf\, \* \ca\, \* \cfx3 \*  \left[\, \frct{ 79970 }{ 9 } 
         + \frct{ 7232 }{ 3 }\, \* \z3 \right]
       + \frct{ 423940 }{ 243 } \, \*  \nf\, \* \cax3\, \* \cf
\nn \\[0.5mm] && \mbox{\hspn}
       + \nf\, \* \cax2\, \* \cfx2 \*  \left[\, \frct{ 142534 }{ 9 } 
         + 832\, \* \z3 \right]
       - \nf\, \* \ncx3\, \* \cf \*  \left[\, \frct{ 39236 }{ 9 }\, \* \z2 
         - 984\, \* \z4 \right]
\nn \\[0.5mm] && \mbox{\hspn}
       - \nfx2\, \* \cfx3 \*  \left[\, \frct{ 7516 }{ 9 } 
         - \frct{ 512 }{ 3 }\, \* \z3 \right]
       - \nfx2\, \* \ca\, \* \cfx2 \*  \left[\, \frct{ 59554 }{ 27 } 
         + 192\, \* \z3 \right]
       + \frct{ 3584 }{ 9 }\, \* \z2 \, \*  \nfx2\, \* \ncx2\, \* \cf
\\[0.5mm] && \mbox{\hspn}
       - \frct{ 34304 }{ 81 } \, \*  \nfx2\, \* \cax2\, \* \cf
       + \frct{ 7424 }{ 81 } \, \*  \nfx3\, \* \cfx2
       + \frct{ 10384 }{ 243 } \, \*  \nfx3\, \* \ca\, \* \cf
       - \frct{ 224 }{ 27 }\, \* \z2 \, \*  \nfx3\, \* \nc\, \* \cf
       - \frct{ 352 }{ 243 } \, \*  \nfx4\, \* \cf
\nn \; . 
\eea
The large-$n_c$ limit of eq.~(\ref{p43p}) can also been also obtained via 
eqs.~(\ref{FhatFact}) to (\ref{FhatAllN}) using the N$^3$LO result 
$P_{\,\rm ns}^{\,+(3)}$ of ref.~\cite{MRUVV}. 

Using the all-$N$ large-$n_c$ expression for $P_{\,\rm ns}^{\,+(3)}$, it is
possible to predict also the $N^{\,-3}$ and $N^{\,-2}$ (N$^6$LL and N$^7$LL) 
contributions of $P_{\,\rm ns}^{\,+(4)}$ in this limit. 
The corresponding coefficients read
\bea
\label{p46p}
  p_{4,6}^{\,+} &\!=\!&
    \cf\, \* \ncx4 \*  \left[\, \frct{ 83997239 }{ 1944 } 
     - \frct{ 2253859 }{ 81 }\, \* \z2 + \frct{ 13220 }{ 9 }\, \* \z3 
     + \frct{ 48070 }{ 3 }\, \* \z4 + 64\, \* \z3\, \* \z2 
\right. \nn \\[0.5mm] && \left. \mbox{}
     - \frct{ 4312 }{ 3 }\, \* \z5 + 176\, \* \zts - 1444\, \* \z6 \right]
  \,+\, \nf\, \* \cf\, \* \ncx3 \*  \left[  - \frct{ 5138330 }{ 243 } 
     + \frct{ 267860 }{ 27 }\, \* \z2 
\right. \nn \\[0.5mm] && \left. \mbox{}
     - \frct{ 4336 }{ 3 }\, \* \z3 
     - \frct{ 28432 }{ 9 }\, \* \z4 + 128\, \* \z3\, \* \z2 
     + \frct{ 1504 }{ 3 }\, \* \z5 - 64\, \* \zts - 248\, \* \z6 \right]
\nn \\[0.5mm] && \mbox{\hspn}
  + \nfx2\, \* \cf\, \* \ncx2 \*  \left[\, \frct{ 760669 }{ 243 } 
     - \frct{ 9076 }{ 9 }\, \* \z2 + \frct{ 2000 }{ 9 }\, \* \z3 
     + \frct{ 1408 }{ 9 }\, \* \z4 \right] 
\nn \\[0.5mm] && \mbox{\hspn}
  + \nfx3\, \* \cf\, \* \nc \*  \left[  - \frct{ 12826 }{ 81 } 
     + \frct{ 2656 }{ 81 }\, \* \z2 - \frct{ 64 }{ 9 }\, \* \z3 \right]
  + \frct{ 128 }{ 81 } \, \*  \nfx4\, \* \cf
\; , \\[3mm]
\label{p47p}
  p_{4,7}^{\,+} &\!=\!&
    \cf\, \* \ncx4 \*  \left[ - \frct{ 141282997 }{ 2592 } 
     + \frct{ 12219019 }{ 324 }\, \* \z2 - \frct{ 125756 }{ 81 }\, \* \z3
     - \frct{ 655423 }{ 27 }\, \* \z4 - \frct{ 5488 }{ 3 }\, \* \z2\, \* \z3 
\right. \nn \\[0.5mm] && \left. \mbox{\hspn}
     + \frct{ 64174 }{ 9 }\, \* \z5 - \frct{ 2056 }{ 3 }\, \* \zts 
     + \frct{ 1874 }{ 3 }\, \* \z6 - 576\, \* \z2\, \* \z5 
     + 1104\, \* \z3\, \* \z4 - 1288\, \* \z7 \right]
\nn \\[0.5mm] && \mbox{\hspn}
  + \nf\, \* \cf\, \* \ncx3 \*  \left[\, \frct{ 2035745 }{ 72 } 
     - \frct{ 141241 }{ 9 }\, \* \z2 + \frct{ 197588 }{ 81 }\, \* \z3 
     + \frct{ 71642 }{ 9 }\, \* \z4 - \frct{ 512 }{ 3 }\, \* \z2\, \* \z3 
\right. \nn \\[0.5mm] && \left. \mbox{}
     - \frct{ 13784 }{ 9 }\, \* \z5 - 64\, \* \z2\, \* \z5 
     + 32\, \* \z3\, \* \z4 + \frct{ 944 }{ 3 }\, \* \zts + 368\, \* \z6
     - 112\, \* \z7 \right]
\nn \\[0.5mm] && \mbox{\hspn}
  + \nfx2\, \* \cf\, \* \ncx2 \*  \left[ - \frct{ 1032713 }{ 243 } 
     + \frct{ 47984 }{ 27 }\, \* \z2 - \frct{ 18992 }{ 27 }\, \* \z3 
     - \frct{ 4744 }{ 9 }\, \* \z4 + 96\, \* \z2\, \* \z3 
     + \frct{ 1072 }{ 9 }\, \* \z5 
\right. \nn \\[0.5mm] && \left. \mbox{}
     - \frct{ 64 }{ 3 }\, \* \zts 
     - \frct{ 248 }{ 3 }\, \* \z6 \right] 
  + \nfx3\, \* \cf\, \* \nc \*  \left[\, \frct{ 41497 }{ 243 } 
     - \frct{ 3376 }{ 81 }\, \* \z2 + \frct{ 3008 }{ 81 }\, \* \z3 
     - \frct{ 176 }{ 27 }\, \* \z4 \right]
\nn \\[0.5mm] && \mbox{\hspn}
  - \nfx4\, \* \cf \*  \left[\, \frct{ 128 }{ 243 } 
     - \frct{ 64 }{ 81 }\, \* \z3 \right]
\; .
\eea
The $\cf \nfx4$ terms in eqs.~(\ref{p43p}) -- (\ref{p47p}) agree with
refs.~\cite{Lnf1,Lnf2}; all other contributions have not been presented 
before.

\begin{figure}[t]
\vspace*{-1mm}
\centerline{\epsfig{file=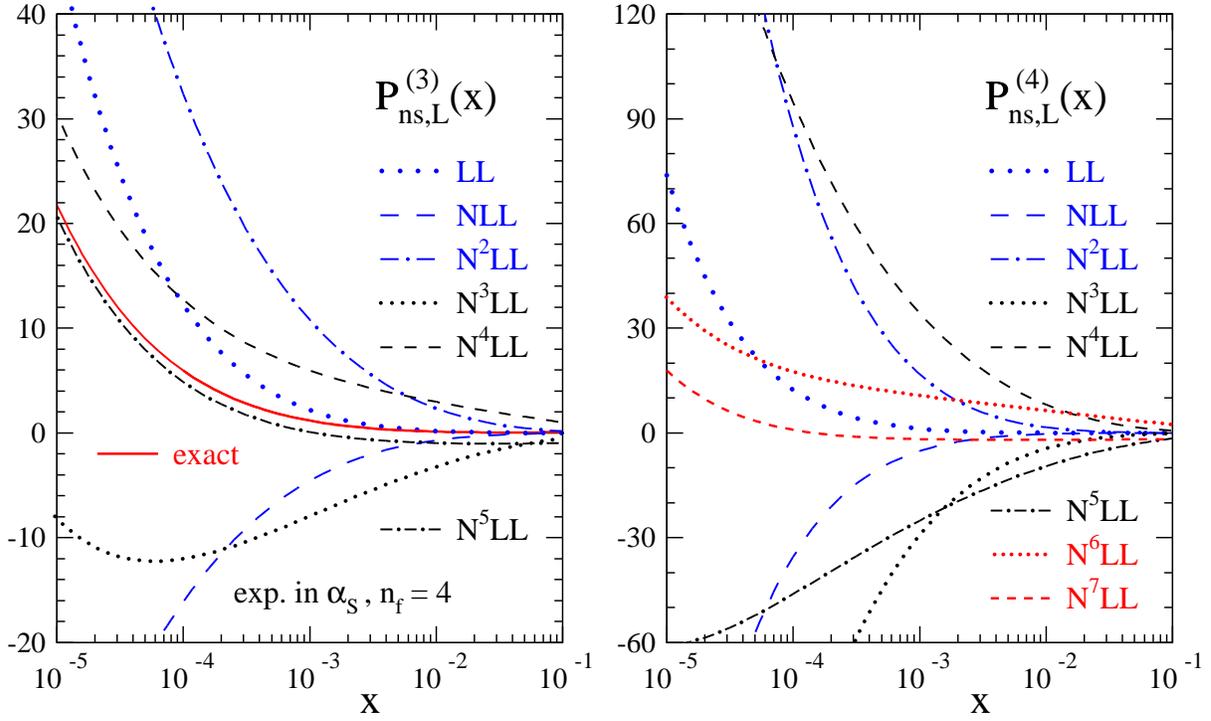,width=16cm,angle=0}}
\vspace{-2mm}
\caption{ \small
 Successive small-$x$ approximations to the N$^n$LO four-flavour splitting
 functions $P_{\rm ns,\,\rm L}^{\,(n)}(x)$ in the large-$\nc$ limit for
 $n=3$ (left), compared to the exact result, and $n=4$ (right). 
 The respective N$^5$LL and N$^{\,7}$LL curves include all terms with 
 $x^{\:\!0} \ln^{\,k} x$ at $k>0$ as specified by eqs.~(\ref{PnsP3nn}), 
 (\ref{PnsP4nn}) and (\ref{p33p}) -- (\ref{p47p}).}
\vspace{-1mm}
\end{figure}

After transformation to $x$-space using eq.~(\ref{Mlog}),
the above results provide all small-$x$ enhanced contributions to the 
non-singlet splitting functions $P_{\rm ns,\,\rm L}^{\,\pm}(x,\as)$ 
in the large-$\nc$ limit at four and five loops. 
These results are shown in fig.~1 for an expansion in powers of $\as$, 
not $\,\ar = \as /(4\pi)$ as used in all formulae, at $\nf = 4$ flavours.
As seen before for the 3-loop splitting functions, see figs.~2 and 3
of ref.~\cite{MVV3}, and coefficient functions, see figs.~7 and 9 of 
ref.~\cite{MVV6} -- see also ref.~\cite{BvN98} -- all logarithms are 
needed for a meaningful approximation at any physically relevant small 
values of~$x$.
The N$^{\,7}$LL small-$x$ limit of $P_{\rm ns,\,\rm L}^{\,(4)}(x)$ 
has already been employed in the first estimate of the 5-loop quark cusp 
anomalous dimension in the large-$\nc$ limit in ref.~\cite{Pns4lowN}. 

In this context, it is instructive to consider the generalization of 
the NNLL small-$x$ resummation (\ref{pns-x}), which can be extended to 
N$^7$LL accuracy in the large-$\nc$ (L) limit:
\bea
\label{pnsL-x}
 \lefteqn{ P_{\,\rm ns,L}^{\,\pm}(x,\as)/(2\,\ar\,\cf) \; = \; 
    \Big\{ 1 \,-\, (\bo - \nc) \,\* \ar\,\*\lix \Big\} \,\*\tI1
 } \nn \\[0.5mm] & & \mbox{\hspn}
  \,+\, \frct{1}{2}\,\* (\bo - \nc)^2 \,\*\arp2 \* \lixp2\,\*\:\tI1
  \,+\, \frct{1}{3}\* (11\,\*\bo + 13\,\*\nc - 18\,\*\z2\,\*\nc) 
        \,\*\ar \*\:\tI0
  \,+\, 2\,\*(2\,\*\z2 -1) \,\*\nc \,\* \ar\,\*\:\tI1
\nn \\[1.5mm] & & \mbox{\hspn}
  \,-\, \frct{1}{6}\,\* (\bo - \nc)^3 \,\*\arp3 \* \lixp3\,\* \tI1
  \,-\, \frct{1}{3}\,\* (11\,\*\bo + 13\,\*\nc - 18\,\*\z2\,\*\nc) 
    \* (\bo - \nc) \,\*\arp2 \* \lix \,\*\:\tI0
 \nn \\ & & \mbox{}
  \,-\, \frct{1}{12}\* ( 136\,\*\bo - 115\,\*\nc ) 
        \,\*\nc \*\arp2\*\lix\,\*\:\tI1
  \,+\, 2\,\*(2\,\*\z3 -1) \,\*\nct\,\*\arp3 \* \lixp3\, \*\:\tI3
\nn \\[1.5mm] & & \mbox{\hspn}
  \,+\, \frct{1}{24}\,\* (\bo - \nc)^4 \,\*\arp4\,\* \lixp4\,\* \tI1
  \,+\, \frct{1}{6}\,\* (11\,\*\bo + 13\,\*\nc - 18\,\*\z2\,\*\nc)
    \* (\bo - \nc)^2 \,\*\arp3 \* \lixp2 \,\*\:\tI0
 \nn \\[0.5mm] & & \mbox{}
  \,+\, \frct{1}{36}\,\*
    \Big( \bos \,\* ( 686 - 72\,\*\z2 )
    - \nc\,\*\bo\,\* ( 181 + 648\,\*\z2 + 144\,\*\z3 )
 \nn \\ & & \mbox{\hspp\hspp}
    + \ncp2 \,\* (647 - 1008\,\*\z2 + 144\,\*\z3 + 1620\,\*\z4 )
    \Big) 
    \,\* \nc\,\* \arp3\,\* \lixp2 \,\*\:\tI1
 \nn \\ & & \mbox{}
  \,+\, \frct{1}{72}\,\*
    \Big( 288\,\*\bos 
    + \nc\,\*\bo\,\* ( 3811 - 912\,\*\z2 + 576\,\*\z3 )
 \nn \\ & & \mbox{\hspp\hspp}
    - \ncp2 \,\* ( 4357 + 852\,\*\z2 + 1584\,\*\z3 - 1008\,\*\z4 )
    \Big) \,\* \arp2 \,\*\:\tI0 
 \nn \\ & & \mbox{}
  \,+\, \frct{2}{3}\,\*
    \Big( \bo \,\* ( 5 - 10\,\*\z2 + 12\,\*\z3 )
        + \nc \,\* ( 4 - 2\,\*\z2 - 6\,\*\z3 )
    \Big) 
    \,\* \ncp2\,\* \arp3\,\* \lixp2 \,\*\:\tI2
 \nn \\ & & \mbox{}
  \,-\, 2\,\* ( 2\,\*\z4 - 1)\,\*\ncp4\,\* \arp4\,\* \lixp4 \,\*\:\tI4
\nn \\[1.5mm] & & \mbox{\hspn}
  \,-\, \frct{1}{120}\,\* (\bo - \nc)^5 \,\*\arp5 \* \lixp5\,\* \tI1
  \,-\, \frct{1}{18}\,\* (11\,\*\bo + 13\,\*\nc - 18\,\*\z2\,\*\nc)
    \* (\bo - \nc)^3 \,\*\arp4 \* \lixp3 \,\*\:\tI0
 \nn \\[0.5mm] & & \mbox{}
  \,-\, \frct{1}{72}\,\*
    \Big( \bos \,\* ( 940 - 96\,\*\z2 )
    + \nc\,\*\bo\,\* ( 367 - 1392 \,\*\z2 - 144\,\*\z3 )
 \nn \\ & & \mbox{\hspp\hspp}
    + \ncp2 \,\* (997 - 1968\,\*\z2 + 144\,\*\z3 + 3240\,\*\z4 )
    \Big) 
    \,\* (\bo - \nc)\,\* \nc\,\* \arp4\,\* \lixp3 \,\*\:\tI1
 \nn \\ & & \mbox{}
  \,-\, \frct{1}{72}\,\*
    \Big( 288\,\*\bot 
    + \nc\,\*\bos\,\* ( 7427 - 2448\,\*\z2 + 864\,\*\z3 )
    - \ncp2\,\*\bo \,\* ( 6730 + 5508\,\*\z2 + 3504\,\*\z3 
 \nn \\ & & \mbox{\hspp\hspp}
    - 5040\,\*\z4 )
    + \ncp3\,\* ( 1175 + 6336\,\*\z2 + 336\,\*\z3 - 5040\,\*\z4 
        + 1728\,\*\z2\,\*\z3 )
    \Big) \,\* \arp3\,\* \lix\,\*\:\tI0 
 \nn \\ & & \mbox{}
  \,-\, \frct{1}{36}\,\*
    \Big( 417\,\*\bos 
    + \nc\,\*\bo\,\* ( 2725 - 288\,\*\z2 + 168\,\*\z3  + 432\,\*\z4 )
 \nn \\ & & \mbox{\hspp\hspp}
    - \ncp2\,\* ( 4507 - 1236\,\*\z3 + 432\,\*\z4 + 2160\,\*\z5 )
    \Big) \,\*\,\nc\,\* \arp3\,\* \lix\,\*\:\tI1 
 \nn \\ & & \mbox{}
  \,-\, \frct{2}{3}\,\*
    \Big( \bo \,\* ( 2 - 10\,\*\z3 + 18\,\*\z4 )
        - \nc \,\* ( 2 - 12\,\*\z2 - 10\,\*\z3 + 9\,\*\z4 + 24\,\*\z2\,\*\z3 )
    \Big) 
    \,\* \ncp3\,\* \arp4\,\* \lixp3 \,\*\:\tI3
 \nn \\ & & \mbox{}
  \,+\, 2\,\* ( 2\,\*\z5 - 1)\,\*\ncp5\,\* \arp5\,\* \lixp5 \,\*\:\tI5
\nn \\[1.5mm] & & \mbox{\hspn}
  \,+\, \frct{1}{720}\,\* (\bo - \nc)^6 \,\*\arp6 \* \lixp6\,\* \tI1
  \,+\, \frct{1}{72}\,\* (11\,\*\bo + 13\,\*\nc - 18\,\*\z2\,\*\nc)
    \* (\bo - \nc)^4 \,\*\arp5\,\* \lixp4 \,\*\:\tI0
 \nn \\[0.5mm] & & \mbox{}
  \,+\; \ldots \; 
  \,-\, 2\,\* ( 2\,\*\z6 - 1)\,\*\ncp6\,\* \arp6\,\* \lixp6 \,\*\:\tI6
\nn \\[1.0mm] & & \mbox{\hspn}
  \,-\, \frct{1}{5040}\,\* (\bo - \nc)^7 \,\*\arp7\,\* \lixp7\,\* \tI1
  \,-\, \frct{1}{360}\,\* (11\,\*\bo + 13\,\*\nc - 18\,\*\z2\,\*\nc)
    \* (\bo - \nc)^5 \,\*\arp6 \* \lixp5 \,\*\:\tI0
 \nn \\[0.5mm] & & \mbox{}
  \,+\; \ldots \; 
  \,+\, 2\,\* ( 2\,\*\z7 - 1)\,\*\ncp7\,\* \arp7\,\* \lixp7 \,\*\:\tI7
\eea
where we have omitted (with `$\ldots$') the lengthy and (at least in the 
present notation) `irregular' parts of the N$^6$LL and N$^7$LL coefficients
for brevity; the ancillary file of this article provides the complete 
expression.
The first two terms at each logarithmic order, which dominate at 
asymptotically small $x$ for a fixed $\as$, can be seen to build up 
an exponential function, viz
\bea
\label{pnsL-xr}
 \lefteqn{ 
   P_{\,\rm ns,L}^{\,\pm}(x,\as)/(2\,\ar\,\cf) \; = \; 
 } \nn \\[-0.5mm] & & \quad 
 \exp \left( - (\bo - \nc)\,\*\ar \* \lix \,\right) \* 
 \left( \tI1 + 
   \frct{1}{3}\,\* (11\,\*\bo + 13\,\*\nc - 18\,\*\z2\,\*\nc)\,\* \ar\*\,\tI0
 \right)
 \: + \; \ldots \:\; .
\eea
This exponential prefactor dampens, and from some unphysically large value
of $\as$ overwhelms, the small-$x$ rise of the modified Bessel functions 
(\ref{Idef}).

So far we have presented the 
$1/N^{\:\!n+1} \,\Leftrightarrow\, x^{\:\!0} \ln^{\:\!n} x$ 
results arising from eqs.~(\ref{FhatAllN}) and (\ref{Pnsto0SL}). 
We~now turn to the double logarithmic
$1/(N\!+\!p)^{n+1} \,\Leftrightarrow\, x^{\:\!p} \ln^{\:\!n} x$ contributions 
which, as discussed in section 2, can be analyzed in a completely analogous 
manner at even $p$ for $P_{\,\rm ns}^{\,+}$ and odd $p$ for 
$P_{\,\rm ns}^{\,-}$, and hence at all integer $p$ for their common 
large-$\nc$ limit $P_{\,\rm ns,L}^{\,\pm}$. 
 
Thus the complete Taylor expansions of the coefficients of $\ln^{\,k} x$ 
can be determined at $k = 6,\, 5,\, 4$ for $P_{\,\rm ns,L}^{\,(3)}$ and at 
$k = 2n, \,\ldots ,\, 2n-3$ for $P_{\,\rm ns,L}^{\,(n)}$ at $n \geq 4$.
In practice, we have performed the necessary computations to $p=70$, which 
was more than sufficient to overconstrain the analytic expressions in terms
of harmonic polylogarithms \cite{HPLs} for which an ansatz with up
to 54 coefficients was used.
This partial reconstruction of the analytic form of $P_{\,\rm ns,L}^{\,\pm}$
can be carried out to any order in $\as$; here we confine ourselves to the
N$^3$LO and N$^4$LO expressions. The former is given by
\bea
\label{Pns3Lx}
&&
  \hspn P_{\,\rm ns,L}^{\,(3)}(x) \;=\;
\xSpaceOuterLB
   + \Lx6  \* \Bigg[
       \ncp3 \,\* \cf  \,\*  \Bigg\{
          \frct{5}{24} \,\* \Big(1
             - \frct{16}{15} \,\* \ooomx
             + x\Big)
          \Bigg\}\Bigg]
\xSpaceOuterLB
   + \Lx5  \* \Bigg[
       \ncp3 \,\* \cf  \,\*  \Bigg\{
          - \frct{4}{3} \,\* \pqq \,\* \H(1)
          + \frct{22}{9} \,\* \Big(1
             - \frct{13}{11} \,\* \ooomx
             + \frct{17}{11} \,\* x\Big)
          \Bigg\}
\xSpaceInnerLB
       + \ncp2 \,\* \cf \,\* \nf  \,\*  \Bigg\{
          - \frct{7}{9} \,\* \Big(1
             - \frct{8}{7} \,\* \ooomx
             + x\Big)
          \Bigg\}\Bigg]
\xSpaceOuterLB
   + \Lx4  \* \Bigg[
       \ncp3 \,\* \cf  \,\*  \Bigg\{
          - \frct{8}{3} \,\* \pqq \,\* \Hh(1,1)
          - \frct{2}{3} \,\* \Big(1
             - 4 \,\* \ooomx
             + x\Big) \,\* \Hh(0,1)
          + \frct{58}{9} \,\* \Big(1
             - \frct{70}{29} \,\* \ooomx
\xSpaceInnerLB
             + \frct{41}{29} \,\* x\Big) \,\* \H(1)
          - \frct{1}{6} \,\* \Big([227
             - 112 \,\* \z2] \,\* \ooomx
             - [251
             - 98 \,\* \z2] \,\* x
             - \frct{1}{3} \,\* [463
             - 294 \,\* \z2]\Big)
          \Bigg\}
\xSpaceInnerLB
       + \ncp2 \,\* \cf \,\* \nf  \,\*  \Bigg\{
          \frct{16}{9} \,\* \pqq \,\* \H(1)
          - \frct{65}{9} \,\* \Big(1
             - \frct{92}{65} \,\* \ooomx
             + \frct{93}{65} \,\* x\Big)
          \Bigg\}
\xSpaceInnerLB
       + \nc \,\* \cf \,\* \nfx2  \,\*  \Bigg\{
          \frct{2}{3} \,\* \Big(1
             - \frct{4}{3} \,\* \ooomx
             + x\Big)
          \Bigg\}\Bigg]
\xSpaceOuterLB
   + \Lx3  \* \Bigg[
       \ncp3 \,\* \cf  \,\*  \Bigg\{
          \frct{56}{3} \,\* \pqq \,\* \Hhh(1,0,1)
          + \frct{160}{3} \,\* \pqq \,\* \Hhh(1,1,1)
          + \frct{4}{3} \,\* \Big(1
             - 4 \,\* \ooomx
             + x\Big) \,\* \Hhh(0,0,1)
\xSpaceInnerLB
          - \frct{50}{9} \,\* \Big(1
             - \frct{94}{25} \,\* \ooomx
             + \frct{109}{25} \,\* x\Big) \,\* \Hh(0,1)
          - \frct{88}{3} \,\* \Big(1
             - \frct{14}{11} \,\* \ooomx
             + x\Big) \,\* \Hhh(0,1,1)
\xSpaceInnerLB
          + \frct{160}{9} \,\* \Big(1
             + \ooomx
             - 2 \,\* x\Big) \,\* \Hh(1,1)
          + \frct{1}{81} \,\* \Big([2987
             - 4212 \,\* \z2
             - 756 \,\* \z3]
             + \frct{1}{2} \,\* [38641
\xSpaceInnerLB
             - 18360 \,\* \z2
             - 1512 \,\* \z3] \,\* x
             - \frct{1}{4} \,\* [55291
             - 27216 \,\* \z2
             - 3456 \,\* \z3] \,\* \ooomx\Big)
\xSpaceInnerLB
          + \frct{1}{54} \,\* \Big([3977
             - 3312 \,\* \z2]
             - 2 \,\* [4745
             - 3312 \,\* \z2] \,\* \ooomx
             + [5513
             - 3312 \,\* \z2] \,\* x\Big) \,\* \H(1)
          \Bigg\}
\xSpaceInnerLB
       + \ncp2 \,\* \cf \,\* \nf  \,\*  \Bigg\{
          - \frct{80}{9} \,\* \pqq \,\* \Hh(1,1)
          - \frct{338}{27} \,\* \Big(1
             - \frct{362}{169} \,\* \ooomx
             + \frct{193}{169} \,\* x\Big) \,\* \H(1)
\xSpaceInnerLB
          + \frct{56}{9} \,\* \Big(1
             - \frct{10}{7} \,\* \ooomx
             + x\Big) \,\* \Hh(0,1)
          + \frct{1}{9} \,\* \Big([725
             - 288 \,\* \z2] \,\* \ooomx
\xSpaceInnerLB
          - \frct{1}{3} \,\* [851
             - 648 \,\* \z2]
             - \frct{1}{3} \,\* [2257
             - 648 \,\* \z2] \,\* x\Big)
          \Bigg\}
       + \cf \,\* \nfx3  \,\*  \Bigg\{
          \frct{8}{81} \,\* \pqq
          \Bigg\}
\xSpaceInnerLB
       + \nc \,\* \cf \,\* \nfx2  \,\*  \Bigg\{
          \frct{8}{27} \,\* \pqq \,\* \H(1)
          + \frct{92}{27} \,\* \Big(1
             - \frct{53}{23} \,\* \ooomx
             + \frct{45}{23} \,\* x\Big)
          \Bigg\}\Bigg]
 \; + \; \dots 
\eea
with $\pqq = 2\,(1-x)^{-1} -1 - x$. 
We have suppressed the argument $x$ of the HPLs for brevity. 
The above LL, NLL and NNLL predictions of the resummation have been used 
in and verified by ref.~\cite{MRUVV}, of which the $\ln^{\,3\!} x$ part 
is a result.  The corresponding N$^4$LO (five-loop) predictions read
\bea
\label{Pns4Lx}
&&
 \hspn P_{\,\rm ns,L}^{\,(4)}(x) \;=\;
\xSpaceOuterLB
 + \Lx8  \,\* \Bigg[
       \ncp4 \,\* \cf  \,\*  \Bigg\{
            \frct{31}{1440} \,\* \Big(1
             - \frct{32}{31} \,\* \ooomx
             + x\Big)
          \Bigg\}\Bigg]
\xSpaceOuterLB
 + \Lx7  \,\* \Bigg[
       \ncp4 \,\* \cf  \,\*  \Bigg\{
          - \frct{2}{9} \,\* \pqq \,\* \H(1)
          + \frct{4}{9} \,\* \Big(1
             - \frct{13}{12} \,\* \ooomx
             + \frct{3}{2} \,\* x\Big)
          \Bigg\}
\xSpaceInnerLB
       + \ncp3 \,\* \cf \,\* \nf  \,\*  \Bigg\{
          - \frct{5}{36} \,\* \Big(1
             - \frct{16}{15} \,\* \ooomx
             + x\Big)
          \Bigg\}\Bigg]
\xSpaceOuterLB
 + \Lx6  \,\* \Bigg[
       \ncp4 \,\* \cf  \,\*  \Bigg\{
          - \frct{4}{3} \,\* \pqq \,\* \Hh(1,1)
          + 2 \,\* \Big(1
             - \frct{8}{3} \,\* \ooomx
             + \frct{5}{3} \,\* x\Big) \,\* \H(1)
          + \frct{1}{3} \,\* \Big(1
             + x\Big) \,\* \Hh(0,1)
\xSpaceInnerLB
          + \frct{1}{36} \,\* \Big([283
             - 150 \,\* \z2]
             - \frct{10}{9} \,\* [305
             - 144 \,\* \z2] \,\* \ooomx
             + \frct{2}{3} \,\* [641
             - 225 \,\* \z2] \,\* x\Big)
          \Bigg\}
\xSpaceInnerLB
       + \ncp3 \,\* \cf \,\* \nf  \,\*  \Bigg\{
            \frct{2}{3} \,\* \pqq \,\* \H(1)
          - \frct{67}{27} \,\* \Big(1
             - \frct{236}{201} \,\* \ooomx
             + \frct{89}{67} \,\* x\Big)
          \Bigg\}
\xSpaceInnerLB
       + \ncp2 \,\* \cf \,\* \nfx2  \,\*  \Bigg\{
            \frct{7}{27} \,\* \Big(1
             - \frct{8}{7} \,\* \ooomx
             + x\Big)
          \Bigg\}\Bigg]
\xSpaceOuterLB
 + \Lx5  \,\* \Bigg[
       \ncp4 \,\* \cf  \,\*  \Bigg\{
            \frct{40}{3} \,\* \pqq \,\* \Hhh(1,0,1)
          + \frct{80}{3} \,\* \pqq \,\* \Hhh(1,1,1)
          - \frct{56}{3} \,\* \Big(1
             - \frct{10}{7} \,\* \ooomx
             + x\Big) \,\* \Hhh(0,1,1)
\xSpaceInnerLB
          - \frct{16}{3} \,\* \Big(1
             - \ooomx
             + x\Big) \,\* \Hhh(0,0,1)
          + \frct{124}{9} \,\* \Big(1
             - \frct{2}{31} \,\* \ooomx
             - \frct{29}{31} \,\* x\Big) \,\* \Hh(1,1)
\xSpaceInnerLB
          + \frct{2}{3} \,\* \Big(1
             + 14 \,\* \ooomx
             - 19 \,\* x\Big) \,\* \Hh(0,1)
          + \frct{11}{27} \,\* \Big([115
             - 72 \,\* \z2] \,\* x
             + \frct{1}{11} \,\* [1049
             - 792 \,\* \z2]
\xSpaceInnerLB
             - \frct{2}{11} \,\* [1157
             - 792 \,\* \z2] \,\* \ooomx\Big) \,\* \H(1)
          + \frct{1}{72} \,\* \Big([8129
             - 4704 \,\* \z2
             - 360 \,\* \z3] \,\* x
\xSpaceInnerLB
             + \frct{1}{3} \,\* [10643
             - 7776 \,\* \z2
             - 1080 \,\* \z3]
             - \frct{2}{9} \,\* [26449
             - 14256 \,\* \z2
             - 1728 \,\* \z3] \,\* \ooomx\Big)
          \Bigg\}
\xSpaceInnerLB
       + \ncp3 \,\* \cf \,\* \nf  \,\*  \Bigg\{
            \frct{74}{9} \,\* \pqq \,\* \H(1)
          - \frct{32}{9} \,\* \pqq \,\* \Hh(1,1)
          + \frct{10}{3} \,\* \Big(1
             - \frct{8}{5} \,\* \ooomx
             + x\Big) \,\* \Hh(0,1)
\xSpaceInnerLB
          - \frct{7}{27} \,\* \Big([169
             - 54 \,\* \z2] \,\* x
             + \frct{1}{7} \,\* [745
             - 378 \,\* \z2]
             - \frct{1}{7} \,\* [1091
             - 432 \,\* \z2] \,\* \ooomx\Big)
          \Bigg\}
\xSpaceInnerLB
       + \ncp2 \,\* \cf \,\* \nfx2  \,\*  \Bigg\{
          - \frct{4}{27} \,\* \pqq \,\* \H(1)
          + \frct{94}{27} \,\* \Big(1
             - \frct{66}{47} \,\* \ooomx
             + \frct{61}{47} \,\* x\Big)
          \Bigg\}
\xSpaceInnerLB
       + \nc \,\* \cf \,\* \nfx3  \,\*  \Bigg\{
          - \frct{4}{27} \,\* \Big(1
             - \frct{4}{3} \,\* \ooomx
             + x\Big)
          \Bigg\}\Bigg]
 \; + \; \dots \;\; . 
\eea

%
\setcounter{equation}{0}
\section{Results for the non-singlet coefficient functions}
\label{sec:resultCns}

The order-by-order resummation of the highest $1/N$ powers of the even-$N$ 
based (`+') non-singlet coefficient functions $C_{a}^{\,+}$ for the
structure functions in eq.~(\ref{evenp}) is `automatically' included in 
the calculations towards the corresponding expressions for the splitting 
function $P_{\,\rm ns}^{\,+}$ reported in the previous section. 
Using the short-hand $F\equiv S^{\,-1/2}$ with $S$ as defined in 
eq.~(\ref{Sdef}), the NNLL results for $C^{\,+}_{2}$ and $C^{\,+}_{L}$ 
analogous to eq.~(\ref{PnsPres}) for $P_{\,\rm ns}^{\,+}$ are found to be
\bea
\label{c2ns-cl}
{\lefteqn{
C_{2}^{\,+}(N) \,=\,}}
\nn\\&& 
F + \frct{1}{192\*\cf}\*N\*\bigg\{-3\*(32\*\cf+11\*\bo)\*(F^{\,-1}-1) + 4\*(18\*\cf+11\*\bo)\*(F-1) 
+ 6\*\bo\*(F^{\,3}-1) 
\nn\\[-0.5mm]&& 
+ 12\*(2\*\cf-\bo)\*(F^{\,5}-1)-5\*\bo\*(F^{\,7}-1)\bigg\}
+\frct{1}{9216\*\cf}\*\ar\*\bigg\{-128\*([333 - 1368\*\,\z2]\*\cfs 
\nn\\[-0.5mm]&&
- [60 - 1728\*\,\z2]\*\ca\*\cf - 540\*\,\z2\*\cas - 87\*\bo\*\cf
- 10\*\bos)\*\frct{1}{\xi}\*(F^{\,-3}-F^{\,-1}+2\*\xi) 
\nn\\[-0.5mm]&&
- ([144000 - 442368\*\,\z2]\*\cfs - [7680 - 552960\*\,\z2]\*\ca\*\cf
- 172800\*\,\z2\*\cas - 16320\*\bo\*\cf 
\nn\\[0.5mm]&&
- 5111\*\bos)\*(F-1) 
+ 8\*(576\*\cfs - 90\*\bo\*\cf - 79\*\bos)\*(F^{\,3}-1)
+ ([5184 - 110592\*\,\z2]\*\cfs 
\nn\\[0.5mm]&&
+ [7680 + 110592\*\,\z2]\*\ca\*\cf - 34560\*\,\z2\*\cas
+ 10368\*\bo\*\cf - 2093\*\bos)\*(F^{\,5}-1)
\nn\\[0.5mm]&&
+ 16\*\bo\*(54\*\cf - 77\*\bo)\*(F^{\,7}-1)
+ (2880\*\cf\*(\cf-\bo) + 181\*\bos)\*(F^{\,9}-1)
\nn\\&&
- 840\*\bo\*(2\*\cf - \bo)\*(F^{\,11}-1)
+ 385\*\bos\*(F^{\,13}-1)\bigg\}
\;,
%
\\[1ex]
%
\label{clns-cl}
{\lefteqn{
C_{L}^{\,+}(N) \,=\,}} 
\nn\\&&
4\,\*\cf\*\ar\*F + \frct{1}{48}\*\ar\*N\*\bigg\{- 3\*(64\*\cf - 5\*\bo)\*(F^{\,-1}-1) - 4\*(6\*\cf + \bo)\*(F-1)
+ 6\*\bo\*(F^{\,3}-1) 
\nn\\[-0.5mm]&&
+ 12\*(2\*\cf - \bo)\*(F^{\,5}-1) - 5\*\bo\*(F^{\,7}-1)\bigg\}
+\frct{1}{2304}\*\ars\*\bigg\{-9216\*\cfs\*\frct{1}{\xi}\*(F^{\,-3}-3\*F^{\,-1}+2) 
\nn\\[-0.5mm]&&
-128\*([153 + 72\*\,\z2]\*\cfs - 60\*\ca\*\cf - 69\*\bo\*\cf
+ 2\*\bos)\*\frct{1}{\xi}\*(F^{\,-3}-F^{\,-1}+2\*\xi) 
\nn\\[-0.5mm]&&
- ([222336 - 73728\*\,\z2]\*\cfs - [38400 - 110592\*\,\z2]\*\ca\*\cf
- 34560\*\,\z2\*\cas - 46080\*\bo\*\cf 
\nn\\[0.5mm]&&
+ 1321\*\bos)\*(F-1) + 8\*(1152\*\cfs - 690\*\bo\*\cf + 53\*\bos)\*(F^{\,3} -1) 
+ ([576 - 110592\*\,\z2]\*\cfs 
\nn\\[0.5mm]&&
+ [7680 + 110592\*\,\z2]\*\ca\*\cf - 34560\*\,\z2\*\cas
+ 9024\*\bo\*\cf - 269\*\bos)\*(F^{\,5}-1) 
\nn\\[0.5mm]&&
+ 16\*\bo\*(114\*\cf - 47\*\bo)\*(F^{\,7}-1) 
+ (2880\*\cf\*(\cf-\bo) + 181\*\bos)\*(F^{\,9}-1) 
\nn\\&&
- 840\*\bo\*(2\*\cf - \bo)\*(F^{\,11}-1) + 385\*\bos\*(F^{\,13}-1) \bigg\}
\,.
\eea
The corresponding result for the $\nu \!-\! \bar{\nu}$ charged-current (CC) 
structure function $F_3$ reads
\bea
\label{c3ns-cl}
{\lefteqn{C_{3}^{\,+}(N) \,=\,}}
\nn\\&&
F + \frct{1}{192\*\cf}\*N\*\bigg\{ - 33\*\bo\*(F^{-1}-1) - 4\*[6\*\cf - 11\*\bo]\*(F-1)
+ 6\*\bo\*(F^3-1) 
\nn\\[-0.5mm]&&
+ 12\*[2\*\cf - \bo]\*(F^5-1) - 5\*\bo\*(F^7-1)\bigg\}
+\frct{1}{9216\*\cf}\*\ar\*\bigg\{ + 128\*(63\*\bo\*\cf 
\nn\\[-0.5mm]&&
+ 10\*\bos + 540\*\z2\*\cas
+ 12\*[5 - 144\*\z2]\*\ca\*\cf - 9\*[49 - 152\*\z2]\*\cfs)\*\frct{1}{\xi}\*(F^{-3}-F^{-1}+2\*\xi) 
\nn\\&&
+ (3456\*\bo\*\cf + 5111\*\bos + 172800\*\z2\*\cas + 7680\*[1 - 72\*\z2]\*\ca\*\cf 
\nn\\[0.5mm]&&
- 1152\*[157 - 384\*\z2]\*\cfs)\*(F-1)
+ 8\*(174\*\bo\*\cf - 79\*\bos)\*(F^3-1) 
+ (14016\*\bo\*\cf 
\nn\\[0.5mm]&&
- 2093\*\bos 
- 34560\*\z2\*\cas + 576\*[1 - 192\*\z2]\*\cfs + 1536\*[5 + 72\*\z2]\*\ca\*\cf)\*(F^5-1) 
\nn\\[0.5mm]&&
+ 16\*(114\*\bo\*\cf - 77\*\bos)\*(F^7-1) 
+ (2880\*\cf\*(\cf-\bo) + 181\*\bos)\*(F^{\,9}-1) 
\nn\\&&
- 840\*(2\*\bo\*\cf - \bos)\*(F^{11}-1)
+ 385\*\bos\*(F^{13}-1)\bigg\}
\; .
\eea

Like the splitting function, these coefficient functions exhibit a structure 
similar to that of their time-like counterparts in ref.~\cite{KVY}, apart 
from the all-important fact that the sign of $\xi$ is different there.
The corresponding $x$-space results can be expressed in terms of generalized 
hypergeometric functions listed in appendix B. These hypergeometric functions 
related to the function $F$ are, however, not present in the non-singlet
physical evolution kernels $K_a^{\,+}$ given by
\bea
  K_a^{\,+}=P^{\,+}+\beta(\ar)\frct{d}{d\ar}\ln C_a^{\,+} \; .
\eea

Eqs.~(\ref{c2ns-cl}) -- (\ref{c3ns-cl}) can be expanded to produce the 
explicit four- and five-loop results
\bea
\label{c2ns4-cl}
\left.c_{2}^{\,+}(N)\right|_{\ar^4} \!&\!=\!& 
390\*\,\cff\,\*N^{\,-8} + \cft\*\left(1052\*\cf - \frct{1822}{3}\*\bo\right)\*N^{\,-7}+\cfs\*\bigg([336-5872\*\,\z2]\*\cfs
\\ && \mbox{\hspn} 
 - 448\*\cf\*\bo
 + \bigg[\frct{2180}{3} + 4992\*\,\z2\bigg]\*\cf\*\ca + \frct{1951}{6}\*\bos - 1560\*\,\z2\*\cas\bigg)\*N^{\,-6}
 \:+\: {\cal O} (N^{\,-5})\,,
\nn \\[2mm]
\label{cLns4-cl}
\left.c_{L}^{\,+}(N)\right|_{\ar^4} \!&\!=\!& 
240\*\,\cff\,\*N^{\,-6} + \cft\*\left(472\*\cf - \frct{992}{3}\*\bo\right)\*N^{\,-5}+\cfs\*\bigg(-[644+4016\*\,\z2]\*\cfs 
\nn\\ && \mbox{\hspn} 
 + 56\*\cf\*\bo
 + [480 + 3840\*\,\z2]\*\cf\*\ca + \frct{460}{3}\*\bos - 1200\*\,\z2\*\cas\bigg)\*N^{\,-4}
 \:+\: {\cal O} (N^{\,-3})\,,
\\[2mm]
\label{c3ns4-cl}
\left.c_{3}^{\,+}(N)\right|_{\ar^4} \!&\!=\!&
390\*\cff\*N^{\,-8} + \cft\*\left(780\*\cf - \frct{1822}{3}\*\bo\right)\*N^{\,-7} + \cfs\*\left(- [496 + 5872\*\z2]\*\cfs \right.
\nn\\ && \mbox{\hspn}
\left. - \frct{8}{3}\*\cf\*\bo
 +\left[\frct{2180}{3} + 4992\*\z2\right]\*\cf\*\ca + \frct{1951}{6}\*\bos - 1560\*\z2\*\cas\right)\*N^{\,-6}
 \:+\: {\cal O} (N^{\,-5}) \qquad
\eea
and
\bea
\label{c2ns5-cl}
\left.c_{2}^{\,+}(N)\right|_{\ar^5} \!&\!=\!& 
2652\*\cfi\*N^{\,-10} + \cff\*\left(8418\*\cf - \frct{17012}{3}\*\bo\right)\*N^{\,-9} + \cft\*\left(\vphantom{\frct{1}{1}}[6438 - 56508\*\z2]\*\cfs\right.
\nn\\ && \mbox{\hspn}
\left.-\frct{23546}{3}\*\cf\*\bo + [6040 + 50688\*\z2]\*\cf\*\ca + \frct{14363}{3}\*\bos - 15840\*\z2\*\cas\right)\*N^{\,-8}
\nn\\ &&
 \:+\: {\cal O} (N^{\,-7})\,,
\\[2mm]
\label{cLns5-cl}
\left.c_{L}^{\,+}(N)\right|_{\ar^5} \!&\!=\!& 
1560\*\cfi\*N^{\,-8} + \cff\*\left(3736\*\cf - \frct{8920}{3}\*\bo\right)\*N^{\,-7} + \cft\*\left(\vphantom{\frct{1}{1}} - \left[2064 + 35648\*\z2\right]\*\cfs\right.
\quad
\nn\\ && \mbox{\hspn}
\left.-1504\*\cf\*\bo + \left[\frct{11120}{3} + 34560\*\z2\right]\*\cf\*\ca + \frct{6574}{3}\*\bos - 10800\*\z2\*\cas\right)\*N^{\,-6}
\nn\\ &&
 \:+\: {\cal O} (N^{\,-5})\,,
\\[2mm]
\label{c3ns5-cl}
\left.c_{3}^{\,+}(N)\right|_{\ar^5} \!&\!=\!& 
2652\*\cfi\*N^{\,-10} + \cff\*\left(6630\*\cf - \frct{17012}{3}\*\bo\right)\*N^{\,-9} + \cft\*\left(\vphantom{\frct{1}{1}}\left[66 - 56508\*\z2\right]\*\cfs\right.
\nn\\ &&
\left.-\frct{11374}{3}\*\cf\*\bo + \left[6040 + 50688\*\z2\right]\*\cf\*\ca + \frct{14363}{3}\*\bos - 15840\*\z2\*\cas\right)\*N^{\,-8}
\nn\\ &&
 \:+\: {\cal O} (N^{\,-7})\,.
\eea

\begin{figure}[p]
\vspace*{-1mm}
\centerline{\epsfig{file=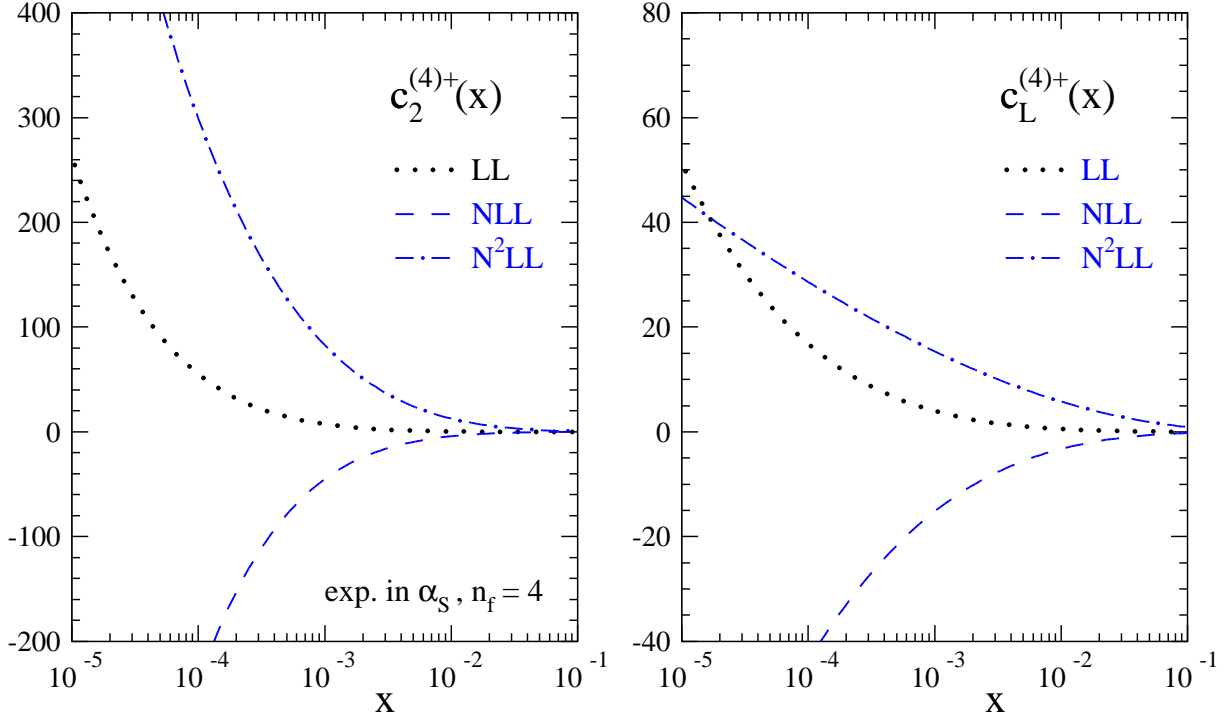,width=16cm,angle=0}}
\vspace{-2mm}
\caption{ \small \label{cns4NNLL}
 The LL, NLL and NNLL small-$x$ approximations to the fourth-order non-singlet
 coefficient functions $c_{2}^{\,(4)+}(x)$ and $c_{L}^{\,(4)+}(x)$ for $\nf=4$
 light flavours. 
 Eqs.~(\ref{c2ns4-cl}) and (\ref{cLns4-cl}) have been transformed to $x$-space 
 using eq.~(\ref{Mlog}) and converted to an expansion in powers of $\as$ by a 
 multiplication with $1/(4\,\pi)^4$.
 }
\vspace{-1mm}
\end{figure}
\begin{figure}[p]
\vspace*{-2mm}
\centerline{\epsfig{file=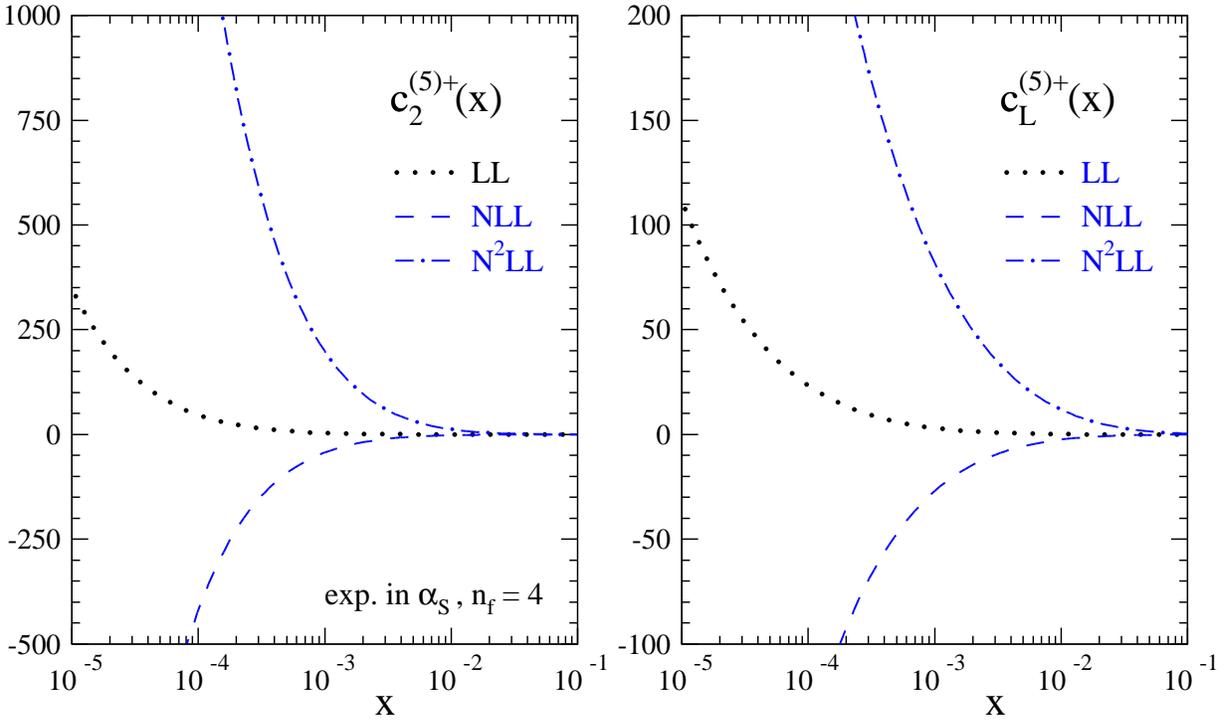,width=16cm,angle=0}}
\vspace{-2mm}
\caption{ \small \label{cns5NNLL}
 As fig.~\ref{cns4NNLL}, but for fifth-order coefficient functions for
 the structure functions $F_2$ and $F_L$ in eq.~(\ref{evenp}).
 }
\vspace{-1mm}
\end{figure}

The LL, NLL and NNLL $x$-space approximations resulting from 
eqs.~(\ref{c2ns4-cl}), (\ref{cLns4-cl}), (\ref{c2ns5-cl}) and (\ref{cLns5-cl})
with eq.~(\ref{Mlog}) are shown for four flavours in figs.~2 and 3.
Unsurprisingly, these results alone do not provide relevant information
about the behaviour of these coefficient functions at any physically
interesting value of $x$. The results in eqs.~(\ref{c2ns4-cl}) --
(\ref{c3ns5-cl}) can become useful, however, once combined with other
partial results such as the large-$x$ limit and a sufficient
number of moments.

As for the non-singlet splitting functions, the resummation can be performed 
for any even or odd power of $x$, and for all powers in the common large-$\nc$
limit $C_{a,\rm L}^{}$ of the coefficient functions $C_{a}^{\,+}$ and 
$C_{a}^{\,-}$. 
Since the  N$^3$LO splitting functions are known in this limit \cite{MRUVV}, 
the all-$x$ coefficients of the four highest small-$x$ logarithms can be 
predicted at any higher order for $a=2$ and $a=3$. The four-loop (N$^4$LO)
results are, using the same abbreviations as in eqs.~(\ref{Pns3Lx}) and 
(\ref{Pns4Lx}) above,
\bea
&&
 \hspn C_{2,\rm L}^{\,(4)} \;=\;
\xSpaceOuterLB
   + \Lx7  \* \Bigg[
       \ncp3 \,\* \cf  \,\*  \Bigg\{
          \frct{65}{448} \,\* \Big(1
             - \frct{16}{15} \,\* \ooomx
             + x\Big)
          \Bigg\}\Bigg]
\xSpaceOuterLB
   + \Lx6  \* \Bigg[
       \ncp3 \,\* \cf  \,\*  \Bigg\{
          - \frct{229}{180} \,\* \pqq \,\* \H(1)
          + \frct{6247}{1296} \,\* \Big(1
             - \frct{35062}{31235} \,\* \ooomx
             + \frct{38867}{31235} \,\* x\Big)
          \Bigg\}
\xSpaceInnerLB
       + \ncp2 \,\* \cf \,\* \nf  \,\*  \Bigg\{
          - \frct{6377}{6480} \,\* \Big(1
             - \frct{8}{7} \,\* \ooomx
             + x\Big)
          \Bigg\}\Bigg]
\xSpaceOuterLB
   + \Lx5  \* \Bigg[
       \ncp3 \,\* \cf  \,\*  \Bigg\{
          - \frct{19}{2} \,\* \pqq \,\* \Hh(1,1)
          + \frct{29137}{1080} \,\* \Big(1
             - \frct{61226}{29137} \,\* \ooomx
             + \frct{35869}{29137} \,\* x\Big) \,\* \H(1)
\xSpaceInnerLB
          + \frct{329}{120} \,\* \Big(1
             - \frct{148}{329} \,\* \ooomx
             + x\Big) \,\* \Hh(0,1)
          + \frct{1}{540} \,\* \Big([54733
             - 9603 \,\* \z2] \,\* x
\xSpaceInnerLB
             + \frct{1}{4} \,\* [152641
             - 38412 \,\* \z2]
             - \frct{1}{24} \,\* [1131721
             - 259200 \,\* \z2] \,\* \ooomx\Big)
          \Bigg\}
\xSpaceInnerLB
       + \ncp2 \,\* \cf \,\* \nf  \,\*  \Bigg\{
          \frct{662}{135} \,\* \pqq \,\* \H(1)
          - \frct{8543}{360} \,\* \Big(1
             - \frct{96332}{76887} \,\* \ooomx
             + \frct{10299}{8543} \,\* x\Big)
          \Bigg\}
\xSpaceInnerLB
       + \nc \,\* \cf \,\* \nfx2  \,\*  \Bigg\{
          \frct{1951}{1080} \,\* \Big(1
             - \frct{4}{3} \,\* \ooomx
             + x\Big)
          \Bigg\}\Bigg]
\xSpaceOuterLB
   + \Lx4  \* \Bigg[
       \ncp3 \,\* \cf  \,\*  \Bigg\{
          \frct{53}{6} \,\* \pqq \,\* \Hhh(1,0,1)
          - \frct{142}{9} \,\* \pqq \,\* \Hhh(1,1,1)
          + \frct{8323}{54} \,\* \Big(1
             + \frct{1584}{41615} \,\* x^{-2}
             - \frct{14864}{8323} \,\* \ooomx
\xSpaceInnerLB
             + \frct{8833}{8323} \,\* x
             - \frct{1728}{8323} \,\* x^2
             + \frct{14256}{41615} \,\* x^3\Big) \,\* \Hh(1,1)
          - \frct{85}{18} \,\* \Big(1
             - \frct{416}{85} \,\* \ooomx
             + x\Big) \,\* \Hhh(0,1,1)
\xSpaceInnerLB
          + \frct{69}{2} \,\* \Big(1
             + \frct{1573}{3726} \,\* \ooomx
             + \frct{85}{69} \,\* x
             - \frct{64}{69} \,\* x^2
             + \frct{176}{115} \,\* x^3\Big) \,\* \Hh(0,1)
          - \frct{53}{18} \,\* \Big(1
             + \frct{48}{53} \,\* \ooomx
\xSpaceInnerLB
             + x\Big) \,\* \Hhh(0,0,1)
          - \frct{88}{15} \,\* \Big(x^{-1}
             + 9 \,\* x^2
             + \frct{5}{9504} \,\* [186223
             - 66672 \,\* \z2] \,\* \ooomx
\xSpaceInnerLB
             - \frct{1}{19008} \,\* [808271
             - 333360 \,\* \z2]
             - \frct{1}{19008} \,\* [1282259
             - 333360 \,\* \z2] \,\* x\Big) \,\* \H(1)
\xSpaceInnerLB
          + \frct{8}{5} \,\* \Big([33
             - 20 \,\* \z2] \,\* x^2
             - \frct{25}{20736} \,\* [406651
             - 190080 \,\* \z2
             - 30240 \,\* \z3] \,\* \ooomx
\xSpaceInnerLB
             + \frct{1}{124416} \,\* [42910871
             - 23142960 \,\* \z2
             - 4034880 \,\* \z3]
\xSpaceInnerLB
             + \frct{1}{124416} \,\* [74167091
             - 29523600 \,\* \z2
             - 4034880 \,\* \z3] \,\* x
             + 33 \,\*
         \z2 \,\* x^3\Big)
          \Bigg\}
\xSpaceInnerLB
       + \ncp2 \,\* \cf \,\* \nf  \,\*  \Bigg\{
          - \frct{422}{27} \,\* \Big(1
             + \frct{24}{211} \,\* x^{-2}
             - 2 \,\* \ooomx
             + \frct{259}{211} \,\* x
             - \frct{144}{211} \,\* x^2
             + \frct{216}{211} \,\* x^3\Big) \,\* \Hh(1,1)
\xSpaceInnerLB
          - \frct{8177}{108} \,\* \Big(1
             - \frct{192}{8177} \,\* x^{-1}
             - \frct{16782}{8177} \,\* \ooomx
             + \frct{10341}{8177} \,\* x
             - \frct{1728}{8177} \,\* x^2\Big) \,\* \H(1)
\xSpaceInnerLB
          + \frct{10}{9} \,\* \Big(1
             - \frct{22}{3} \,\* \ooomx
             - \frct{11}{5} \,\* x
             + \frct{48}{5} \,\* x^2
             - \frct{72}{5} \,\* x^3\Big) \,\* \Hh(0,1)
          - \frct{16}{3} \,\* \Big([3
             - 2 \,\* \z2] \,\* x^2
\xSpaceInnerLB
             + \frct{5}{10368} \,\* [92215
             - 23364 \,\* \z2]
             + \frct{1}{5184} \,\* [338959
             - 56250 \,\* \z2] \,\* x
\xSpaceInnerLB
             - \frct{1}{20736} \,\* [1323835
             - 295488 \,\* \z2] \,\* \ooomx
             + 3 \,\* \z2 \,\* x^3\Big)
          \Bigg\}
       + \cf \,\* \nfx3  \,\*  \Bigg\{
          \frct{119}{162} \,\* \pqq
          \Bigg\}
\xSpaceInnerLB
       + \nc \,\* \cf \,\* \nfx2  \,\*  \Bigg\{
          - \frct{671}{162} \,\* \pqq \,\* \H(1)
          + \frct{111}{4} \,\* \Big(1
             - \frct{41134}{26973} \,\* \ooomx
             + \frct{11249}{8991} \,\* x\Big)
          \Bigg\}\Bigg]
\; , \\
&&
 \hspn C_{3,\rm L}^{\,(4)} \;=\;
\xSpaceOuterLB
  + \Lx7  \* \Bigg[
       \ncp3 \,\* \cf  \,\*  \Bigg\{
          \frct{65}{448} \,\* \Big(1
             - \frct{16}{15} \,\* \ooomx
             + x\Big)
          \Bigg\}\Bigg]
\xSpaceOuterLB
  + \Lx6  \* \Bigg[
       \ncp3 \,\* \cf  \,\*  \Bigg\{
          - \frct{229}{180} \,\* \pqq \,\* \H(1)
          + \frct{30929}{6480} \,\* \Big(1
             - \frct{35062}{30929} \,\* \ooomx
             + \frct{38561}{30929} \,\* x\Big)
          \Bigg\}
\xSpaceInnerLB
       + \ncp2 \,\* \cf \,\* \nf  \,\*  \Bigg\{
          - \frct{6377}{6480} \,\* \Big(1
             - \frct{8}{7} \,\* \ooomx
             + x\Big)
          \Bigg\}\Bigg]
\xSpaceOuterLB
  + \Lx5  \* \Bigg[
       \ncp3 \,\* \cf  \,\*  \Bigg\{
          - \frct{19}{2} \,\* \pqq \,\* \Hh(1,1)
          + \frct{28057}{1080} \,\* \Big(1
             - \frct{61226}{28057} \,\* \ooomx
             + \frct{34789}{28057} \,\* x\Big) \,\* \H(1)
\xSpaceInnerLB
          + \frct{329}{120} \,\* \Big(1
             - \frct{148}{329} \,\* \ooomx
             + x\Big) \,\* \Hh(0,1)
          + \frct{1}{360} \,\* \Big([35675
             - 6402 \,\* \z2] \,\* x
\xSpaceInnerLB
             + \frct{1}{2} \,\* [49055
             - 12804 \,\* \z2]
             - \frct{1}{36} \,\* [1131721
             - 259200 \,\* \z2] \,\* \ooomx\Big)
          \Bigg\}
\xSpaceInnerLB
       + \ncp2 \,\* \cf \,\* \nf  \,\*  \Bigg\{
          \frct{662}{135} \,\* \pqq \,\* \H(1)
          - \frct{24961}{1080} \,\* \Big(1
             - \frct{96332}{74883} \,\* \ooomx
             + \frct{30229}{24961} \,\* x\Big)
          \Bigg\}
\xSpaceInnerLB
       + \nc \,\* \cf \,\* \nfx2  \,\*  \Bigg\{
          \frct{1951}{1080} \,\* \Big(1
             - \frct{4}{3} \,\* \ooomx
             + x\Big)
          \Bigg\}\Bigg]
\xSpaceOuterLB
  + \Lx4  \* \Bigg[
       \ncp3 \,\* \cf  \,\*  \Bigg\{
          \frct{53}{6} \,\* \pqq \,\* \Hhh(1,0,1)
          - \frct{142}{9} \,\* \pqq \,\* \Hhh(1,1,1)
          + \frct{7963}{54} \,\* \Big(1
             - \frct{528}{7963} \,\* x^{-1}
             - \frct{14864}{7963} \,\* \ooomx
\xSpaceInnerLB
             + \frct{7609}{7963} \,\* x
             + \frct{528}{7963} \,\* x^2\Big) \,\* \Hh(1,1)
          - \frct{85}{18} \,\* \Big(1
             - \frct{416}{85} \,\* \ooomx
             + x\Big) \,\* \Hhh(0,1,1)
          + \frct{65}{2} \,\* \Big(1
\xSpaceInnerLB
             + \frct{121}{270} \,\* \ooomx
             + \frct{49}{65} \,\* x
             + \frct{176}{585} \,\* x^2\Big) \,\* \Hh(0,1)
          - \frct{53}{18} \,\* \Big(1
             + \frct{48}{53} \,\* \ooomx
             + x\Big) \,\* \Hhh(0,0,1)
\xSpaceInnerLB
          + \frct{1}{648} \,\* \Big([159931
             - 66672 \,\* \z2]
             - 2 \,\* [186223
             - 66672 \,\* \z2] \,\* \ooomx
\xSpaceInnerLB
             + [235015
             - 66672 \,\* \z2] \,\* x\Big) \,\* \H(1)
          - \frct{5}{2592} \,\* \Big([406651
             - 190080 \,\* \z2
\xSpaceInnerLB
             - 30240 \,\* \z3] \,\* \ooomx
             - \frct{1}{30} \,\* [7672909
             - 4526856 \,\* \z2
             - 806976 \,\* \z3]
\xSpaceInnerLB
             - \frct{1}{30} \,\* [14040109
             - 5927400 \,\* \z2
             - 806976 \,\* \z3] \,\* x
             - \frct{25344}{5} \,\* \z2 \,\* x^2\Big)
          \Bigg\}
\xSpaceInnerLB
       + \ncp2 \,\* \cf \,\* \nf  \,\*  \Bigg\{
          - \frct{422}{27} \,\* \Big(1
             - \frct{24}{211} \,\* x^{-1}
             - 2 \,\* \ooomx
             + x
             + \frct{24}{211} \,\* x^2\Big) \,\* \Hh(1,1)
\xSpaceInnerLB
          + \frct{10}{9} \,\* \Big(1
             - \frct{22}{3} \,\* \ooomx
             + x
             - \frct{8}{5} \,\* x^2\Big) \,\* \Hh(0,1)
          - \frct{8065}{108} \,\* \Big(1
             - \frct{16782}{8065} \,\* \ooomx
\xSpaceInnerLB
             + \frct{9461}{8065} \,\* x\Big) \,\* \H(1)
          - \frct{1}{972} \,\* \Big([208597
             - 58410 \,\* \z2]
             + \frct{1}{2} \,\* [633641
             - 115956 \,\* \z2] \,\* x
\xSpaceInnerLB
             - \frct{1}{4} \,\* [1323835
             - 295488 \,\* \z2] \,\* \ooomx
             + 1728 \,\* \z2 \,\* x^2\Big)
          \Bigg\}
       + \cf \,\* \nfx3  \,\*  \Bigg\{
          \frct{119}{162} \,\* \pqq
          \Bigg\}
\xSpaceInnerLB
       + \nc \,\* \cf \,\* \nfx2  \,\*  \Bigg\{
          - \frct{671}{162} \,\* \pqq \,\* \H(1)
          + \frct{8239}{324} \,\* \Big(1
             - \frct{41134}{24717} \,\* \ooomx
             + \frct{10497}{8239} \,\* x\Big)
          \Bigg\}\Bigg]
\; . 
\eea
We have also determined the corresponding five-loop results. 
Since they would become useful mainly in the context of research towards 
N$^5$LO all-$x$ expressions, which we do not 
expect in the foreseeable future, we skip these here for brevity but provide
them in the ancillary file of this paper.

In the case of $C_L$, the fixed order results are restricted to NNLO even 
in the large-$\nc$ limit, since the four-loop coefficient function would be 
required for N$^3$LO accuracy, recall the discussion below eq.~(\ref{c23L0}). 
Hence only the highest three small-$x$ logarithms can be resummed using 
eq.~(\ref{FhatAllN}). The resulting N$^3$LO and N$^4$LO predictions read
\bea
&&
 \hspn C_{L,\rm L}^{\,(4)} \;=\;
\xSpaceOuterLB
   + \Lx6  \* \Bigg[
       \ncp3 \,\* \cf  \,\*  \Bigg\{
          \frct{17}{180} \,\* x
          \Bigg\}\Bigg]
\xSpaceOuterLB
   + \Lx5  \* \Bigg[
       \ncp3 \,\* \cf  \,\*  \Bigg\{
          - \frct{1}{4} \,\* \Big(1
             - \frct{3386}{135} \,\* x\Big)
          + 2 \,\* x \,\* \H(1)
          \Bigg\}
       + \ncp2 \,\* \cf \,\* \nf  \,\*  \Bigg\{
          - \frct{167}{135} \,\* x
          \Bigg\}\Bigg]
\xSpaceOuterLB
   + \Lx4  \* \Bigg[
       \ncp3 \,\* \cf  \,\*  \Bigg\{
          - \frct{116}{15} \,\* \Big(1
             + \frct{88}{29} \,\* x^{-1}
             - \frct{5869}{1044} \,\* x
             + \frct{132}{29} \,\* x^2\Big) \,\* \H(1)
\xSpaceInnerLB
          + \frct{1687}{1080} \,\* \Big(1
             + \frct{1152}{1687} \,\* [33
             - 20 \,\* \z2] \,\* x^2
             + \frct{1}{10122} \,\* [867587
             - 32940 \,\* \z2] \,\* x
             + \frct{38016}{1687} \,\* \z2 \,\* x^3\Big)
\xSpaceInnerLB
          + \frct{352}{15} \,\* \Big(x^{-2}
             - \frct{5}{11} \,\* x^{-1}
             + \frct{5}{4} \,\* x
             - \frct{10}{11} \,\* x^2
             + \frct{3}{2} \,\* x^3\Big) \,\* \Hh(1,1)
          + 20 \,\* \Big(x
             - \frct{16}{15} \,\* x^2
             + \frct{44}{25} \,\* x^3\Big) \,\* \Hh(0,1)
          \Bigg\}
\xSpaceInnerLB
       + \ncp2 \,\* \cf \,\* \nf  \,\*  \Bigg\{
          - \frct{34}{27} \,\* \Big(1
             + \frct{48}{17} \,\* [3
             - 2 \,\* \z2] \,\* x^2
             + \frct{1}{408} \,\* [16361
             + 576 \,\* \z2] \,\* x
             + \frct{144}{17} \,\* \z2 \,\* x^3\Big)
\xSpaceInnerLB
          - \frct{64}{9} \,\* \Big(x^{-2}
             - \frct{1}{2} \,\* x^{-1}
             + \frct{1}{2} \,\* x
             - x^2
             + \frct{3}{2} \,\* x^3\Big) \,\* \Hh(1,1)
          + \frct{64}{9} \,\* \Big(x^{-1}
             - \frct{31}{24} \,\* x
             + \frct{3}{2} \,\* x^2\Big) \,\* \H(1)
\xSpaceInnerLB
          - \frct{32}{9} \,\* \Big(x
             - 2 \,\* x^2
             + 3 \,\* x^3\Big) \,\* \Hh(0,1)
          \Bigg\}
       + \nc \,\* \cf \,\* \nfx2  \,\*  \Bigg\{
          \frct{376}{81} \,\* x
          \Bigg\}\Bigg]
\eea
and 
\bea
&&
 \hspn C_{L,\rm L}^{\,(5)} \;=\;
\xSpaceOuterLB
   + \Lx8  \* \Bigg[
       \ncp4 \,\* \cf  \,\*  \Bigg\{
          \frct{149}{26880} \,\* x
          \Bigg\}\Bigg]
\xSpaceOuterLB
   + \Lx7  \* \Bigg[
       \ncp4 \,\* \cf  \,\*  \Bigg\{
          - \frct{13}{672} \,\* \Big(1
             - \frct{61168}{1755} \,\* x\Big)
          + \frct{13}{42} \,\* x \,\* \H(1)
          \Bigg\}
       + \ncp3 \,\* \cf \,\* \nf  \,\*  \Bigg\{
          - \frct{3043}{22680} \,\* x
          \Bigg\}\Bigg]
\xSpaceOuterLB
   + \Lx6  \* \Bigg[
       \ncp4 \,\* \cf  \,\*  \Bigg\{
          - \frct{971}{450} \,\* \Big(1
             + \frct{2552}{971} \,\* x^{-1}
             - \frct{106633}{17478} \,\* x
             + \frct{3828}{971} \,\* x^2\Big) \,\* \H(1)
\xSpaceInnerLB
          + \frct{82109}{64800} \,\* \Big(1
             + \frct{16704}{82109} \,\* [33
             - 20 \,\* \z2] \,\* x^2
             + \frct{1}{246327} \,\* [6243947
             - 181980 \,\* \z2] \,\* x
\xSpaceInnerLB
             + \frct{551232}{82109} \,\* \z2 \,\* x^3\Big)
          + \frct{1276}{225} \,\* \Big(x^{-2}
             - \frct{5}{11} \,\* x^{-1}
             + \frct{1725}{1276} \,\* x
             - \frct{10}{11} \,\* x^2
             + \frct{3}{2} \,\* x^3\Big) \,\* \Hh(1,1)
\xSpaceInnerLB
          + \frct{46}{9} \,\* \Big(x
             - \frct{116}{115} \,\* x^2
             + \frct{957}{575} \,\* x^3\Big) \,\* \Hh(0,1)
          \Bigg\}
       + \ncp3 \,\* \cf \,\* \nf  \,\*  \Bigg\{
          + \frct{232}{135} \,\* \Big(x^{-1}
             - \frct{1115}{696} \,\* x
             + \frct{3}{2} \,\* x^2\Big) \,\* \H(1)
\xSpaceInnerLB
          - \frct{232}{135} \,\* \Big(x^{-2}
             - \frct{1}{2} \,\* x^{-1}
             + \frct{1}{2} \,\* x
             - x^2
             + \frct{3}{2} \,\* x^3\Big) \,\* \Hh(1,1)
          - \frct{116}{135} \,\* \Big(x
             - 2 \,\* x^2
             + 3 \,\* x^3\Big) \,\* \Hh(0,1)
\xSpaceInnerLB
          - \frct{1669}{3240} \,\* \Big(1
             + \frct{2784}{1669} \,\* [3
             - 2 \,\* \z2] \,\* x^2
             + \frct{1}{5007} \,\* [115745
             + 2088 \,\* \z2] \,\* x
             + \frct{8352}{1669} \,\* \z2 \,\* x^3\Big)
          \Bigg\}
\xSpaceInnerLB
       + \ncp2 \,\* \cf \,\* \nfx2  \,\*  \Bigg\{
          \frct{10891}{9720} \,\* x
          \Bigg\}\Bigg]
\; .
\eea

%
\setcounter{equation}{0}
\section{Results for flavour-singlet quantities}
\label{sec:result-sg}
%

We now turn to the singlet case, and first present the results for the 
splitting functions $P_{\,\rm ik}$. The diagonal (i$\,=\,$k) quantities can 
be written as sums of `non-singlet' (ns)  and pure-singlet (ps)~pieces,
\bea
\label{Pii-ps}
  P_{\rm qq}(N) &\!\equiv\!& P_{\rm ns}^{\,+}(N) + P_{\rm qq}^{\:\rm ps}(N)\;, 
\nn\\[0.5mm]
  P_{\rm gg}(N) &\!\equiv\!& P_{\rm gg}^{\,+}(N) + P_{\rm gg}^{\:\rm ps}(N)\;,
\eea
where $P_{\rm gg}^{\,+}$ is an non-singlet--like quantity, i.e., $P_{\rm gg}$
in the limit $C_F = 0$ (cf.~ref.~\cite{MV07}). At leading-logarithmic (LL) 
accuracy,
\beq
\label{Pgg-nsLL}
  P_{\rm gg}^{\,+}(N) \;=\; -\,\frct{1}{2}\,N (S'-1)\;,
\eeq
where 
\beq
\label{SPdef}
   S' \,=\, (1-4\,\xi')^{1/2}
\:\: \mbox{ with } \quad
   \xi' \;=\; -\,\frct{4\:\!\ca\:\!\ar}{N^2}
   \;\equiv\; -\,\frct{\ca\:\!\as}{\pi\, N^{\:\!2}}
\;\: .
\eeq
Consequently the LL $P_{gg}^{+}(x)$ is an oscillatory function, in notable
contrast to the corresponding quark quantity in eqs.~(\ref{PnsPres}) and 
(\ref{pns-x}). 
The other LL contributions are found to be 
\bea
\label{eq:S_pqq_LL}
  P_{\rm qq}^{\,\rm ps\,(n)}(N) &\!=\!& 
    \cat{n}\,\frac{2^{\:\!n+1}}{N^{\:\!2n+1}} \,
    \sum_{i=0}^{\floor{\frac{n-1}{2}}}\:\sum_{k=0}^{n-1-2i}
    (-2)^{i+1+k}\,(\nf\cf)^{i+1}\cax{k}\cfx{\rho}\,
    \binom{k+i}{k}\binom{\rho+i+1}{\rho}\;,\\[1mm]
\label{eq:S_pgg_LL}
  P_{\rm gg}^{\,\rm ps\,(n)}(N) &\!=\!& 
    \cat{n}\,\frac{2^{\:\!n+1}}{N^{\:\!2n+1}} \,
    \sum_{i=0}^{\floor{\frac{n-1}{2}}}\:\sum_{k=0}^{n-1-2i}
    (-2)^{i+1+k}\,(\nf\cf)^{i+1}\cax{k}\cfx{\rho}\,
    \binom{k+i+1}{k}\binom{\rho+i}{\rho}\;,\\[1mm]
\label{eq:S_pqg_LL}
  P_{\rm qg}^{\,(n)}(N) &\!=\!& 
    \nf\cat{n}\,\frac{2^{\:\!n+1}}{N^{\:\!2n+1}} \,
    \sum_{i=0}^{\floor{\frac{n}{2}}}\:\sum_{k=0}^{n-2i}
    (-2)^{i+k}\,(\nf\cf)^i\cax{k}\cfx{\delta}\,
    \binom{k+i}{k}\binom{\delta+i}{\delta}\;,\\[2mm]
\label{eq:S_pgq_LL}
  P_{\rm gq}^{\,(n)}(N) &\!=\!& 
    -\,\frct{2\,\cf}{\nf}\, P_{qg}^{(n)}(N)\;,
\eea
where the relation between $P_{\rm qg}$ and $P_{\rm gq}$ holds 
only at LL. In these expressions, $\floor{\ldots}$ denotes the Gau\ss\
bracket (floor function),
$\rho \,=\,  n - k - 2\,i - 1$, $\delta \,=\, n - k - 2\,i$,
and $\cat{n}$ are the Catalan numbers,
\bea
\label{catalan}
  \cat{n} \;=\; \frct{(2n)!}{n!\,(n+1)!} \;.
\eea
The first (second) binomial factors 
in eqs.~(\ref{eq:S_pqq_LL}) -- (\ref{eq:S_pqg_LL}) may be
interpreted as counting the number of ways $k$ ($\rho / \delta$)
gluons can be emitted from the gluon (quark) propagators, the number
of which is directly related to the number of factors $\nf\cf$.
In this interpretation, the quark and gluon emissions contribute to
these logarithms equally, with emission strengths proportional
to the colour factors.

Incidentally, the Catalan numbers~(\ref{catalan}) are the
coefficients of the Taylor expansion of $S$ in the LL non-singlet splitting
function, suggesting the singlet results could be generalizations of
their simpler non-singlet counterparts.  However, we have not found 
closed-form expressions.  The NLL and NNLL results are considerably more
complicated, and even expressions in the form of the sum representations above
are not available.  To achieve this could require defining the LL results
in terms of new special functions, and suitably generalizing them for
the sub-leading results.  Potential complications also arise from the
observation that the NLL and NNLL results have terms with denominators
similar to that of the $N$-space logarithmic functions in the SIA case 
\cite{KVY}, which could indicate the presence of logarithmic functions 
that depend on the special functions yet to be~found.

The NNLL double-logarithmic resummation of the $1/N$ pole terms leads
to the following results for the four-loop (N$^3$LO) splitting functions 
($P_{\,\rm ns}^{\,+(3)}(N)$ has been given in eq.~(\ref{PnsP4nn}) above)
\bea
   P_{\rm qq}^{\,(3)}(N) \!&\!=\!&\! P_{\,\rm ns}^{\,+(3)}(N) +
   \nf\, \* \cf\, \* \bigg\{ N^{-7} \, \*  \bigg( - 640\, \* \cax2\,
          + 640\, \* \cf\, \* \ca\,
          - 480\, \* \cfx2\,
          + 320\, \* \nf\, \* \cf\,\bigg)\,
   \nn \\[0.5mm] & & \mbox{\hspn}
       + N^{-6} \, \*  \bigg( - \frct{2176}{3}\: \* \cax2\,
          + \frct{3424}{3}\: \* \cf\, \* \ca\,
          - \frct{1024}{3}\: \* \cfx2\,
          - 256\, \* \nf\, \* \cf\,\bigg)\,
   \nn \\[0.5mm] & & \mbox{\hspn}
       + N^{-5} \, \*  \bigg( - 288\, \* \nf\, \* \ca\,
          + \frct{15232}{9}\: \* \nf\, \* \cf\,
          - \frct{32}{9}\: \* \nfx2\,
          - \frct{8}{3}\: \* (519 - 524\, \* \z2)\, \* \cfx2\,
   \nn \\[0.5mm] & & \mbox{\hspn}
          + \frct{8}{9}\: \* (541 - 1332\, \* \z2)\, \* \cf\, \* \ca\,
          - \frct{8}{9}\: \* (1709 - 192\, \* \z2)\, \* \cax2\,\bigg)\,\bigg\}
+ {\cal O}(N^{-4})\;,
\\[3mm]
   P_{\rm qg}^{\,(3)}(N) \!&\!=\!&\!
   \nf\,\*\bigg\{ N^{-7} \, \*  \bigg( - 640\, \* \cax3\,
          + 320\, \* \cf\, \* \cax2\,
          - 160\, \* \cfx2\, \* \ca\,
          + 80\, \* \cfx3\,
          + 640\, \* \nf\, \* \cf\, \* \ca\,
   \nn \\[0.5mm] & & \mbox{\hspn}
          - 320\, \* \nf\, \* \cfx2\,\bigg)\,
       + N^{-6} \, \*  \bigg( - \frct{416}{3}\: \* \cax3\,
          + 192\, \* \cf\, \* \cax2\,
          - \frct{632}{3}\: \* \cfx2\, \* \ca\,
          - \frct{32}{3}\: \* \cfx3\,
   \nn \\[0.5mm] & & \mbox{\hspn}
          - \frct{320}{3}\: \* \nf\, \* \cax2\,
          - \frct{1408}{3}\: \* \nf\, \* \cf\, \* \ca\,
          + 432\, \* \nf\, \* \cfx2\,\bigg)\,
       + N^{-5} \, \*  \bigg( - \frct{32}{9}\: \* \nfx2\, \* \ca\,
          + \frct{2224}{27}\: \* \nfx2\, \* \cf\,
   \nn \\[0.5mm] & & \mbox{\hspn}
          - \frct{32}{27}\: \* (148 + 81\, \* \z2)\, \* \nf\, \* \cax2\,
          + \frct{2}{3}\: \* (557 - 1448\, \* \z2)\, \* \cfx3\,
          - \frct{40}{27}\: \* (1711 + 108\, \* \z2)\, \* \cax3\,
   \nn \\[0.5mm] & & \mbox{\hspn}
          + \frct{8}{9}\: \* (2951 + 300\, \* \z2)\, \* \nf\, \* \cf\, \* \ca\,
          + \frct{8}{27}\: \* (6427 - 3960\, \* \z2)\, \* \cf\, \* \cax2\,
          - \frct{2}{27}\: \* (6707 
   \nn \\[0.5mm] & & \mbox{\hspn}
            - 19368\, \* \z2)\, \* \cfx2\, \* \ca\,
          - \frct{4}{27}\: \* (13583 - 3600\, \* \z2)\, 
            \* \nf\, \* \cfx2\,\bigg)\,\bigg\}
+ {\cal O}(N^{-4})\;,
\\[3mm]
   P_{\rm gq}^{\,(3)}(N) \!&\!=\!&\!
   \cf\,\*\bigg\{ N^{-7} \, \*  \bigg( 1280\, \* \cax3\,
          - 640\, \* \cf\, \* \cax2\,
          + 320\, \* \cfx2\, \* \ca\,
          - 160\, \* \cfx3\,
          - 1280\, \* \nf\, \* \cf\, \* \ca\,
   \nn \\[0.5mm] & & \mbox{\hspn}
          + 640\, \* \nf\, \* \cfx2\,\bigg)\,
       + N^{-6} \, \*  \bigg( \frct{4160}{3}\: \* \cax3\,
          - 1280\, \* \cf\, \* \cax2\,
          + \frct{2800}{3}\: \* \cfx2\, \* \ca\,
          - 320\, \* \cfx3\,
   \nn \\[0.5mm] & & \mbox{\hspn}
          + \frct{640}{3}\: \* \nf\, \* \cax2\,
          - 640\, \* \nf\, \* \cf\, \* \ca\,
          + \frct{800}{3}\: \* \nf\, \* \cfx2\,\bigg)\,
       + N^{-5} \, \*  \bigg( \frct{64}{9}\: \* \nfx2\, \* \ca\,
          - \frct{12256}{27}\: \* \nfx2\, \* \cf\,
   \nn \\[0.5mm] & & \mbox{\hspn}
          - \frct{4}{3}\: \* (25 - 1248\, \* \z2)\, \* \cfx3\,
          + \frct{64}{27}\: \* (542 + 81\, \* \z2)\, \* \nf\, \* \cax2\,
          - \frct{16}{3}\: \* (817 + 164\, \* \z2)\, \* \nf\, \* \cf\, \* \ca\,
   \nn \\[0.5mm] & & \mbox{\hspn}
          + \frct{16}{27}\: \* (1969 + 936\, \* \z2)\, \* \cf\, \* \cax2\,
          + \frct{16}{27}\: \* (3871 + 2340\, \* \z2)\, \* \cax3\,
          + \frct{8}{27}\: \* (7747 
   \nn \\[0.5mm] & & \mbox{\hspn}
            - 2448\, \* \z2)\, \* \nf\, \* \cfx2\,
          - \frct{4}{27}\: \* (8633 + 12672\, \* \z2)\, \* 
            \cfx2\, \* \ca\,\bigg)\,\bigg\}
+ {\cal O}(N^{-4})\;,
\\[3mm]
   P_{\rm gg}^{\,(3)}(N) \!&\!=\!&\!
       N^{-7} \, \*  \bigg( 1280\, \* \cax4\,
          - 1920\, \* \nf\, \* \cf\, \* \cax2\,
          + 640\, \* \nf\, \* \cfx2\, \* \ca\,
          - 160\, \* \nf\, \* \cfx3\,
          + 320\, \* \nfx2\, \* \cfx2\,\bigg)\,
   \nn \\[0.5mm] & & \mbox{\hspn}
       + N^{-6} \, \*  \bigg( \frct{640}{3}\: \* \cax4\,
          + \frct{1280}{3}\: \* \nf\, \* \cax3\,
          + \frct{1856}{3}\: \* \nf\, \* \cf\, \* \cax2\,
          + \frct{256}{3}\: \* \nf\, \* \cfx2\, \* \ca\,
          + \frct{64}{3}\: \* \nf\, \* \cfx3\,
   \nn \\[0.5mm] & & \mbox{\hspn}
          - \frct{640}{3}\: \* \nfx2\, \* \cf\, \* \ca\,
          - \frct{1472}{3}\: \* \nfx2\, \* \cfx2\,\bigg)\,
       + N^{-5} \, \*  \bigg(          \frct{128}{3}\: \* \nfx2\, \* \cax2\,
          - \frct{4768}{9}\: \* \nfx2\, \* \cf\, \* \ca\,
   \nn \\[0.5mm] & & \mbox{\hspn}
          + \frct{19904}{9}\: \* \nfx2\, \* \cfx2\,
          - \frct{32}{9}\: \* \nfx3\, \* \cf\,
          + \frct{128}{3}\: \* (20 + 9\, \* \z2)\, \* \nf\, \* \cax3\,
          + 32\, \* (137 + 64\, \* \z2)\, \* \cax4\,
   \nn \\[0.5mm] & & \mbox{\hspn}
          - \frct{8}{3}\: \* (195 - 148\, \* \z2)\, \* \nf\, \* \cfx3\,
          + \frct{8}{9}\: \* (1997 - 756\, \* \z2)\, \* \nf\, \* \cfx2\, \* \ca\,
   \nn \\[0.5mm] & & \mbox{\hspn}
          - \frct{8}{3}\: \* (2751 + 688\, \* \z2)\, \* \nf\, \* \cf\, \* \cax2\,\bigg)\,
+ {\cal O}(N^{-4})\;.
\eea
which we expect to become relevant in the near future in combination with
the fixed-$N$ moments in ref.~\cite{Pij3lowN} and other constraints. 
Their N$^4$LO counterparts read
\bea
   P_{\rm qq}^{\,(4)}(N) \!&\!=\!&\! P_{\,\rm ns}^{\,+(4)}(N) +
   \nf\, \* \cf\, \* \bigg\{ N^{-9} \, \*  \bigg( 7168\, \* \cax3\,
          - 7168\, \* \cf\, \* \cax2\,
          + 5376\, \* \cfx2\, \* \ca\,
          - 3584\, \* \cfx3\,
   \nn \\[0.5mm] & & \mbox{\hspn}
          - 7168\, \* \nf\, \* \cf\, \* \ca\,
          + 5376\, \* \nf\, \* \cfx2\,\bigg)\,
       + N^{-8} \, \*  \bigg(          7936\, \* \cax3\,
          - \frct{38720}{3}\: \* \cf\, \* \cax2\,
          + \frct{41984}{3}\: \* \cfx2\, \* \ca\,
   \nn \\[0.5mm] & & \mbox{\hspn}
          - \frct{12272}{3}\: \* \cfx3\,
          + \frct{1792}{3}\: \* \nf\, \* \cax2\,
          - 1088\, \* \nf\, \* \cf\, \* \ca\,
          - \frct{6656}{3}\: \* \nf\, \* \cfx2\,
          + \frct{896}{3}\: \* \nfx2\, \* \cf\,\bigg)\,
   \nn \\[0.5mm] & & \mbox{\hspn}
       + N^{-7} \, \*  \bigg( \frct{256}{9}\: \* \nfx2\, \* \ca\,
          - \frct{20480}{9}\: \* \nfx2\, \* \cf\,
          + \frct{32}{3}\: \* (442 + 105\, \* \z2)\, \* \nf\, \* \cax2\,
          + \frct{32}{9}\: \* (7054 
   \nn \\[0.5mm] & & \mbox{\hspn}
            + 243\, \* \z2)\, \* \cax3\,
          - \frct{4}{3}\: \* (9109 - 19668\, \* \z2)\, \* \cfx3\,
          + \frct{4}{9}\: \* (9211 - 72108\, \* \z2)\, \* \cfx2\, \* \ca\,
   \nn \\[0.5mm] & & \mbox{\hspn}
          - \frct{16}{9}\: \* (16829 + 1602\, \* \z2)\, \* \nf\, \* \cf\, \* \ca\,
          - \frct{8}{9}\: \* (24337 - 22320\, \* \z2)\, \* \cf\, \* \cax2\,
   \nn \\[0.5mm] & & \mbox{\hspn}
          + \frct{8}{9}\: \* (33715 - 9216\, \* \z2)\, \* \nf\, \* \cfx2\,\bigg)\,\bigg\}
+ {\cal O}(N^{-6})\;,
\\[3mm]
   P_{\rm qg}^{\,(4)}(N) \!&\!=\!&\!
   \nf\,\*\bigg\{ N^{-9} \, \*  \bigg( 7168\, \* \cax4\,
          - 3584\, \* \cf\, \* \cax3\,
          + 1792\, \* \cfx2\, \* \cax2\,
          - 896\, \* \cfx3\, \* \ca\,
          + 448\, \* \cfx4\,
   \nn \\[0.5mm] & & \mbox{\hspn}
          - 10752\, \* \nf\, \* \cf\, \* \cax2\,
          + 7168\, \* \nf\, \* \cfx2\, \* \ca\,
          - 2688\, \* \nf\, \* \cfx3\,
          + 1792\, \* \nfx2\, \* \cfx2\,\bigg)\,
   \nn \\[0.5mm] & & \mbox{\hspn}
       + N^{-8} \, \*  \bigg( \frct{4096}{3}\: \* \cax4\,
          - 2368\, \* \cf\, \* \cax3\,
          + \frct{7840}{3}\: \* \cfx2\, \* \cax2\,
          - \frct{6064}{3}\: \* \cfx3\, \* \ca\,
          + \frct{584}{3}\: \* \cfx4\,
   \nn \\[0.5mm] & & \mbox{\hspn}
          + 1792\, \* \nf\, \* \cax3\,
          + 4736\, \* \nf\, \* \cf\, \* \cax2\,
          - \frct{6272}{3}\: \* \nf\, \* \cfx2\, \* \ca\,
          + \frct{7424}{3}\: \* \nf\, \* \cfx3\,
          - \frct{1792}{3}\: \* \nfx2\, \* \cf\, \* \ca\,
   \nn \\[0.5mm] & & \mbox{\hspn}
          - \frct{11648}{3}\: \* \nfx2\, \* \cfx2\,\bigg)\,
       + N^{-7} \, \*  \bigg(           128\, \* \nfx2\, \* \cax2\,
          - \frct{52672}{27}\: \* \nfx2\, \* \cf\, \* \ca\,
          + \frct{427424}{27}\: \* \nfx2\, \* \cfx2\,
   \nn \\[0.5mm] & & \mbox{\hspn}
          - \frct{128}{9}\: \* \nfx3\, \* \cf\,
          + \frct{2}{3}\: \* (2915 - 13216\, \* \z2)\, \* \cfx4\,
          + \frct{16}{27}\: \* (5216 + 3375\, \* \z2)\, \* \nf\, \* \cax3\,
   \nn \\[0.5mm] & & \mbox{\hspn}
          - \frct{2}{27}\: \* (29293 - 244260\, \* \z2)\, \* \cfx3\, \* \ca\,
          + \frct{16}{27}\: \* (59326 + 8199\, \* \z2)\, \* \cax4\,
   \nn \\[0.5mm] & & \mbox{\hspn}
          + \frct{4}{27}\: \* (73415 - 115992\, \* \z2)\, \* \cfx2\, \* \cax2\,
          - \frct{8}{27}\: \* (81626 - 37539\, \* \z2)\, \* \cf\, \* \cax3\,
   \nn \\[0.5mm] & & \mbox{\hspn}
          - \frct{4}{27}\: \* (114685 - 57816\, \* \z2)\, \* \nf\, \* \cfx3\,
          + \frct{8}{27}\: \* (118813 - 41067\, \* \z2)\, \* \nf\, \* \cfx2\, \* \ca\,
   \nn \\[0.5mm] & & \mbox{\hspn}
          - \frct{8}{27}\: \* (181400 + 18351\, \* \z2)\, \* \nf\, \* \cf\, \* \cax2\,\bigg)\,\bigg\}
+ {\cal O}(N^{-6})\;,
\\[3mm]
   P_{\rm gq}^{\,(4)}(N) \!&\!=\!&\!
\cf\,\*\bigg\{ N^{-9} \, \*  \bigg( - 14336\, \* \cax4\,
          + 7168\, \* \cf\, \* \cax3\,
          - 3584\, \* \cfx2\, \* \cax2\,
          + 1792\, \* \cfx3\, \* \ca\,
   \nn \\[0.5mm] & & \mbox{\hspn}
          - 896\, \* \cfx4\,
          + 21504\, \* \nf\, \* \cf\, \* \cax2\,
          - 14336\, \* \nf\, \* \cfx2\, \* \ca\,
          + 5376\, \* \nf\, \* \cfx3\,
          - 3584\, \* \nfx2\, \* \cfx2\,\bigg)\,
   \nn \\[0.5mm] & & \mbox{\hspn}
       + N^{-8} \, \*  \bigg(          - 16128\, \* \cax4\,
          + \frct{43904}{3}\: \* \cf\, \* \cax3\,
          - \frct{31808}{3}\: \* \cfx2\, \* \cax2\,
          + 6944\, \* \cfx3\, \* \ca\,
          - 2240\, \* \cfx4\,
   \nn \\[0.5mm] & & \mbox{\hspn}
          - 3584\, \* \nf\, \* \cax3\,
          + \frct{46592}{3}\: \* \nf\, \* \cf\, \* \cax2\,
          - \frct{51968}{3}\: \* \nf\, \* \cfx2\, \* \ca\,
          + 4928\, \* \nf\, \* \cfx3\,
          + \frct{3584}{3}\: \* \nfx2\, \* \cf\, \* \ca\,
   \nn \\[0.5mm] & & \mbox{\hspn}
          + \frct{7168}{3}\: \* \nfx2\, \* \cfx2\,\bigg)\,
       + N^{-7} \, \*  \bigg(          - 256\, \* \nfx2\, \* \cax2\,
          + \frct{318208}{27}\: \* \nfx2\, \* \cf\, \* \ca\,
          - \frct{750464}{27}\: \* \nfx2\, \* \cfx2\,
   \nn \\[0.5mm] & & \mbox{\hspn}
          + \frct{256}{9}\: \* \nfx3\, \* \cf\,
          - \frct{112}{3}\: \* (42 - 437\, \* \z2)\, \* \cfx4\,
          - \frct{112}{27}\: \* (191 + 1017\, \* \z2)\, \* \cf\, \* \cax3\,
   \nn \\[0.5mm] & & \mbox{\hspn}
          + \frct{64}{27}\: \* (8005 - 5517\, \* \z2)\, \* \nf\, \* \cfx3\,
          - \frct{8}{27}\: \* (13313 + 104940\, \* \z2)\, \* \cfx3\, \* \ca\,
   \nn \\[0.5mm] & & \mbox{\hspn}
          - \frct{32}{27}\: \* (14392 + 3375\, \* \z2)\, \* \nf\, \* \cax3\,
          + \frct{8}{27}\: \* (17711 + 77652\, \* \z2)\, \* \cfx2\, \* \cax2\,
   \nn \\[0.5mm] & & \mbox{\hspn}
          - \frct{32}{27}\: \* (37616 + 17019\, \* \z2)\, \* \cax4\,
          - \frct{16}{27}\: \* (63557 - 20547\, \* \z2)\, \* 
            \nf\, \* \cfx2\, \* \ca\,
   \nn \\[0.5mm] & & \mbox{\hspn}
          + \frct{16}{27}\: \* (149746 + 32031\, \* \z2)\, \* 
            \nf\, \* \cf\, \* \cax2\,\bigg)\,\bigg\}
+ {\cal O}(N^{-4})\,,
\\[3mm]
   P_{\rm gg}^{\,(4)}(N) \!&\!=\!&\!
       N^{-9} \, \*  \bigg( - 14336\, \* \cax5\,
          + 28672\, \* \nf\, \* \cf\, \* \cax3\,
          - 10752\, \* \nf\, \* \cfx2\, \* \cax2\,
          + 3584\, \* \nf\, \* \cfx3\, \* \ca\,
   \nn \\[0.5mm] & & \mbox{\hspn}
          - 896\, \* \nf\, \* \cfx4\,
          - 10752\, \* \nfx2\, \* \cfx2\, \* \ca\,
          + 3584\, \* \nfx2\, \* \cfx3\, \bigg)\,
\,
       + N^{-8} \, \*  \bigg( - \frct{8960}{3}\: \* \cax5\,
   \nn \\[0.5mm] & & \mbox{\hspn}
          - \frct{17920}{3}\: \* \nf\, \* \cax4\,
          - \frct{14848}{3}\: \* \nf\, \* \cf\, \* \cax3\,
          - \frct{15040}{3}\: \* \nf\, \* \cfx2\, \* \cax2\,
          + \frct{9536}{3}\: \* \nf\, \* \cfx3\, \* \ca\,
   \nn \\[0.5mm] & & \mbox{\hspn}
          - \frct{1168}{3}\: \* \nf\, \* \cfx4\,
          + 5376\, \* \nfx2\, \* \cf\, \* \cax2\,
          + 10048\, \* \nfx2\, \* \cfx2\, \* \ca\,
          - \frct{11264}{3}\: \* \nfx2\, \* \cfx3\,
          - \frct{896}{3}\: \* \nfx3\, \* \cfx2\, \bigg)\,
   \nn \\[0.5mm] & & \mbox{\hspn}
       + N^{-7} \, \*  \bigg( 
          - \frct{640}{9}\: \* [907 + 396\, \* \z2]\, \* \cax5\,
          - \frct{640}{9}\: \* [164 + 81\, \* \z2]\, \* \nf\, \* \cax4\,
          - \frct{2560}{3}\: \* \nfx2\, \* \cax3\,
   \nn \\[0.5mm] & & \mbox{\hspn}
          + \frct{224}{9}\: \* [5438 + 1431\, \* \z2]\, \* 
            \nf\, \* \cf\, \* \cax3\,
          - \frct{4}{3}\: \* [2171 - 7692\, \* \z2]\, \* \nf\, \* \cfx4\,
   \nn \\[0.5mm] & & \mbox{\hspn}
          - \frct{8}{9}\: \* [43463 - 14940\, \* \z2]\, \* \nf\, \* \cfx2\, \* \cax2\,
          + \frct{4}{9}\: \* [18349 - 39132\, \* \z2]\, \* \nf\, \* \cfx3\, \* \ca\,
   \nn \\[0.5mm] & & \mbox{\hspn}
          + \frct{32}{9}\: \* [3274 + 495\, \* \z2]\, \* \nfx2\, \* \cf\, \* \cax2\,
          - \frct{16}{9}\: \* [38371 + 3978\, \* \z2]\, \* \nfx2\, \* \cfx2\, \* \ca\,
   \nn \\[0.5mm] & & \mbox{\hspn}
          + \frct{8}{9}\: \* [26605 - 5184\, \* \z2]\, \* \nfx2\, \* \cfx3\,
          + 256\, \* \nfx3\, \* \cf\, \* \ca\,
          - \frct{14080}{9}\: \* \nfx3\, \* \cfx2\,
      \bigg)\,
+ {\cal O}(N^{-6})\;.
\eea

Finally we present the corresponding results for the singlet structure
functions $F_2$ and $F_L$. The~quark coefficient functions can be written 
as sum of non-singlet and pure-singlet pieces,
\beq
\label{Ca-ps}
   C_{a,\rm q}(N) \;=\; C_{a}^{\,+}(N) + C_{a,\rm ps}(N) \; .
\eeq
The leading-logarithmic resummations of the pure-singlet and gluon coefficient
functions are
\bea
\label{eq:S_c2q_LL}
   C_{2,\rm ps}^{\,(n)}(N) \!&\!=\!&\!
   \dat{n}\frac{2^{n}}{N^{2n}}\sum_{i=0}^{\floor{\frac{n-2}{2}}}
   \sum_{k=0}^{n-2-2i}(-2)^{i+1+k}(\nf\cf)^{i+1}\cax{k}\cfx{\rho'}
   \binom{k+i}{k}\binom{\rho'+i+1}{\rho'}
\: , \\
\label{eq:S_c2g_LL}
   C_{2,\rm g}^{\,(n)}(N) \!&\!=\!&\!
   \nf\dat{n}\frac{2^{n}}{N^{2n}}\sum_{i=0}^{\floor{\frac{n-1}{2}}}
   \sum_{k=0}^{n-1-2i}(-2)^{i+k}(\nf\cf)^i\cax{k}\cfx{\delta'}
   \binom{k+i}{k}\binom{\delta'+i}{\delta'}
\: , \\
\label{eq:S_cLq_LL}
  C_{L,\rm ps}^{\,(n)}(N) \!&\!=\!&\!
  \dat{n-1}\frac{2^{n+1}}{N^{2n-2}}\sum_{i=0}^{\floor{\frac{n-2}{2}}}
  \sum_{k=0}^{n-2-2i}(-2)^{i+1+k}(\nf\cf)^{i+1}\cax{k}\cfx{\rho'}
  \binom{k+i}{k}\binom{\rho'+i+1}{\rho'}
\: , \quad \\
\label{eq:S_cLg_LL}
   C_{L,\rm g}^{\,(n)}(N) \!&\!=\!&\!
   \nf\dat{n-1}\frac{2^{n+1}}{N^{2n-2}}\sum_{i=0}^{\floor{\frac{n-1}{2}}}
   \sum_{k=0}^{n-1-2i}(-2)^{i+k}(\nf\cf)^i\cax{k}\cfx{\delta'}
   \binom{k+i}{k}\binom{\delta'+i}{\delta'}\;,
\eea
where $\rho' = n-k-2i-2$, $\delta' = n-k-2i-1$, and the $\dat{n}$ are 
defined as 
\bea
\label{eq:quad}
  \dat{n} &\!=\!&
  \frac{1}{n!}\,\prod_{k=0^{\phantom{1}\!\!\!}}^{n-1} 
  \; \left(1+4k\right) \;.
\eea
These are the coefficients of the Taylor expansion of 
\beq
  F \:=\: S^{\,-1/2} \:=\: (1-4\,\xi)^{-1/4} \;, 
\eeq
i.e., those of the leading-logarithmic contributions to the non-singlet 
coefficient functions (\ref{c2ns-cl}) -- (\ref{c3ns-cl}). Similarly to the 
splitting functions above, the first (second) binomial factor could be
interpreted as the number of ways $k$ ($\rho'$/$\delta'$) gluons can
be emitted from the gluon (quark) propagators. Again, analytic NLL
and NNLL results, which could require generalization of the LL singlet
coefficient functions, are not available at this point. 
Corresponding results for the scalar-exchange structure function $F_\phi$ 
in eq.~(\ref{evenp}), which is of only theoretical relevance, can be found 
in appendix~C.

Hence, at least for the time being, the coefficients for the NLL and NNLL 
contributions to these coefficient functions are only known at each order
separately. Since we do not expect yet higher orders to be become relevant
in the foreseeable future, we finally present these results at the fourth 
and fifth order in $\ar = \as/(4\,\pi)$. The former read
\bea
\label{c2q-as4}
  c_{2,\rm q}^{\,(4)}(N) \!&\!=\!&\! c_2^{\,+}(N)\big|_{\ar^{\,4}} +
   \nf\, \* \cf\, \* \bigg\{ N^{-8} \, \*  \bigg( - 3120\, \* \cax2\,
          + 3120\, \* \cf\, \* \ca\,
          - 2340\, \* \cfx2\,
          + 1560\, \* \nf\, \* \cf\,\bigg)\,
   \nn \\[0.5mm] & & \mbox{\hspn}
       + N^{-7} \, \*  \bigg(          - \frct{60872}{9}\: \* \cax2\,
          + \frct{86228}{9}\: \* \cf\, \* \ca\,
          - \frct{7798}{3}\: \* \cfx2\,
          + \frct{5216}{9}\: \* \nf\, \* \ca\,
          - \frct{16688}{9}\: \* \nf\, \* \cf\,\bigg)\,
   \nn \\[0.5mm] & & \mbox{\hspn}
       + N^{-6} \, \*  \bigg(          \frct{9848}{27}\: \* \nf\, \* \ca\,
          - \frct{952}{9}\: \* \nfx2\,
          - \frct{1}{3}\: \* (16611 - 21752\, \* \z2)\, \* \cfx2\,
   \nn \\[0.5mm] & & \mbox{\hspn}
          + \frct{8}{27}\: \* (24251 - 20439\, \* \z2)\, \* \cf\, \* \ca\,
          + \frct{2}{27}\: \* (124393 - 14688\, \* \z2)\, \* \nf\, \* \cf\,
   \nn \\[0.5mm] & & \mbox{\hspn}
          - \frct{2}{27}\: \* (242611 - 22752\, \* \z2)\, \* \cax2\,\bigg)\,\bigg\}
+ {\cal O}(N^{-5})\;,
\\[3mm]
  c_{2,\rm g}^{\,(4)}(N) \!&\!=\!&\!
\label{c2g-as4}
   \nf\, \* \bigg\{ N^{-8} \, \*  \bigg( - 3120\, \* \cax3\,
          + 1560\, \* \cf\, \* \cax2\,
          - 780\, \* \cfx2\, \* \ca\,
          + 390\, \* \cfx3\,
          + 3120\, \* \nf\, \* \cf\, \* \ca\,
   \nn \\[0.5mm] & & \mbox{\hspn}
          - 1560\, \* \nf\, \* \cfx2\,\bigg)\,
       + N^{-7} \, \*  \bigg(          - \frct{35132}{9}\: \* \cax3\,
          + \frct{30052}{9}\: \* \cf\, \* \cax2\,
          - \frct{21101}{9}\: \* \cfx2\, \* \ca\,
   \nn \\[0.5mm] & & \mbox{\hspn}
          + \frct{889}{3}\: \* \cfx3\,
          + \frct{536}{9}\: \* \nf\, \* \cax2\,
          - \frct{2056}{3}\: \* \nf\, \* \cf\, \* \ca\,
          + \frct{13778}{9}\: \* \nf\, \* \cfx2\,
          - \frct{2608}{9}\: \* \nfx2\, \* \cf\,\bigg)\,
   \nn \\[0.5mm] & & \mbox{\hspn}
       + N^{-6} \, \*  \bigg( - \frct{248}{27}\: \* \nfx2\, \* \ca\,
          + \frct{20300}{27}\: \* \nfx2\, \* \cf\,
          + \frct{52}{3}\: \* (771 - 41\, \* \z2)\, \* \nf\, \* \cf\, \* \ca\,
   \nn \\[0.5mm] & & \mbox{\hspn}
          + \frct{1}{6}\: \* (2453 - 23816\, \* \z2)\, \* \cfx3\,
          - \frct{4}{27}\: \* (2882 + 1647\, \* \z2)\, \* \nf\, \* \cax2\,
   \nn \\[0.5mm] & & \mbox{\hspn}
          + \frct{67}{27}\: \* (3265 - 1512\, \* \z2)\, \* \cf\, \* \cax2\,
          - \frct{1}{27}\: \* (19957 - 145440\, \* \z2)\, \* \cfx2\, \* \ca\,
   \nn \\[0.5mm] & & \mbox{\hspn}
          - \frct{8}{27}\: \* (25579 - 9972\, \* \z2)\, \* \nf\, \* \cfx2\,
          - \frct{10}{27}\: \* (48911 + 846\, \* \z2)\, \* \cax3\,\bigg)\,\bigg\}
+ {\cal O}(N^{-5})\;,
\\[3mm]
  c_{L,\rm q}^{\,(4)}(N) \!&\!=\!&\!c_L^{\,+}(N)\big|_{\ar^{\,4}} +
\label{cLq-as4}
   \nf\, \* \cf\, \* \bigg\{ N^{-6} \, \*  \bigg( - 1920\, \* \cax2\,
          + 1920\, \* \cf\, \* \ca\,
          - 1440\, \* \cfx2\,
          + 960\, \* \nf\, \* \cf\,\bigg)\,
   \nn \\[0.5mm] & & \mbox{\hspn}
       + N^{-5} \, \*  \bigg(          - \frct{24640}{9}\: \* \cax2\,
          + \frct{37408}{9}\: \* \cf\, \* \ca\,
          - \frct{2048}{3}\: \* \cfx2\,
          + \frct{2176}{9}\: \* \nf\, \* \ca\,
          - \frct{13024}{9}\: \* \nf\, \* \cf\,\bigg)\,
   \nn \\[0.5mm] & & \mbox{\hspn}
       + N^{-4} \, \*  \bigg(          - \frct{5696}{27}\: \* \nf\, \* \ca\,
          - \frct{32}{3}\: \* (49 - 361\, \* \z2)\, \* \cfx2\,
          - \frct{224}{27}\: \* (698 - 207\, \* \z2)\, \* \cax2\,
   \nn \\[0.5mm] & & \mbox{\hspn}
          - \frct{8}{27}\: \* (4913 + 11988\, \* \z2)\, \* \cf\, \* \ca\,
          + \frct{16}{27}\: \* (8461 - 1188\, \* \z2)\, \* \nf\, \* \cf\,
          - \frct{128}{3}\: \* \nfx2\,
          \bigg)\,\bigg\}
   \nn \\[1mm] & & \mbox{\hspn}
+ {\cal O}(N^{-3})
\eea
and
\bea
  c_{L,\rm g}^{\,(4)}(N) \!&\!=\!&\!
\label{cLg-as4}
   \nf\, \* \bigg\{ N^{-6} \, \*  \bigg( - 1920\, \* \cax3\,
          + 960\, \* \cf\, \* \cax2\,
          - 480\, \* \cfx2\, \* \ca\,
          + 240\, \* \cfx3\,
          + 1920\, \* \nf\, \* \cf\, \* \ca\,
   \nn \\[0.5mm] & & \mbox{\hspn}
          - 960\, \* \nf\, \* \cfx2\,\bigg)\,
       + N^{-5} \, \*  \bigg(          - \frct{8800}{9}\: \* \cax3\,
          + \frct{9248}{9}\: \* \cf\, \* \cax2\,
          - \frct{8296}{9}\: \* \cfx2\, \* \ca\,
   \nn \\[0.5mm] & & \mbox{\hspn}
          - \frct{16}{3}\: \* \cfx3\,
          - \frct{704}{9}\: \* \nf\, \* \cax2\,
          - \frct{4640}{3}\: \* \nf\, \* \cf\, \* \ca\,
          + \frct{13648}{9}\: \* \nf\, \* \cfx2\,
          - \frct{1088}{9}\: \* \nfx2\, \* \cf\,\bigg)\,
   \nn \\[0.5mm] & & \mbox{\hspn}
       + N^{-4} \, \*  \bigg( 
          - \frct{4}{3}\: \* (115 + 1964\, \* \z2)\, \* \cfx3\,
          - \frct{32}{27}\: \* (263 + 162\, \* \z2)\, \* \nf\, \* \cax2\,
   \nn \\[0.5mm] & & \mbox{\hspn}
          + \frct{16}{3}\: \* (1231 - 118\, \* \z2)\, \* \nf\, \* \cf\, \* \ca\,
          - \frct{32}{27}\: \* (6314 - 459\, \* \z2)\, \* \cax3\,
          + \frct{11776}{27}\: \* \nfx2\, \* \cf\,
   \nn \\[0.5mm] & & \mbox{\hspn}
          + \frct{4}{27}\: \* (6487 + 24048\, \* \z2)\, \* \cfx2\, \* \ca\,
          + \frct{8}{27}\: \* (8785 - 7722\, \* \z2)\, \* \cf\, \* \cax2\,
          - \frct{64}{27}\: \* \nfx2\, \* \ca\,
   \nn \\[0.5mm] & & \mbox{\hspn}
          - \frct{8}{27}\: \* (14249 - 5256\, \* \z2)\, \* \nf\, \* \cfx2\,\bigg)\,\bigg\}
+ {\cal O}(N^{-3})\,.
\eea
The non-singlet parts of eqs.~(\ref{c2q-as4}) and (\ref{cLq-as4}) have been 
given in eqs.~(\ref{c2ns4-cl}) and (\ref{cLns4-cl}) above. The highest three
$1/N$ poles of the corresponding 5-loop coefficient functions are given by
\bea
\label{c2q-as5}
  c_{2,\rm q}^{\,(5)}(N) \!&\!=\!&\! c_2^{\,+}(N)\big|_{\ar^{\,5}} +
   \nf\, \* \cf\, \* \bigg\{ N^{-10} \, \*  \bigg( 42432\, \* \cax3\,
          - 42432\, \* \cf\, \* \cax2\,
   \nn \\[0.5mm] & & \mbox{\hspn}
          + 31824\, \* \cfx2\, \* \ca\,
          - 21216\, \* \cfx3\,
          - 42432\, \* \nf\, \* \cf\, \* \ca\,
          + 31824\, \* \nf\, \* \cfx2\,\bigg)\,
   \nn \\[0.5mm] & & \mbox{\hspn}
       + N^{-9} \, \*  \bigg( \frct{5366608}{45}\: \* \cax3\,
          - \frct{7102528}{45}\: \* \cf\, \* \cax2\,
          + \frct{2208812}{15}\: \* \cfx2\, \* \ca\,
          - \frct{511648}{15}\: \* \cfx3\,
   \nn \\[0.5mm] & & \mbox{\hspn}
          - \frct{81248}{9}\: \* \nf\, \* \cax2\,
          - \frct{1361056}{45}\: \* \nf\, \* \cf\, \* \ca\,
          - \frct{243376}{15}\: \* \nf\, \* \cfx2\,
          + \frct{72448}{9}\: \* \nfx2\, \* \cf\,\bigg)\,
   \nn \\[0.5mm] & & \mbox{\hspn}
       + N^{-8} \, \*  \bigg( 
          - \frct{4}{5}\: \* (74593 - 180392\, \* \z2)\, \* \cfx3\,
          + \frct{2}{135}\: \* (7465355 - 11586096\, \* \z2)\, \* \cfx2\, \* \ca\,
   \nn \\[0.5mm] & & \mbox{\hspn}
          - \frct{16}{135}\: \* (3126887 - 924570\, \* \z2)\, \* \cf\, \* \cax2\,
          + \frct{16}{135}\: \* (3063709 - 69039\, \* \z2)\, \* \cax3\,
   \nn \\[0.5mm] & & \mbox{\hspn}
          + \frct{16}{135}\: \* (1390214 - 523683\, \* \z2)\, \* \nf\, \* \cfx2\,
          - \frct{16}{45}\: \* (429100 - 30021\, \* \z2)\, \* \nf\, \* \cf\, \* \ca\,
   \nn \\[0.5mm] & & \mbox{\hspn}
          - \frct{16}{135}\: \* (102961 - 37125\, \* \z2)\, \* \nf\, \* \cax2\,
          - \frct{2472352}{135}\: \* \nfx2\, \* \cf\,
          + \frct{17696}{9}\: \* \nfx2\, \* \ca\,
   \bigg)\,\bigg\}
   \nn \\[1mm] & & \mbox{\hspn}
+ {\cal O}(N^{-7})\;,
\\[3mm]
  c_{2,\rm g}^{\,(5)}(N) \!&\!=\!&\!
\label{c2g-as5}
    \nf\, \* \bigg\{ N^{-10} \, \*  \bigg( 42432\, \* \cax4\,
          - 21216\, \* \cf\, \* \cax3\,
          + 10608\, \* \cfx2\, \* \cax2\,
          - 5304\, \* \cfx3\, \* \ca\,
   \nn \\[0.5mm] & & \mbox{\hspn}
          + 2652\, \* \cfx4\,
          - 63648\, \* \nf\, \* \cf\, \* \cax2\,
          + 42432\, \* \nf\, \* \cfx2\, \* \ca\,
          - 15912\, \* \nf\, \* \cfx3\,
          + 10608\, \* \nfx2\, \* \cfx2\,\bigg)\,
   \nn \\[0.5mm] & & \mbox{\hspn}
       + N^{-9} \, \*  \bigg(          \frct{3616288}{45}\: \* \cax4\,
          - \frct{2668904}{45}\: \* \cf\, \* \cax3\,
          + \frct{1762432}{45}\: \* \cfx2\, \* \cax2\,
          - \frct{1089206}{45}\: \* \cfx3\, \* \ca\,
   \nn \\[0.5mm] & & \mbox{\hspn}
          + \frct{50356}{15}\: \* \cfx4\,
          - \frct{17600}{9}\: \* \nf\, \* \cax3\,
          - \frct{1868624}{45}\: \* \nf\, \* \cf\, \* \cax2\,
          + \frct{1566016}{45}\: \* \nf\, \* \cfx2\, \* \ca\,
   \nn \\[0.5mm] & & \mbox{\hspn}
          + \frct{334184}{45}\: \* \nf\, \* \cfx3\,
          + \frct{81248}{9}\: \* \nfx2\, \* \cf\, \* \ca\,
          - \frct{1239104}{45}\: \* \nfx2\, \* \cfx2\,\bigg)\,
       + N^{-8} \, \*  \bigg( \frct{12464}{27}\: \* \nfx2\, \* \cax2\,
   \nn \\[0.5mm] & & \mbox{\hspn}
          - \frct{1105856}{135}\: \* \nfx2\, \* \cf\, \* \ca\,
          - \frct{8848}{9}\: \* \nfx3\, \* \cf\,
          + \frct{16}{135}\: \* (39757 + 61020\, \* \z2)\, \* \nf\, \* \cax3\,
   \nn \\[0.5mm] & & \mbox{\hspn}
          + \frct{1}{15}\: \* (59357 - 673108\, \* \z2)\, \* \cfx4\,
          - \frct{4}{135}\: \* (365911 - 3205476\, \* \z2)\, \* \cfx3\, \* \ca\,
   \nn \\[0.5mm] & & \mbox{\hspn}
          - \frct{4}{15}\: \* (648293 - 174498\, \* \z2)\, \* \cf\, \* \cax3\,
          + \frct{2}{45}\: \* (1977587 - 1966200\, \* \z2)\, \* \cfx2\, \* \cax2\,
   \nn \\[0.5mm] & & \mbox{\hspn}
          + \frct{8}{135}\: \* (2407760 - 1256427\, \* \z2)\, \* \nf\, \* \cfx2\, \* \ca\,
          + \frct{4}{135}\: \* (3500111 - 241380\, \* \z2)\, \* \nfx2\, \* \cfx2\,
   \nn \\[0.5mm] & & \mbox{\hspn}
          - \frct{2}{135}\: \* (4630465 - 3452868\, \* \z2)\, \* \nf\, \* \cfx3\,
          + \frct{4}{135}\: \* (11350279 + 666720\, \* \z2)\, \* \cax4\,
   \nn \\[0.5mm] & & \mbox{\hspn}
          - \frct{4}{135}\: \* (12031717 + 12510\, \* \z2)\, \* \nf\, \* \cf\, \* \cax2\,\bigg)\,\bigg\}
+ {\cal O}(N^{-7})\;,
\\[3mm]
  c_{L,\rm q}^{\,(5)}(N) \!&\!=\!&\! c_L^{\,+}(N)\big|_{\ar^{\,5}} +
   \nf\, \* \cf\, \* \bigg\{ N^{-8} \, \*  \bigg( 24960\, \* \cax3\,
          - 24960\, \* \cf\, \* \cax2\,
   \nn \\[0.5mm] & & \mbox{\hspn}
          + 18720\, \* \cfx2\, \* \ca\,
          - 12480\, \* \cfx3\,
          - 24960\, \* \nf\, \* \cf\, \* \ca\,
          + 18720\, \* \nf\, \* \cfx2\,\bigg)\,
   \nn \\[0.5mm] & & \mbox{\hspn}
       + N^{-7} \, \*  \bigg(  
          + \frct{436192}{9}\: \* \cax3\,
          - \frct{602368}{9}\: \* \cf\, \* \cax2\,
          + \frct{198248}{3}\: \* \cfx2\, \* \ca\,
          - \frct{33904}{3}\: \* \cfx3\,
   \nn \\[0.5mm] & & \mbox{\hspn}
          - \frct{30400}{9}\: \* \nf\, \* \cax2\,
          - \frct{27328}{9}\: \* \nf\, \* \cf\, \* \ca\,
          - \frct{56272}{3}\: \* \nf\, \* \cfx2\,
          + \frct{33920}{9}\: \* \nfx2\, \* \cf\,\bigg)\,
   \nn \\[0.5mm] & & \mbox{\hspn}
       + N^{-6} \, \*  \bigg(  
          - \frct{56}{3}\: \* (359 - 4436\, \* \z2)\, \* \cfx3\,
          - \frct{8}{27}\: \* (52207 + 341604\, \* \z2)\, \* \cfx2\, \* \ca\,
   \nn \\[0.5mm] & & \mbox{\hspn}
          - \frct{16}{27}\: \* (180227 - 109944\, \* \z2)\, \* \cf\, \* \cax2\,
          + \frct{16}{27}\: \* (218827 - 20610\, \* \z2)\, \* \cax3\,
   \nn \\[0.5mm] & & \mbox{\hspn}
          - \frct{32}{9}\: \* (18590 - 2271\, \* \z2)\, \* \nf\, \* \cf\, \* \ca\,
          + \frct{8}{27}\: \* (278473 - 110304\, \* \z2)\, \* \nf\, \* \cfx2\,
   \nn \\[0.5mm] & & \mbox{\hspn}
          + \frct{160}{27}\: \* (278 + 513\, \* \z2)\, \* \nf\, \* \cax2\,
          + \frct{6208}{9}\: \* \nfx2\, \* \ca\,
          - \frct{305536}{27}\: \* \nfx2\, \* \cf\,
   \bigg)\,\bigg\}
+ {\cal O}(N^{-5})
\eea
and
\bea
  c_{L,\rm g}^{\,(5)}(N) \!&\!=\!&\!
\label{cLg-as5}
   \nf\, \* \bigg\{ N^{-8} \, \*  \bigg( 24960\, \* \cax4\,
          - 12480\, \* \cf\, \* \cax3\,
          + 6240\, \* \cfx2\, \* \cax2\,
          - 3120\, \* \cfx3\, \* \ca\,
   \nn \\[0.5mm] & & \mbox{\hspn}
          + 1560\, \* \cfx4\,
          - 37440\, \* \nf\, \* \cf\, \* \cax2\,
          + 24960\, \* \nf\, \* \cfx2\, \* \ca\,
          - 9360\, \* \nf\, \* \cfx3\,
          + 6240\, \* \nfx2\, \* \cfx2\,\bigg)\,
   \nn \\[0.5mm] & & \mbox{\hspn}
       + N^{-7} \, \*  \bigg( \frct{230272}{9}\: \* \cax4\,
          - \frct{178352}{9}\: \* \cf\, \* \cax3\,
          + \frct{139120}{9}\: \* \cfx2\, \* \cax2\,
          - \frct{93860}{9}\: \* \cfx3\, \* \ca\,
   \nn \\[0.5mm] & & \mbox{\hspn}
          + \frct{2140}{3}\: \* \cfx4\,
          + \frct{7040}{9}\: \* \nf\, \* \cax3\,
          + \frct{22528}{9}\: \* \nf\, \* \cf\, \* \cax2\,
          - \frct{1856}{9}\: \* \nf\, \* \cfx2\, \* \ca\,
          + \frct{96152}{9}\: \* \nf\, \* \cfx3\,
   \nn \\[0.5mm] & & \mbox{\hspn}
          + \frct{30400}{9}\: \* \nfx2\, \* \cf\, \* \ca\,
          - \frct{164144}{9}\: \* \nfx2\, \* \cfx2\,\bigg)\,
       + N^{-6} \, \*  \bigg(          \frct{3424}{27}\: \* \nfx2\, \* \cax2\,
          - \frct{200672}{27}\: \* \nfx2\, \* \cf\, \* \ca\,
   \nn \\[0.5mm] & & \mbox{\hspn}
          - \frct{3104}{9}\: \* \nfx3\, \* \cf\,
          - \frct{2}{3}\: \* (1261 + 42056\, \* \z2)\, \* \cfx4\,
          + \frct{32}{27}\: \* (4967 + 4212\, \* \z2)\, \* \nf\, \* \cax3\,
   \nn \\[0.5mm] & & \mbox{\hspn}
          + \frct{16}{27}\: \* (101897 - 7344\, \* \z2)\, \* \nfx2\, \* \cfx2\,
          + \frct{8}{27}\: \* (191519 - 130122\, \* \z2)\, \* \nf\, \* \cfx2\, \* \ca\,
   \nn \\[0.5mm] & & \mbox{\hspn}
          + \frct{8}{27}\: \* (458999 + 12744\, \* \z2)\, \* \cax4\,
          - \frct{8}{27}\: \* (542135 - 18342\, \* \z2)\, \* \nf\, \* \cf\, \* \cax2\,
   \nn \\[0.5mm] & & \mbox{\hspn}
          + 4\, \* (5905 - 13404\, \* \z2)\, \* \cfx2\, \* \cax2\,
          + \frct{16}{27}\: \* (11939 + 102024\, \* \z2)\, \* \cfx3\, \* \ca\,
   \\[0.5mm] & & \mbox{\hspn}
          - \frct{8}{3}\: \* (22553 - 9334\, \* \z2)\, \* \cf\, \* \cax3\,
          - \frct{20}{27}\: \* (46153 - 35208\, \* \z2)\, \* \nf\, \* \cfx3\,
   \bigg)\,\bigg\}
+ {\cal O}(N^{-5})\;, \nn
\eea
where the non-singlet contributions can be found in eqs.~(\ref{c2ns5-cl}) 
and (\ref{cLns5-cl}).


\setcounter{equation}{0}
\section{Summary and outlook}
\label{sec:summary}

We have presented a comprehensive study of high-energy double logarithms 
appearing at the \mbox{$n$-th} order in perturbation theory in the splitting 
functions for the evolution of the parton distribution functions and in the 
coefficient functions of inclusive deep-inelastic scattering.
These have the structure $\asn \, x^{\,p} \ln^{\,2n - n_0^{} - k} x\,$, where
$p \geq 0$ in gauge-boson--exchange DIS, and the parameter $n_0^{}$ depends 
on the quantity under consideration.
$k$ denotes the logarithmic accuracy: $k=0$ provides the leading-logarithmic
(LL) terms, $k=1$ the next-to-leading logarithmic (NLL) contributions etc.

For the flavour non-singlet quantities, the dominant contributions start at 
$p = 0$ with an offset of $n_0^{}=2$ for the splitting functions $P^{\,\pm}$ 
and the coefficient function $C^{\,\pm}_L$, and of $n_0^{} = 1$ for the 
coefficient functions $C_2^{\,\pm}$ and $C_3^{\,\pm}$.
The structure of the unfactorized $n$-th order partonic structure functions 
in dimensional regularization has been employed to perform a NNLL resummation 
of these small-$x$ logarithms to all orders in full QCD for the splitting 
function $P^{\,+}$ and the coefficient functions $C^{\,+}_2$, $C^{\,+}_3$ and 
$C^{\,+}_L$; the former can be expressed in terms of modified Bessel functions.
Using, in addition, the known structure of the singular terms for $P^{\,+}$ in 
Mellin $N$-space as $N\to 0$, the all-order resummation of the leading $\ln x$ 
terms has been pushed up to N$^7$LL accuracy in the large-$n_c$ limit, where 
the functions $P^{\,+}$ and $\,P^{\,-}$ coincide. For the coefficient functions 
$C^{\,+}_2$ and $C^{\,+}_3$ the large-$n_c$ resummation has been extended to 
N$^3$LL accuracy.
In all cases, explicit fixed-order expansions up to the fifth order in 
perturbation theory have been presented for future reference.

In the flavour-singlet sector of standard DIS, the dominant small-$x$ 
contributions at the $n$-th order are proportional to one inverse power 
of $x$ enhanced by single logarithms, i.e., the splitting functions 
$P_{\,\rm ik}$ and the singlet coefficient functions $C_2$ and $C_L$ at 
$n$ loops are of the form $x^{\,-1} \ln^{\,m} x\,$ with 
$m = 1,\, \ldots, \,n - n_0^{\,\prime}$ in the small-$x$ limit.
Our study has not added any new information on these terms or their 
all-order resummation, which is a long-standing and prominent, yet in 
its full generality still open problem in QCD.
Instead, we have considered the double logarithms appearing with powers 
$x^{\:\!p}$ for even $p \geq 0$, which correspond to expansions in Mellin 
$N$-space around $(N+p)=0$.
In these cases, we can again use dimensional regularization and exploit
the structure of the unfactorized partonic structure functions at the 
$n$-th order in perturbation theory.
We have presented NNLL small-$x$ ($p=0$) predictions up to five loops in 
full QCD. We can compute order-by-order to a very high power of $\as$ in 
this framework, but we have not been able to find an all-order form which 
generates these results beyond the leading logarithms.

The results, certainly in the singlet sector, are not of immediate 
phenomenological relevance in DIS.
However, they elucidate the analytic structure underlying the expressions 
at fixed order in perturbation theory. In addition, they provide important 
information for complete analytic computations of those quantities.
In this regard, the results for the non-singlet splitting functions have 
already been used in the determination of the all-$N$ expression of the 
large-$\nc$ non-singlet splitting function at four loops, based on 
a limited number of Mellin moments together with constraints on its
endpoint behaviour and its the functional form.
This application also allowed for important independent checks on the 
methods employed in the present article for the study of the high-energy 
(single and) double logarithms, and in turn provided input coefficients
for the N$^3$LL resummations.
 
While the intriguing structures of the resummations performed contribute 
to a much improved theoretical understanding, the chosen approach has also
clear limitations. Most notably the leading $p=0$ contributions to the
non-singlet splitting function $P^{\,-}$ and the main $\nu\! +\! \bar{\nu}$
charged-current structure function $F_3$ are not accessible beyond the
large-$\nc$ limit. Progress in this direction will require new methods in 
addition to those developed and considered here. 
The closed-form all-order resummation of the small-$x$ double logarithms in 
the singlet sector remains an open problem, pending the identification of the
proper set of functions, which complement the modified Bessel functions found
to suffice in the non-singlet sector. 
Finally, the systematic study of DIS with an exchanged scalar and the 
implications for the flavour-singlet coefficient functions is a subject we 
have touched upon only briefly, with a few results presented in the appendix.
These issues deserve further thorough investigation, which we leave for
the future.

\vspace{1mm}
\noindent
An ancillary files with our results in {\sc Form} format is available 
from {\tt http://arXiv.org}.

\vspace{-3mm}
\subsection*{Acknowledgements}
\vspace{-2mm}

This work has been supported 
by the UK {\it Science \& Technology Facilities Council}$\,$ (STFC) 
grants ST/J000493/1, ST/L000431/1 and ST/T000988/1 (University of Liverpool) 
and ST/T00102X/1 (University of Sussex),
and by the {\it Deutsche Forschungsgemeinschaft} (DFG) through the Research 
Unit FOR 2926, {\it Next Generation pQCD for Hadron Structure: Preparing for 
the EIC}, project number 40824754 and DFG grant MO~1801/4-1.

\pagebreak


\renewcommand{\theequation}{A.\arabic{equation}}
\setcounter{equation}{0}
\section*{Appendix A: expansions about $\,\bm N=0\,$ to order $\bm\asth$}
 
Here we present the expansions of the fixed-order results used for the 
resummation of the $x^{\,0} \ln^{\,k} x$ contributions to the even-$N$ based
splitting and coefficient functions.
The N$^{\rm n}$LL predictions are fixed by the corresponding N$^{\rm n}$LO 
results, hence only the $N^{-4}$ coefficients of the fourth-order splitting 
functions $P^{\,(3)}$ are missing for the N$^3$LL resummation of $F_2$, $F_3$ 
and $F_\phi$.
In view of a future determination of these quantities, the results below are 
given at N$^3$LL accuracy. 
 
The corresponding expressions for the LO, NLO and N$^2$LO non-singlet splitting 
functions read
\bea
  P^{\,+\,(0)} \!&\!=\!&  
  \cf\*\bigg\{
        2\,\*\NMx1 
  \,+\, 1 
  \,+\, [2-4\*\z2]\,\* N 
  \,-\, [2-4\*\z3]\,\* \NPx2 
       \bigg\}
  \:+\: {\cal O}(\NPx3)
\; , \nn\\[1mm]
  P^{\,+\,(1)}  \!&\!=\!& 
  \cf\*\bigg\{
         4\*\,\cf\*\NMx3 
  \,+\, (4\*\,\cf-2\*\bo)\,\*\NMx2
    + \bigg(
        \frct{20}{3}\,\*\ca 
    - [4+8\*\z2]\,\*\cf 
    + \frct{22}{3}\,\*\bo
      \bigg) \* \NMx1
\nn\\&&\hphantom{\cf\bigg\{}\mbox{\hspn}
    + \bigg(
       - \bigg[\frct{17}{3} + 12\,\*\z3\bigg]\,\*\ca
       + \bigg[\frct{19}{2} + 16\,\*\z3\bigg]\,\*\cf
       - \frct{29}{6}\,\*\bo
    \bigg)
      \bigg\}
  \:+\: {\cal O}(\NPx1)
\; , \nn\\[1mm]
  P^{\,+\,(2)}  \!&\!=\!& 
  \cf\*\bigg\{
       16\*\,\cfs\,\* \NMx5 
  \,+\, (24\*\,\cfs - 12\*\,\cf\*\bo)\,\*\NMx4 
     - \bigg(
       60\*\z2\,\*\cas 
     - \bigg[\frct{80}{3} + 192\*\z2\bigg]\*\cf\*\ca 
\nn\\&&\hphantom{\cf\bigg\{}\mbox{\hspn}
     - [8- 208\*\z2]\*\,\cfs 
     - \frct{64}{3}\,\*\,\cf\*\bo 
     - 2\*\bos\bigg)\* \NMx3
      + \bigg(
        [14 + 48\,\*\z2]\,\* \cas
\nn\\&&\hphantom{\cf\bigg\{}\mbox{\hspn}
      + [30 + 192\,\*\z2 + 96\,\*\z3]\,\* \cfs
      - \bigg[-\frct{38}{3} + 216\,\*\z2 + 48\,\*\z3\bigg]\,\*\ca\,\*\cf
      - \frct{22}{3}\,\*\bos
\nn\\&&\hphantom{\cf\bigg\{}\mbox{\hspn}
      - \bigg[\frct{50}{3} - 12\,\*\z2\bigg]\,\*\ca\,\*\bo
      - \frct{44}{3}\,\*\cf\,\*\bo
      \bigg)\* \NMx2
      \bigg\}
  \:+\: {\cal O}(\NMx1)\,.
\eea
The input coefficient functions for $\fhat{2}$ in eq.~(\ref{FhatFact}) 
in Laurent expansions analogous to eq.~(\ref{PexpN0})~are
\bea
c_2^{\,+\,(1,0)} \!&\!=\!& 
\cf\*\bigg\{2\*N^{-2} + 3\*N^{-1} - [5+2\*\z2]
	- [4 - 5\,\*\z2 + 2\,\*\z3]\,\*N
\bigg\}
+ {\cal O}(N^2)\,,\nn\\
c_2^{\,+\,(1,1)} \!&\!=\!&
\cf\*\bigg\{-2\*N^{-3}-3\*N^{-2} + [5+3\*\z2]\*N^{-1}
	- [10 - \frct{7}{2}\,\*\z2]
\bigg\}
+ {\cal O}(N^1) \,, \nn\\
c_2^{\,+\,(1,2)} \!&\!=\!&
\cf\*\bigg\{2\*N^{-4} +3\*N^{-3} - [5+3\*\z2]\*N^{-2}
	+ [10 -\frct{9}{2}\,\*\z2 + \frct{14}{3}\,\*\z3]\,\*N^{-1}
\bigg\}
+ {\cal O}(N^{0})\,, \nn\\
c_2^{\,+\,(2,0)} \!&\!=\!& 
\cf\*\bigg\{10\*\,\cf\*N^{-4} + (- 5\*\bo + 18\*\cf)\*N^{-3} +( 10\*\,\ca - [17+24\*\z2]\*\cf +6\*\bo)\*N^{-2}
\nn\\&&\hphantom{\cf\bigg\{}\mbox{\hspn}
	+ \bigg(
		  \bigg[\frct{3}{2} - 8\,\*\z2 + 56\,\*\z3\bigg]\,\*\cf
		- \bigg[\frct{119}{9} + 12\,\*\z3\bigg]\,\*\ca
		- \bigg[\frct{89}{18} - 4\,\*\z2\bigg]\,\*\bo
	 \bigg)\,\*N^{-1}
\bigg\} 
\nn\\&&\vphantom{\bigg\{}\mbox{\hspn}
+ {\cal O}(N^{0})\,,\nn\\
c_2^{\,+\,(2,1)} \!&\!=\!&
\cf\*\bigg\{-26\*\,\cf\*N^{-5} +(13\*\bo - 50\*\cf)\*N^{-4}
	+ \bigg([47+68\*\z2]\*\cf - \frct{70}{3}\*\ca - \frct{32}{3}\*\bo\bigg)\*N^{-3}
\nn\\&&\hphantom{\cf\bigg\{}\mbox{\hspn}
	+ \bigg(
		  [34 + 24\,\*\z3]\,\*\ca
		- [49 - 54\,\*\z2 + 128\,\*\z3]\,\*\cf
		+ [10 - 14\,\*\z2]\,\*\bo
	\bigg)\*N^{-2}
\bigg\} 
\nn\\&&\vphantom{\bigg\{}\mbox{\hspn}
+ {\cal O}(N^{-1})\,, \nn\\
c_2^{\,+\,(3,0)} \!&\!=\!& 
\cf\*\bigg\{60\*\,\cfs\*N^{-6}
	+ \bigg(-\frct{182}{3}\*\cf\*\bo + 134\*\cfs\bigg)\*N^{-5}
	+ \bigg(\bigg[\frct{260}{3} + 384\*\z2\bigg]\*\cf\*\ca
\nn\\&&\hphantom{\cf\bigg\{}\mbox{\hspn}
		- 120\*\z2\*\cas
		- [30+524\*\z2]\*\cfs + \frct{46}{3}\*\bos + \frct{5}{3}\*\cf\*\bo
	\bigg)\*N^{-4}
\nn\\&&\hphantom{\cf\bigg\{}\mbox{\hspn}
	+ \bigg(
		  \bigg[\frct{112}{3} + 80\,\*\z2\bigg]\,\*\cas
		+ \bigg[\frct{1315}{27} + \frct{266}{3}\,\*\z2\bigg]\,\*\cf\,\*\bo
		- \bigg[\frct{580}{9} - 24\,\*\z2\bigg]\,\*\ca\,\*\bo
\nn\\&&\hphantom{\cf\bigg\{}\mbox{\hspn}
		- \bigg[\frct{113}{3} - \frct{598}{3}\,\*\z2 - \frct{1292}{3}\,\*\z3\bigg]\,\*\cfs
		+ \bigg[\frct{950}{27} - 384\,\*\z2 - 128\,\*\z3\bigg]\,\*\ca\,\*\cf
\nn\\&&\hphantom{\cf\bigg\{}\mbox{\hspn}
		- \frct{248}{9}\,\*\bos
	\bigg)\*N^{-3}
\bigg\}
+ {\cal O}(N^{-2})\,.
\eea
The corresponding expansion coefficients for the longitudinal structure
function are given by
\begin{eqnarray}
  c_L^{\,+\,(1,0)} \!&\!=\!& 
  \cf\*\bigg\{4\*N^{0} - 4\*N + 4\*N^2\bigg\}
+ {\cal O}(N^{3})\,,\nn\\
  c_L^{\,+\,(1,1)} \!&\!=\!&
\cf\*\bigg\{4\*N^{0} - [4 - 4\*\z2]\*N\bigg\}
+ {\cal O}(N^{2})\,,\nn\\
c_L^{\,+\,(1,2)} \!&\!=\!&
\cf\*[8 - 2\*\z2]\*N^{0}
+ {\cal O}(N)\,,\nn\\
c_L^{\,+\,(2,0)} \!&\!=\!&
  \cf\*\bigg\{8\*\,\cf\*N^{-2} + (12\*\cf - 4\*\bo)\*N^{-1}
  + \bigg( \frct{40}{3}\*\ca - [74+8\*\z2]\*\cf + \frct{38}{3}\*\bo\bigg)\*N^{0} \bigg\}
\nn\\&&\vphantom{\bigg\{}\mbox{\hspn}
  + {\cal O}(N)\,,\nn\\
  c_L^{\,+\,(2,1)} \!&\!=\!&
\cf\*\bigg\{- 8\*\,\cf\*N^{-3} + ( 4\*\bo - 4\*\cf)\*N^{-2}+ \bigg([70 + 20\*\z2]\*\cf - \frct{40}{3}\*\ca - \frct{50}{3}\*\bo \bigg)\*N^{-1} \bigg\}
\nn\\&&\vphantom{\bigg\{}\mbox{\hspn}
  + {\cal O}(N^{0})\,, \nn\\
  c_L^{\,+\,(3,0)} \!&\!=\!& 
  \cf\*\bigg\{40\*\,\cfs\*N^{-4} + (64\*\cfs - 36\*\cf\*\bo)\*N^{-3}
  + \bigg(\bigg[\frct{200}{3} + 384\*\z2\bigg]\*\cf\*\ca  - 120\*\z2\*\cas
\nn\\&&\hphantom{\cf\bigg\{}\mbox{\hspn}
  - [168+416\*\z2]\*\cfs + \frct{112}{3}\*\cf\*\bo + 8\*\bos \bigg)\*N^{-2}\bigg\}
  + {\cal O}(N^{-1})\,,
\end{eqnarray}
and the input coefficient functions for the even-$N$ based $\fhat{3}$ read
\bea
c_3^{\,+\,(1,0)} \!&\!=\!& 
\cf\*\bigg\{2\*N^{-2} + N^{-1} - [7+2\*\z2] - [2 - 5\,\*\z2 + 2\,\*\z3]\,\*N \bigg\}
+ {\cal O}(N^2)\,,\nn\\
c_3^{\,+\,(1,1)} \!&\!=\!&
\cf\*\bigg\{-2\*N^{-3}-N^{-2} + [1+3\*\z2]\*N^{-1} - \bigg[14 - \frct{3}{2}\,\*\z2\bigg] \bigg\}
+ {\cal O}(N) \,, \nn\\
c_3^{\,+\,(1,2)} \!&\!=\!&
\cf\*\bigg\{2\*N^{-4} + N^{-3} - [1+3\*\z2]\*N^{-2} - \bigg[\frct{3}{2}\,\*\z2 - \frct{14}{3}\,\*\z3\bigg]\,\*N^{-1} \bigg\}
+ {\cal O}(N^{0})\,, \nn\\
c_3^{\,+\,(2,0)} \!&\!=\!& 
	\cf\*\bigg\{10\*\,\cf\*N^{-4} + (10\*\cf - 5\*\bo)\*N^{-3} + ( 10\*\,\ca - [33+24\*\z2]\*\cf +10\*\bo)\*N^{-2}
\nn\\&&\hphantom{\cf\*\bigg\{}{}
	+\bigg(
	\bigg[\frct{29}{2} + 4\,\*\z2 + 56\,\*\z3\bigg]\,\*\cf
	- \bigg[\frct{179}{9} + 12\,\*\z3\bigg]\,\*\ca
	- \bigg[\frct{131}{18} - 4\,\*\z2\bigg]\,\*\bo
	\bigg)\,\*N^{-1}
	\bigg\} 
\nn\\&&{}
+ {\cal O}(N^{0})\,,\nn\\
c_3^{\,+\,(2,1)} \!&\!=\!&
	\cf\*\bigg\{-26\*\,\cf\*N^{-5} +(13\*\bo - 26\*\cf)\*N^{-4}
	+\bigg([71+68\*\z2]\*\cf -\frct{70}{3}\*\ca - \frct{68}{3}\*\bo\bigg)\*N^{-3}
\nn\\&&\hphantom{\cf\*\bigg\{}{}
	+\bigg(
		 [54 + 24\,\*\z3]\,\*\ca
		- [120 - 8\,\*\z2 + 128\,\*\z3]\,\*\cf
		+ [25 - 14\,\*\z2]\,\*\bo
	\bigg)\,\*N^{-2}
\bigg\} 
\nn\\&&\vphantom{\bigg\{}{}
+ {\cal O}(N^{-1})\,, \nn\\
c_3^{\,+\,(3,0)} \!&\!=\!& 
	\cf\*\bigg\{60\*\,\cfs\*N^{-6}
	+\bigg(90\*\cfs - \frct{182}{3}\*\cf\*\bo\bigg)\*N^{-5}
	+\bigg(\bigg[\frct{260}{3}+384\*\z2\bigg]\*\cf\*\ca -120\*\z2\*\cas
\nn\\&&\hphantom{\cf\*\bigg\{}{}
	- [142+524\*\z2]\*\cfs + \frct{46}{3}\*\bos + \frct{143}{3}\*\cf\*\bo \bigg)\*N^{-4}
	+ \bigg(
		  \bigg[\frct{112}{3} + 140\,\*\z2\bigg]\,\*\cas
\nn\\&&\hphantom{\cf\*\bigg\{}{}
		- \bigg[\frct{47}{3} - \frct{1438}{3}\,\*\z2 - \frct{1292}{3}\,\*\z3\bigg]\,\*\cfs
		- \bigg[\frct{670}{27} + 576\,\*\z2 + 128\,\*\z3\bigg]\,\*\ca\,\*\cf
\nn\\&&\hphantom{\cf\*\bigg\{}{}
		+ \bigg[\frct{1909}{27} + \frct{266}{3}\,\*\z2\bigg]\,\*\cf\,\*\bo
		- \bigg[\frct{580}{9} - 24\,\*\z2\bigg]\,\*\ca\,\*\bo
		- \frct{356}{9}\,\*\bos
	\bigg)\,\*N^{-3}
	\bigg\}
\nn\\&&\vphantom{\bigg\{}{}
+ {\cal O}(N^{-2})\,.
\eea

Next we present the input quantities for the singlet cases.  
The `diagonal' quantities can be separated into non-singlet and
pure-singlet pieces.
The quark cases in the following expressions are written in terms of the 
non-singlet parts presented above. Here we present the input expressions
only to N$^2$LL accuracy.
The singlet splitting functions, expanded about $N=0$ are given by
\begin{eqnarray}
P_{\,\rm qq}^{\,(0)} \!&\!=\!&\! P^{\,+\,(0)}\,, \nn\\
P_{\,\rm qq}^{\,(1)} \!&\!=\!&\! P^{\,+\,(1)} 
+ \nf\*\cf\*\Big\{-8\*N^{-3} - 4\*N^{-2} - 8\*N^{-1}\Big\}
+ {\cal O}(N^0)\,, \nn\\
P_{\,\rm qq}^{\,(2)} \!&\!=\!&\! P^{\,+\,(2)} 
+ \nf\*\cf\*\bigg\{(64\*\ca-64\*\cf\,)\*N^{-5} 
    + \bigg(\frct{232}{3}\*\ca-24\*\cf-\frct{16}{3}\*\nf\bigg)\*N^{-4}
\nn\\&&\hphantom{ P^{\,+\,(2)} + \nf\*\cf\*\bigg\{}\mbox{\hspn}
	+ \bigg(\bigg[\frct{404}{9}+8\*\z2\bigg]\*\ca
	-[160-96\*\z2]\*\cf + \frct{232}{9}\*\nf\bigg)\*N^{-3}\bigg\}
+ {\cal O}(N^{-2})\,, \nn\\
P_{\,\rm qg}^{\,(0)} \!&\!=\!&\!
\nf\*\bigg\{2\*N^{-1} - 2 + 3\*N \bigg\}
+ {\cal O}(N^{2})\,, \nn\\
P_{\,\rm qg}^{\,(1)} \!&\!=\!&\! 
\nf\*\bigg\{(4\*\cf-8\*\ca)\*N^{-3}-(6\*\cf+4\*\ca)\*N^{-2}+([28-8\*\z2]\*\cf-8\*\ca)\*N^{-1}\bigg\}
+ {\cal O}(N^{0})\,, \nn\\
P_{\,\rm qg}^{\,(2)} \!&\!=\!&\! 
	\nf\*\bigg\{(64\*\cas -32\*\ca\*\cf + 16\*\cfs - 32\*\cf\*\nf)\*N^{-5}
\nn\\&&\hphantom{\nf\bigg\{}\mbox{\hspn}
	+ \bigg(\frct{56}{3}\*\cas-\frct{44}{3}\*\cf\*\ca-12\*\cfs+\frct{16}{3}\*\ca\*\nf
	+\frct{152}{3}\*\cf\*\nf\bigg)\*N^{-4}
\nn\\&&\hphantom{\nf\bigg\{}\mbox{\hspn}
	+ \bigg(\bigg[\frct{1724}{9}-12\*\z2\bigg]\*\cas-\bigg[\frct{1171}{9}
	-96\*\z2\bigg]\*\ca\*\cf+[89-56\*\z2]\*\cfs+\frct{64}{9}\*\ca\*\nf
\nn\\&&\hphantom{\nf\bigg\{}\mbox{\hspn}
	-\frct{1370}{9}\*\cf\*\nf\bigg)\*N^{-3}\bigg\}
+ {\cal O}(N^{-2})\,, \nn\\
P_{\,\rm gq}^{\,(0)} \!&\!=\!&\! 
\cf\*\bigg\{-4\*N^{-1} - 2 -6\*N \bigg\}
+ {\cal O}(N^{2})\,, \nn\\
P_{\,\rm gq}^{\,(1)} \!&\!=\!&\!
\cf\*\bigg\{(16\*\ca-8\*\cf)\*N^{-3} + (16\*\ca-8\*\cf)\*N^{-2}
\nn\\&&\hphantom{\cf\bigg\{}\mbox{\hspn}
+ \bigg(14\*\cf+\frct{128}{9}\*\nf-\bigg[\frct{332}{9}-16\*\z2\bigg]\*\ca\bigg)\*N^{-1}\bigg\}
+ {\cal O}(N^{0})\,, \nn\\
P_{\,\rm gq}^{\,(2)} \!&\!=\!&\! 
	\cf\*\bigg\{(-128\*\cas+64\*\ca\*\cf-32\*\cfs+64\*\cf\*\nf)\*N^{-5}
\nn\\&&\hphantom{\cf\bigg\{}\mbox{\hspn}
	+\bigg(-\frct{400}{3}\*\cas+\frct{376}{3}\*\cf\*\ca
	-48\*\cfs
	-\frct{32}{3}\*\ca\*\nf-\frct{16}{3}\*\cf\*\nf\bigg)\*N^{-4}
\nn\\&&\hphantom{\cf\bigg\{}\mbox{\hspn}
	+ \bigg(-\bigg[\frct{280}{9}+104\*\z2\bigg]\*\cas
	-\frct{2446}{9}\*\ca\*\cf
	+[42+48\*\z2]\*\cfs
	-\frct{992}{9}\*\ca\*\nf
\nn\\&&\hphantom{\cf\bigg\{}\mbox{\hspn}
	+\frct{2380}{9}\*\cf\*\nf\bigg)\*N^{-3}\bigg\}
+ {\cal O}(N^{-2})\,, \nn\\
P_{\,\rm gg}^{\,(0)} \!&\!=\!&\! {}
-4\*\ca\*N^{-1} + \bigg(\frct{5}{3}\*\ca-\frct{2}{3}\*\nf\bigg) - [7+4\*\z2]\*\ca\*N
+ {\cal O}(N^{2})\,, \nn\\
P_{\,\rm gg}^{\,(1)} \!&\!=\!&\!
(16\*\cas-8\*\cf\*\nf)\*N^{-3} + \bigg(\frct{4}{3}\*\cas+\frct{8}{3}\*\ca\*\nf+12\*\cf\*\nf\bigg)\*N^{-2}
\nn\\&&{}
+ \bigg(\bigg[\frct{74}{9}+16\*\z2\bigg]\*\cas+\frct{76}{9}\*\ca\*\nf
-32\*\cf\*\nf\bigg)\*N^{-1}
+ {\cal O}(N^{0})\,, \nn\\
P_{\,\rm gg}^{(2)} \!&\!=\!&\! 
	(-128\*\cath+128\*\cf\*\ca\*\nf-32\*\cfs\*\nf)\*N^{-5}
\nn\\&&{}
	+\bigg(-16\*\cath-32\*\cas\*\nf-\frct{232}{3}\*\ca\*\cf\*\nf
	+ 24\*\cfs\*\nf 
	+ \frct{16}{3}\*\cf\*\nfs \bigg)\*N^{-4}
\nn\\&&{}
	+\bigg(-\bigg[\frct{2612}{9}+160\*\z2\bigg]\*\cath
	- \bigg[\frct{208}{3}+24\*\z2\bigg]\*\cas\*\nf
	+ \bigg[\frct{3548}{9}+96\*\z2\bigg]\*\ca\*\cf\*\nf
\nn\\&&{}
	-[120-32\*\z2]\*\cfs\*\nf-\frct{16}{9}\*\ca\*\nfs+\frct{184}{9}\*\cf\*\nfs \bigg)\*N^{-3}
+ {\cal O}(N^{-2})\,.
\end{eqnarray}
The corresponding expansion coefficients of the $D$-dimensional coefficient
functions for $F_2$ are
\begin{eqnarray}
c_{2,\rm q}^{\,(1,l)} \!&\!=\!&\! c_2^{+(1,l)}\,,\qquad (l=0,1,2)\,, \nn \\
c_{2,\rm q}^{\,(2,0)} \!&\!=\!&\! c_2^{\,+(2,0)} 
+ \nf\*\cf\*\bigg\{-20\*N^{-4}-2\*N^{-3}-[56-16\*\z2]\*N^{-2}\bigg\}
+ {\cal O}(N^{-1}) \nn\\
c_{2,\rm q}^{\,(2,1)} \!&\!=\!&\! c_2^{\,+\,(2,1)} +
 \nf\*\cf\*\bigg\{52\*N^{-5}+2\*N^{-4}+[160-56\*\z2]\*N^{-3}\bigg\}
+ {\cal O}(N^{-2}) \nn\\
c_{2,\rm q}^{\,(3,0)} \!&\!=\!&\! c_2^{+(3,0)} +
	\nf\*\cf\*\bigg\{240\*(\ca-\cf)\*N^{-6}
	+ \bigg(\frct{3416}{9}\*\ca-\frct{440}{3}\*\cf-\frct{368}{9}\*\nf\bigg)\*N^{-5}
\nn\\&&\hphantom{c_2^{+(3,0)} + \nf\,\*\cf\bigg\{}\mbox{\hspn}
	+ \bigg(\bigg[\frct{16984}{27}
	-\frct{320}{3}\*\z2\bigg]\*\ca
	-\bigg[572-\frct{1328}{3}\*\z2\bigg]\*\cf
	+\frct{1784}{27}\*\nf\bigg)\*N^{-4}\bigg\}
\nn\\&&\hphantom{c_2^{+(3,0)} +}{}
+ {\cal O}(N^{-3})\,.
\end{eqnarray}
and
\begin{eqnarray}
c_{2,\rm g}^{\,(1,0)}\!&\!=\!&\!
\nf\*\bigg\{2\*N^{-2}-2\*N^{-1}+[6-2\*\z2]\bigg\}
+ {\cal O}(N)\,, \nn\\
c_{2,\rm g}^{\,(1,1)}\!&\!=\!&\!
\nf\*\bigg\{-2\*N^{-3}+2\*N^{-2}-[6-3\*\z2]\*N^{-1}\bigg\}
+ {\cal O}(N^0)\,, \nn\\
c_{2,\rm g}^{\,(1,2)}\!&\!=\!&\!
\nf\*\bigg\{2\*N^{-4}-2\*N^{-3}+[6-3\*\z2]\*N^{-2}\bigg\}
+ {\cal O}(N^{-1})\,, \nn\\
c_{2,\rm g}^{(2,0)}\!&\!=\!&\!
	\nf\*\bigg\{(10\*\cf-20\*\ca)\*N^{-4}
	-(2\*\ca+3\*\cf)\*N^{-3}
	+([-58+8\*\z2]\*\ca
\nn\\&&\hphantom{\nf\bigg\{}\mbox{\hspn}
	+[16-16\*\z2]\*\cf)\*N^{-2} \Big\}
+ {\cal O}(N^{-1})\,, \nn\\
c_{2,\rm g}^{\,(2,1)}\!&\!=\!&\!
	\nf\*\bigg\{(52\*\ca-26\*\cf)\*N^{-5}
	+(2\*\ca+3\*\cf)\*N^{-4}
	+([166-32\*\z2]\*\ca
\nn\\&&\hphantom{\nf\bigg\{}\mbox{\hspn}
	-[20-44\*\z2]\*\cf)\*N^{-3} \Big\}
+ {\cal O}(N^{-2})\,, \nn\\
c_{2,\rm g}^{\,(3,0)}\!&\!=\!&\!
	\nf\*\bigg\{(240\*\cas-120\*\cf\*\ca+60\*\cfs-120\*\cf\*\nf)\*N^{-6}
\nn\\&&\hphantom{\nf\bigg\{}\mbox{\hspn}
	+\bigg(\frct{1436}{9}\*\cas-\frct{1636}{9}\*\cf\*\ca+\frct{44}{3}\*\cfs-\frct{8}{9}\*\ca\*\nf
	+\frct{1636}{9}\*\cf\*\nf\bigg)\*N^{-5}
\nn\\&&\hphantom{\nf\bigg\{}\mbox{\hspn}
	+\bigg(\bigg[\frct{27338}{27}-56\*\z2\bigg]\*\cas
	+\bigg[-\frct{4589}{27}+\frct{656}{3}\*\z2\bigg]\*\cf\*\ca
	+\bigg[\frct{178}{3}-\frct{524}{3}\*\z2\bigg]\*\cfs
\nn\\&&\hphantom{\nf\bigg\{}\mbox{\hspn}
	+ \frct{532}{27}\*\ca\*\nf +\bigg[-\frct{17782}{27}+88\*\z2\bigg]\*\cf\*\nf \bigg)\*N^{-4} \bigg\}
+ {\cal O}(N^{-3})\,.
\end{eqnarray}
The input coefficients $c_{\phi,\rm i}^{}$ for $\fhat{\phi}$, which provides
the resummation of $P_{\,\rm qg}$ and $P_{\,\rm gg}$, are given by
\begin{eqnarray}
c_{\phi,\rm q}^{\,(1,0)}\!&\!=\!&\!
\cf\*\bigg\{-4\*N^{-2}-4\*N^{-1}+[5+4\*\z2]\bigg\}
+ {\cal O}(N)\,, \nn\\
c_{\phi,\rm q}^{\,(1,1)}\!&\!=\!&\!
\cf\*\bigg\{4\*N^{-3}+4\*N^{-2}+[1-6\*\z2]\*N^{-1}\bigg\}
+ {\cal O}(N^0)\,, \nn\\
c_{\phi,\rm q}^{\,(1,2)}\!&\!=\!&\!
\cf\*\bigg\{-4\*N^{-4}-4\*N^{-3}-[1-6\*\z2]\*N^{-2}\bigg\}
 + {\cal O}(N^{-1})\,, \nn\\
c_{\phi,\rm q}^{\,(2,0)}\!&\!=\!&\!
	\cf\*\bigg\{(40\*\ca-20\*\cf)\*N^{-4}
	+\bigg(\frct{344}{3}\*\ca-28\*\cf-\frct{32}{3}\*\nf\bigg)\*N^{-3}
\nn\\&&\hphantom{\cf\bigg\{}\mbox{\hspn}
	+([16-16\*\z2]\*\ca+[21+32\*\z2]\*\cf
	+ 12\*\nf)\*N^{-2} \bigg\}
+ {\cal O}(N^{-1})\,, \nn\\
c_{\phi,\rm q}^{\, (2,1)}\!&\!=\!&\!
	\cf\*\bigg\{(-104\*\ca+52\*\cf)\*N^{-5}+(-328\*\ca+76\*\cf+32\*\nf)\*N^{-4}
\nn\\&&\hphantom{\cf\bigg\{}\mbox{\hspn}
	+\bigg(\bigg[-\frct{1196}{9}+80\*\z2\bigg]\*\ca
	-[25+104\*\z2]\*\cf -\frct{196}{9}\*\nf\bigg)\*N^{-3} \bigg\}
+ {\cal O}(N^{-2})\,, \nn\\
c_{\phi,\rm q}^{\,(3,0)}\!&\!=\!&\!
	\cf\*\bigg\{(-480\*\cas+240\*\cf\*\ca-120\*\cfs+240\*\cf\*\nf)\*N^{-6}
\nn\\&&\hphantom{\cf\bigg\{}\mbox{\hspn}
	+\bigg(-\frct{13960}{9}\*\cas+\frct{8960}{9}\*\cf\*\ca
	-224\*\cfs+\frct{1072}{9}\*\ca\*\nf-\frct{440}{9}\*\cf\*\nf\bigg)\*N^{-5}
\nn\\&&\hphantom{\cf\bigg\{}\mbox{\hspn}
	+\bigg(\bigg[-\frct{69928}{27}+\frct{208}{3}\*\z2\bigg]\*\cas+\bigg[\frct{2338}{27}
	-\frct{1120}{3}\*\z2\bigg]\*\cf\*\ca+\bigg[\frct{308}{3}+328\*\z2\bigg]\*\cfs
\nn\\&&\hphantom{\cf\bigg\{}\mbox{\hspn}
	+ \frct{6592}{27}\*\ca\*\nf +\bigg[\frct{17456}{27}-176\*\z2\bigg]\*\cf\*\nf
	-32\*\nfs\bigg)\*N^{-4} \bigg\}
+ {\cal O}(N^{-3})
\end{eqnarray}
and
\begin{eqnarray}
c_{\phi,\rm g}^{\,(1,0)}\!&\!=\!&\!
-4\*\ca\*N^{-2} -\bigg(\frct{23}{3}\*\ca-\frct{2}{3}\*\nf\bigg)\*N^{-1} + \bigg(\bigg[\frct{118}{9}+4\*\z2\bigg]\*\ca-\frct{16}{9}\*\nf\bigg)
+ {\cal O}(N)\,, \nn\\
c_{\phi,\rm g}^{\,(1,1)}\!&\!=\!&\!
4\*\ca\*N^{-3} +\bigg(\frct{23}{3}\*\ca-\frct{2}{3}\*\nf\bigg)\*N^{-2} + \bigg(\bigg[-\frct{64}{9}-6\*\z2\bigg]\*\ca+\frct{16}{9}\*\nf\bigg)\*N^{-1}
 + {\cal O}(N^0)\,, \nn\\
c_{\phi,\rm g}^{\,(1,2)}\!&\!=\!&\!
-4\*\ca\*N^{-4} +\bigg(-\frct{23}{3}\*\ca+\frct{2}{3}\*\nf\bigg)\*N^{-3} + \bigg(\bigg[\frct{64}{9}+6\*\z2\bigg]\*\ca-\frct{16}{9}\*\nf\bigg)\*N^{-2}
 + {\cal O}(N^{-1})\,, \nn\\
c_{\phi,\rm g}^{\,(2,0)}\!&\!=\!&\!
(40\*\cas - 20\*\cf\*\nf)\*N^{-4} +(78\*\cas - 4\*\ca\*\nf + 14\*\cf\*\nf)\*N^{-3}+ \bigg( \frct{833}{9}\*\cas - \frct{22}{9}\*\ca\*\nf 
\nn\\&&
+[-34+ 16\*\z2]\*\cf\*\nf + \frct{8}{9}\*\nfs\bigg)\*N^{-2}
+ {\cal O}(N^{-1})\,, \nn\\
c_{\phi,\rm g}^{\,(2,1)}\!&\!=\!&\!
	(-104\*\cas + 52\*\cf\*\nf)\*N^{-5}
	+ \bigg(-\frct{698}{3}\*\cas + \frct{44}{3}\*\ca\*\nf - 30\*\cf\*\nf \bigg)\*N^{-4}
\nn\\&&{}
	+\bigg(\bigg[-\frct{2857}{9} 
	+ 32\*\z2\bigg]\*\cas + \frct{142}{9}\*\ca\*\nf + [94 - 56\*\z2]\*\cf\*\nf - \frct{8}{3}\*\nfs \bigg)\*N^{-3}
 + {\cal O}(N^{-2})\,, \nn\\
c_{\phi,\rm g}^{\,(3,0)}\!&\!=\!&\! 
	-120\*\bigg(4\*\cath - \cf\*\nf\*(4\*\ca - \cf)\bigg)\*N^{-6}+\bigg(-\frct{10000}{9}\*\cath
	+ \frct{352}{9}\*\nf\*\cas + \frct{3140}{9}\*\cf\*\ca\*\nf
\nn\\&&{}
	 + \frct{44}{3}\*\nf\*\cfs - \frct{536}{9}\*\nfs\*\cf \bigg)\*N^{-5}
	 +\bigg(  -\bigg[\frct{59902}{27} + 224\*\z2\bigg]\*\cath
	 + \bigg[\frct{560}{27} - 48\*\z2\bigg]\*\nf\*\cas
\nn\\&&{}
	+ \bigg[\frct{16622}{27} - \frct{256}{3}\*\z2\bigg]\*\nf\*\cf\*\ca
	-\bigg[162- \frct{616}{3}\*\z2\bigg]\*\nf\*\cfs - \frct{328}{27}\*\nfs\*\ca
	+ \frct{3508}{27}\*\nfs\*\cf \bigg)\*N^{-4}
\nn\\&&\vphantom{\bigg\{}{}
+ {\cal O}(N^{-3})\,. 
\end{eqnarray}
Finally the input coefficients of $c_{L,\rm q}^{}$ and $c_{L,\rm g}^{}$ read
\begin{eqnarray}
c_{L,\rm q}^{\,(1,l)} \!&\!=\!&\! c_L^{+(1,l)}\,,\qquad (l=0,1,2)\,, \nn \\
c_{L,\rm q}^{\,(2,0)} \!&\!=\!&\! c_L^{+(2,0)} 
+\nf\,\*\cf\*\bigg\{- 16\*N^{-2} + \bigg(\frct{144}{9}+ 16\*\z2\bigg)\Big\}
+ {\cal O}(N)\,, \nn\\
c_{L,\rm q}^{(2,1)} \!&\!=\!&\! c_L^{+(2,1)}
+\nf\,\*\cf\*\bigg\{16\*N^{-3} - \frct{96}{3}\*N^{-2}+ \bigg[\frct{648}{9} - 40\*\z2\bigg]\*N^{-1}\bigg\}
 + {\cal O}(N^0)\,, \nn\\
c_{L,\rm q}^{(3,0)} \!&\!=\!&\! c_L^{+(3,0)} 
	+\nf\,\*\cf\*\bigg\{(160\*\ca - 160\*\cf)\*N^{-4}
	+ \bigg(\frct{496}{3}\*\ca -16\*\cf - \frct{64}{3}\*\nf \bigg)\*N^{-3}
\nn\\&&\hphantom{c_L^{+(3,0)} + \nf\*\cf\*\bigg\{}{} \mbox{\hspn}
	+ \bigg(\bigg[\frct{400}{9} - 112\*\z2\bigg]\*\ca -[80 - 256\*\z2]\*\cf  + \frct{512}{9}\*\nf\bigg)\*N^{-2}\bigg\}
\nn\\&&\hphantom{c_L^{+(3,0)} +}\vphantom{\bigg\{}{}
+ {\cal O}(N^{-1})\,.
\end{eqnarray}
and
\begin{eqnarray}
c_{L,\rm g}^{\,(1,0)} \!&\!=\!&\!
\nf\*\bigg\{4-6\*N+7\*N^{2}\bigg\}
+ {\cal O}(N^{3})\,, \nn\\
c_{L,\rm g}^{\,(1,1)} \!&\!=\!&\!
\nf\*\bigg\{8 - [12-4\*\z2]\*N\bigg\}
+ {\cal O}(N^{2})\,, \nn\\
c_{L,\rm g}^{\,(1,2)} \!&\!=\!&\!
\nf\*\bigg\{[16-2\*\z2]\bigg\}
+ {\cal O}(N)\,, \nn\\
c_{L,\rm g}^{\,(2,0)} \!&\!=\!&\!
\nf\*\bigg\{-(16\*\ca-8\*\cf)\*N^{-2}- 8\*\cf\*N^{-1}+ ([16 + 16\*\z2]\*\ca - [4 + 8\*\z2]\*\cf) \bigg\} 
+ {\cal O}(N)\,, \nn\\
c_{L,\rm g}^{\,(2,1)} \!&\!=\!&\!
	\nf\*\bigg\{(16\*\ca-8\*\cf)\*N^{-3}-(32\*\ca-16\*\cf)\*N^{-2}
\nn\\&&\hphantom{\nf\*\bigg\{}{} \mbox{\hspn}
	+([72 - 40\*\z2]\*\ca - [12 - 20\*\z2]\*\cf)\*N^{-1}\bigg\}
+ {\cal O}(N^0)\,, \nn\\
c_{L,\rm g}^{\,(3,0)} \!&\!=\!&\!
	\nf\*\bigg\{(160\*\cas - 80\*\cf\*\ca + 40\*\cfs - 80\*\nf\*\cf)\*N^{-4}
\nn\\&&\hphantom{\nf\*\bigg\{}{} \mbox{\hspn}
	+ \bigg( \frct{56}{3}\*\cas - \frct{152}{3}\*\cf\*\ca - 20\*\cfs + \frct{16}{3}\*\nf\*\ca
	+ \frct{464}{3}\*\nf\*\cf\bigg)\*N^{-3}
\nn\\&&\hphantom{\nf\*\bigg\{}{} \mbox{\hspn}
	+ \bigg( \bigg[\frct{3640}{9}-120\*\z2\bigg]\*\cas + \bigg[\frct{308}{9} + 144\*\z2\bigg]\*\cf\*\ca
	- [16 + 96\*\z2]\*\cfs 
\nn\\&&\hphantom{\nf\*\bigg\{}{} \mbox{\hspn}
    + \frct{80}{9}\*\nf\*\ca
	-\bigg[\frct{3416}{9} - 64\*\z2\bigg]\*\nf\*\cf\bigg)\*N^{-2}\bigg\}
+ {\cal O}(N^{-1})\,.
\end{eqnarray}


\renewcommand{\theequation}{B.\arabic{equation}}
\setcounter{equation}{0}
\section*{Appendix B: hypergeometric functions for the 
non-singlet coefficient functions}


The hypergeometric functions relevant for the non-singlet 
$x$-space coefficient functions are
\bea
\label{cns-x}
m\,a\,\int_0^1\! dx \: x^{\,N-1}\ln \frct{1}{x}\phantom{.}_1F_2\bigg(\frct{m}{4}+1;\,2,\frct{3}{2}\,;\,a\,\ln^2 \frct{1}{x}\bigg)\, &=& \bigg(1-\frct{4\,a}{N^2}\bigg)^{-m/4}-1
\; , \nn\\[0.5mm]
m\,a\,\int_0^1\! dx \: x^{\,N-1}\phantom{.}_1F_2\bigg(\frct{m}{4}+1;\,2,\frct{1}{2}\,;\,a\,\ln^2 \frct{1}{x}\bigg) &=& N\bigg\{\bigg(1-\frct{4\,a}{N^2}\bigg)^{-m/4}-1\bigg\}
\; , \nn\\[1mm]
\frct{1}{2}\,m(m+4)\,a^2\,\int_0^1\! dx \: x^{\,N-1}\ln \frct{1}{x}\phantom{.}_1F_2\bigg(\frct{m}{4}+2;\,3,\frct{3}{2}\,;\,a\,\ln^2 \frct{1}{x}\bigg) &=& N^2\bigg\{\bigg(1-\frct{4\,a}{N^2}\bigg)^{-m/4}-1-\frct{m\,a}{N^2}\bigg\}
\nn\\[1mm]
\eea
with $a \;=\; 2\,\cf\,\ar$.


\renewcommand{\theequation}{C.\arabic{equation}}
\setcounter{equation}{0}
\section*{Appendix C: leading-logarithmic coefficient functions
for $\bm F_\phi$}


In this final appendix, we present some analytic results for the 
scalar-exchange coefficient functions $C_{\phi,\rm q}$ and $C_{\phi\,g}$.
Using the decomposition 
\beq
  C_{\phi,\rm g}(N) \;=\; C_{\phi}^{+}(N) + C_{\phi,\rm g}^{\,\rm ps}(N)
\; , 
\eeq
which is analogous to that of the gluon-gluon splitting function in the
second line of eq.~(\ref{Pii-ps}), the $F_\phi$ counterparts of the LL
expressions (\ref{eq:S_c2q_LL}) -- (\ref{eq:S_cLg_LL}) for $C_2$ and $C_L^{}$
are given by
\bea
  C_{\phi}^{\,+\,(n)}(N) \!&\!=\!&\! \frac{(-4\,\ca)^n}{N^{\:\!2n}}\,\dat{n}
\;, \\ 
  C_{\phi,\rm g}^{\,\rm ps\,(n)}(N) \!&\!=\!&\!
  \dat{n}\,\frac{2^{\:\!n}}{N^{\:\!2n}}
  \sum_{i=0}^{\floor{\frac{n-2}{2}}}\sum_{k=0}^{n-2-2i}
  (-2)^{i+1+k}(\nf\cf)^{i+1}\cax{k}\cfx{\rho'}
  \binom{k+i+1}{k}\binom{\rho'+i}{\rho'}
\;, 
\eea
and 
\bea
\label{eq:S_cphi_LL}
  C_{\phi,\rm q}^{(n)}(N) \;\; &\!=\!&\!
  -\cf\dat{n}\,\frac{2^{\:\!n+1}}{N^{\:\!2n}}
  \sum_{i=0}^{\floor{\frac{n-1}{2}}}\sum_{k=0}^{n-1-2i}
  (-2)^{i+k}(\nf\cf)^i\cax{k}\cfx{\delta'}\binom{k+i}{k}
  \binom{\delta'+i}{\delta'}
\; , \quad
\eea
where $\delta'$ and $\rho'$ have been given below eq.~(\ref{eq:S_cLg_LL}). 

Unlike the splitting functions and coefficient functions for $F_2$ and $F_L$, 
the coefficient functions for $\phi$-exchange DIS exhibit double logarithms 
also in BFKL-limit, i.e., they include contributions of the form
$\asn \, x^{\,-1} \ln^{\,2n - n_0^{} - k} x\,$.
We have considered these in what might be called `extended quantum
gluodynamics' (eQGD), i.e., QCD in the limit $C_F = 0$, and found
an analogue to eq.~(\ref{FhatAllN}) for the expansion about $N=1$
which generates the resummation of these double logarithms.

In terms of
\beq
  S'' \; =\; (1-4\,\xi'')^{-1} 
\;\; \mbox{ with }\;\; 
  \xi'' \;=\; \ca\ar/\Nb^2 
\;\; \mbox{ where }\;\; 
  \Nb \:\equiv\: N-1
\; , 
\eeq
the resummed coefficient function, including the finite NNLL contributions
at $\mathcal{O}(\ar)$, is found to~be
\bea
C_{\phi}(\Nb)\big|_{\cf=0}\!&\!=\!&\! 
   (S''-1) + \frct{1}{12\,\ca}\Nb\bigg\{ 44\,\ca(S''-1)+(3\,\bo-22\,\ca)(S''^2-1)-3\,\bo(S''^3-1)\bigg\}\nonumber\\
&&\mbox{\hspn}
+\frct{1}{3\,\ca}\,\ar\bigg\{\ca\bigg(5\,\bo + 4\,\ca[1 - 3\,\z2]\bigg) + \ca\bigg(\ca\bigg[\frct{653}{6}-12\,\z2\bigg]-\bo\bigg)(S''-1)\nonumber\\
&&+ \ca\bigg(\frct{23}{2}\,\bo - \ca\bigg[\frct{201}{2}+12\,\z2\bigg]\bigg)(S''^2-1)
\nonumber\\&&
+ \bigg(\frct{121}{3}\,\cas-22\,\ca\,\bo + \frct{3}{4}\,\bo^2\bigg)(S''^3-1)\nonumber\\
&&+\bo\bigg(\frct{33}{2}\,\ca-3\,\bo\bigg)(S''^4-1)\,+\,\frct{9}{4}\,\bo^2(S''^5-1)\bigg\}
\; ,
\eea
where the two terms in the first line provide the LL and NLL contributions,
and the remaining four lines the NNLL result. This resummation does, of 
course, also return the corresponding $x^{\:\!-1}$ double-logarithms in 
$P_{\,\rm gg}\big|_{\cf=0}$. These are found to vanish, as they have to,
see refs.~\cite{Jarosz82,CataniFM90,CataniH94}.


%
{\small
\setlength{\baselineskip}{0.35cm}

}

\end{document}